\documentclass[titlepage,11pt]{article}
\usepackage[USenglish]{babel}
\usepackage{a4wide}
\usepackage[all]{xy}
\usepackage{amssymb}
\usepackage{amsmath}
\usepackage{latexsym}
\usepackage{pifont}
\usepackage{pdfsync}
\usepackage{color}
\usepackage{enumerate}
\usepackage[parfill]{parskip}    
\usepackage{xr}
\usepackage{makeidx}
\usepackage[utf8]{inputenc} 
\usepackage{ifthen}
\title{Sets, the Axiom of Choice, \emph{and all that}: A Tutorial}
\author{Ernst-Erich Doberkat\\Chair for Software
  Technology\\Technische Universität Dortmund\\\texttt{ernst-erich.doberat@udo.edu}}
\date{\today}
\newcommand{\labelImpl}[2]{\ensuremath{\ref{#1}~\Rightarrow~\ref{#2}}}

\newcommand{\Klasse}[2]{\left[#1\right]_{#2}}
\newcommand{\Faktor}[2]{{#1}/{#2}}
\newcommand{\fMap}[1]{\eta_{#1}}
\newcommand{\Bild}[2]{{#1}\left[#2\right]}
\newcommand{\InvBild}[2]{\Bild{#1^{-1}}{#2}}

\newcommand{\Folge}[1]{(#1_n)_{n \in \Nat}}

\newcommand{\spaceFont}[1]{\mathfrak{#1}}

\newcommand{\SubProbSenza}{\spaceFont{S}}
\newcommand{\ProbSenza}{\spaceFont{P}}
\newcommand{\PowerSet}[1]{\ensuremath{\mathcal{P}\left(#1\right)}}

\newcommand{\Closure}[1]{\ensuremath{\mathsf{cl}(#1)}}

\newtheorem{definition}{Definition}[section]
\newcommand{\BeginDefinition}[1]{%
  \begin{definition}\label{#1}
}
\newcommand{\EndDefinition}{\end{definition}}

\newtheorem{example}[definition]{Example}
\newcommand{\BeginExample}[1]{%
  \begin{example}\label{#1}\rm
}
\newcommand{\EndExample}{--- \end{example}}

\newtheorem{observation}[definition]{Observation}
\newcommand{\BeginObservation}[1]{
  \begin{observation}\label{#1}\rm
}
\newcommand{\EndObservation}{--- \end{observation}}

\newtheorem{theorem}[definition]{Theorem}
\newcommand{\BeginTheorem}[1]{%
  \begin{theorem}\label{#1}
}
\newcommand{\EndTheorem}{\end{theorem}}

\newtheorem{corollary}[definition]{Corollary}
\newcommand{\BeginCorollary}[1]{
  \begin{corollary}\label{#1}
}

\newtheorem{proposition}[definition]{Proposition}
\newcommand{\BeginProposition}[1]{%
  \begin{proposition}\label{#1}
}
\newcommand{\EndProposition}{\end{proposition}}

\newcommand{\EndCorollary}{\end{corollary}}
\newtheorem{lemma}[definition]{Lemma}
\newcommand{\BeginLemma}[1]{%
  \begin{lemma}\label{#1}
}
\newcommand{\EndLemma}{\end{lemma}}

\newtheorem{claim}{Claim}
\newcommand{\BeginClaim}[1]{%
  \begin{claim}\label{#1}
}
\newcommand{\EndClaim}{\end{claim}}

\newenvironment{proof}{\textbf{Proof\ }}{\ensuremath{\QED}}
\newcommand{\BeginProof}{\begin{proof}}
\newcommand{\EndProof}{\end{proof}}

\newenvironment{remark}{\textbf{Remark:\ }}{}
\newcommand{\BeginRemark}{\begin{remark}}
\newcommand{\EndRemark}{\QED\end{remark}}
\newcommand{\QED}{%
\ensuremath{\dashv}
}

\newcommand{\Real}{\mathbb{R}}
\newcommand{\pReal}{\mathbb{R}_{+}}
\newcommand{\Nat}{\mathbb{N}}
\newcommand{\Rational}{\mathbb{Q}}

\def\theta{\vartheta}
\def\dots{\ldots}

\def\Ganz{\mathbb{Z}}
\makeatletter
\def\@axx#1{\ensuremath{\mathbb{(#1)}}}
\def\AC{\@axx{AC}}
\def\WO{\@axx{WO}}
\def\ZL{\@axx{ZL}}
\def\MP{\@axx{MP}}
\def\MI{\@axx{MI}}
\def\AD{\@axx{AD}}
\makeatother
\def\Natt{\ensuremath{{\mathcal N}}}
\def\Natn{\Nat_0}

\def\Seq{\ensuremath{\mathcal{S}}}
\def\Seqg{\Seq_g}
\def\Sequ{\Seq_u}
\newcommand{\isEquiv}[3]{\ensuremath{{#1}\ {#3}\ {#2}}}
\newenvironment{theExercises}
{\begin{exercise}\rm}
{\end{exercise}}
\newtheorem{exercise}{Exercise}
\newcommand{\BeginExercise}[1]{%
  \begin{theExercises}\label{#1}
}
\newcommand{\EndExercise}{\end{theExercises}}
\newcommand{\Exercise}[2]{\BeginExercise{#1}{#2}\EndExercise}
\newcommand{\Solution}[2]{\paragraph{Solution for
    Exercise~\ref{#1}}{#2}}
\def\endEx{{\Large\ding{44}}}
\renewcommand{\EndExample}{\endEx \end{example}}
\def\diam#1{\mathrm{diam}(#1)}
\def\frown{\smallfrown}
\def\smallBox#1{\parbox[t]{.49\linewidth}{\vspace{1pt}#1}}

\def\AxiomBox#1#2{\setlength{\fboxrule}{1.5pt}
  \begin{center}\par
    \fbox{\begin{minipage}{.6\linewidth}
        \begin{center}
          \ensuremath{\mathbb{(#1)}} #2
        \end{center}
      \end{minipage}}
  \end{center}

}
\def\CatFont{\mathbf}

\renewcommand{\spaceFont}[1]{\CatFont{#1}}

\makeatletter
\def\@theProbSenza#1{\ensuremath{\mathbb{#1}}}
\def\SubProbSenza{\@theProbSenza{S}}
\def\ProbSenza{\@theProbSenza{P}}
\def\FinSenza{\@theProbSenza{M}}
\def\SigmaSenza{\@theProbSenza{M}_{\ensuremath{\sigma}}}
\makeatother

\makeatletter
\def\@beta{\pmb{\beta}}
\def\betaSenza#1{\@beta_{#1}}

\makeatother

\def\DefSect{
\ifthenelse{\boolean{isBook}}{
\def\Section{\chapter}
\def\Subsection{\section}
\def\Subsubsection{\subsection}
}{
\def\Section{\section}
\def\Subsection{\subsection}
\def\Subsubsection{\subsubsection}
}
}

\def\Rand#1{\begin{color}{red}\fbox{#1}\end{color}}
\def\Closure#1{\ensuremath{{#1}^{a}}}
\def\Interior#1{\ensuremath{{#1}^{o}}}
\makeatletter
\def\@Tut#1#2{\cite[#1]{#2}}
\def\CategCite#1{\@Tut{#1}{EED-Categs}}
\def\SetCite#1{\@Tut{#1}{EED-Tut_sets}}
\makeatother

\makeatletter
\def\@norm#1#2#3{\ensuremath{||{#1}||_{#2}^{#3}}}
\def\infNorm#1#2{\@norm{#1}{\infty}{#2}}
\def\aNorm#1{\@norm{#1}{}{}}
\makeatother

\makeatletter
\def\@conv#1{\stackrel{#1}{\longrightarrow}} 
\def\aeC{\@conv{a.e.}} 
\def\nmC{\@conv{i.m.}}
\makeatother

\makeatletter
\newcommand{\@theL}[3]{\ensuremath{{#3}_{#1}{#2}}}
\newcommand{\cLp}[2][p]{\@theL{#1}{(#2)}{{\cal L}}}
\newcommand{\rLp}[2][p]{\@theL{#1}{(#2)}{L}}
\newcommand\cLpS[1][p]{\@theL{#1}{}{{\cal L}}}
\newcommand\rLpS[1][p]{\@theL{#1}{}{L}}
\makeatother

\renewcommand{\Bild}[2]{{#1}\bigl[#2\bigr]}
\renewcommand{\Folge}[2][n]{\ensuremath{({#2}_{#1})_{#1\in\Nat}}}

\newboolean{isBook}
\setboolean{isBook}{false}
\DefSect
\def\Rand#1{}
\def\phi{\varphi}
\usepackage{fancyhdr}
\makeindex
\begin{document}
\maketitle
\begin{abstract}
  This tutorial deals with the application of the Axiom of Choice in
  one of its popular disguises to objects which are of some interest in
  computer science (like lattices, Boolean algebras, filters and
  ideals, games). We discuss some common variants of this axiom such
  as Zorn's Lemma, Tuckey's Maximality Principle, the Well-Ordering
  Theorem and the Maximal Ideal Theorem; each equivalence gives
  applications its due attention. We show that the Axiom of Choice can
  be used to demonstrate the existence of non-measurable sets in the
  real line. This is an occasion to introduce some measure theory
  within the context of Boolen $\sigma$-algebras. 

  Games are introduced as well, and the Axiom of Determinacy is
  discussed, giving rise to show that this axiom can be used to demonstrate
  that each subset of the real line is measurable. Hence we use games
  as a tool for proofs. We try to shed some light on the slightly
  complicated and irritating interplay between these two axioms.

  We assume the basic knowledge of mathematics that is introduced by a
  one year course for beginning computer scientists at a German
  university. A grain of mathematical maturity may help as well. Some
  exercises  are offered, and solutions are suggested as
  well.
\end{abstract}

\newpage
\tableofcontents\newpage
\def\Folder{Sets}
\Section{The Axiom of Choice and Some Of Its Equivalents}
\label{sec:axiom-choice-its}

Sets are a universal tool for computer scientists, the tool which has
been imported as a \emph{lingua franca} from mathematics. Program
development, for example, starts sometimes from a mathematical
description of the things to be done, the specification and the data
structures and --- you guess it --- sets are the language, in which
these first designs are usually written down. There is even a programming
language called \texttt{SETL} based on sets~\cite{SETL-NYU}, this language served as a
prototyping tool and was essentially motivated by the ambition to make
the road from a formal description of an object to its 
representation through an executable program as short as possible~\cite{EED-Fox}. 

In fact, it turned out that programming in what might be called
executable set theory has the advantage of
being able to experiment with the objects at hand, leading, for
example to the first implementation of the programming language
\texttt{Ada}, the implementation of which was deemed for quite a long
time as nearly impossible. On the other hand it turned out that sets
may be a feature nice to have in a programming language, but that they
are probably not always the appropriate universal data structure for
engineering program systems; this is witnessed by the fact that some
languages, like for example \texttt{Haskell}\cite{Real-World-Haskell}, have set-like constructs
such as list comprehension, but they do not implement sets fully. As the
case may be, sets are important objects when arguing about programs,
they constitute an important component of the tool kit which a serious
computer scientist should have ready in his backpack.

When surveying the computer science literature,
we see that sets and the corresponding constructs like maps, power
sets, orders etc. are being used freely, but there is usually no
concern regarding the axiomatic basis of these objects~---~sets are
being used, albeit in a fairly naive way. This should not surprise anybody,
because they are just tools and most of the time not the objects of
consideration themselves. A tool should be handy and come to the use of a computer
scientist as soon as needed, but it really should not bring with
it complications of its own. Fairly early in the education of the computer
scientist, however, she or he encounters the phenomenon of recursion,
be it as a recursive function, be it as a recursive definition. And
here of course the question arises immediately, why the corresponding
constructs work, specifically, how one can be sure that a particular
recursive program is actually terminating. The same question, probably
a little bit more focused, appears in techniques which are related to
term rewriting. Here one inquires whether a particular chain of
replacements will actually lead to a
result in a finite amount of time. People in term rewriting have found a way of writing this
down, namely a terminating condition which is closely related to some
well-ordering. This means that we do not have infinitely long chains;
this is of course a very similar condition to the one that is encountered when
talking about the termination of a recursive procedure: Here we do not
want to have infinitely long chains of procedure or method calls.
This suggests structural similarities between the
invocation of a recursive method terminates, and  rewriting a term. If
you think about it, mathematical induction enters this family of
observations, the members of which show a considerable similarity.

When we investigate the background in front of which all this
happens, we find that we need to look at well-orderings. These are
orderings which forbid the existence of infinitely long decreasing chains.  It
turns out that the mathematical ideas expressed here are fairly
closely connected to ordinal numbers. It is not difficult to construct a
bridge from orderings and well-orders to the question whether it is
actually possible to find a well-order for each and every set. The
bridge a computer scientist might traverse is loosely described as follows:
Because we want to be able to deal with arbitrary objects, and because
we want to run programs over these arbitrary objects, it should be
possible to construct terminating recursive methods for those
objects. But in order to do that, we should make sure that no infinite
chains of method invocations may occur, which in turn poses the
question whether or not we can impose an order on these objects that
renders infinite chains impossible (admittedly somewhat indirectly, because the
order is imposed actually by procedure calls). But here we are~---~we
want to know whether such a construction is possible; mathematically this leads
to the possibility of well-ordering each and every set.

This question is of course fairly easy to answer when we talk about
finite scenarios, but sometimes it is mandatory to consider infinite
objects as well. The world may be finite, but our models of the world
are  not always. Hence the question arises whether we can take an
arbitrarily large set and construct a well-ordering for this set. As
it turns out, this question is very closely connected to another
question, which at first glance does not really look similar at all:
We are given a collection of non-empty sets, are we able to select
from each set exactly one element? This question has vexed
mathematicians for more than one century now. One of the interesting
points which indicates that things are probably a little bit more
complicated than they look is the observation that the possibility of
well-ordering arbitrary set is equivalent to the question of
selection, which came to be known as the Axiom of Choice. It turned
out during the discussion in mathematics that there is really a whole
and very full bag of other properties, which are equivalent to this
axiom. We will see that the Axiom of Choice is
equivalent to some well-known proof principles like Zorn's Lemma
or Tuckey's Maximality Principle. Because this discussion
relating the Axiom of Choice and similar constructions has been raging
in mathematics for more than one century now, we can not hope to be
able to even completely list all those things which we have to leave
out. Nevertheless we try to touch upon some topics, which appear to be
important for developing mathematical structures within computer
science. 

Since the Axiom of Choice and its variants touch upon those topics in
mathematics much in use in computer science, this gives us the
opportunity to select some of these topics and discuss them,
independently and in the light of the use of the Axiom of Choice. We
discuss for example lattices, introduce ideals and filters and pose
the maximality question: Is it always possible to extend a filter to a
maximal filter? It turns out that the answer is in the affirmative,
and this has some interesting applications for the structure theory
of, for example, Boolean algebras. Because of this we are able to
discuss one of the true classics in this area, namely Stone's
Representation Theorem, which says that every Boolean algebra is
isomorphic to an algebra of sets. Another interesting application of
Zorn's Lemma is Alexander's Theorem, which shows that one may restrict
one's attention to covering a topological space with subbase elements
for establishing compactness of the space. Because we have then
compactness at our disposal, we establish also compactness of the
space of all prime ideals of a Boolean algebra. Quite apart from these
question, which are oriented towards order structures, we establish
the Hahn-Banach Theorem, which shows that a dominated linear
functional can be extended from a linear subspace to the entire space
in a dominated way.

A particular class of Boolean algebras are closed under countable
infima and suprema, these algebras are called $\sigma$-algebras. Since
these algebras are interesting in particular when it comes to
probabilistic models in computer science, we treat these
$\sigma$-algebras in some detail, in particular with respect to
measures and their extensions. The general situation in application is
sometimes that one has the generator of a Boolean $\sigma$-algebra and
a set function which behaves decently on this generator, and one wants
to extend this set function to the whole $\sigma$-algebra. This gives
rise to a fairly rich and interesting construction, which in turn has
some connections to the question of the Axiom of Choice. The extension
process extends the measure far beyond the Boolean $\sigma$-algebra
generated by the family under consideration, and the question arises
how far this extension really goes. This may be of interest, e.g., if
one wants to measure some set, so that one has to know whether this
set can be measured at all, hence whether it is actually in the domain
of the extended measure. The Axiom of Choice helps in demonstrating
that this is not always possible. It can be shown that there are sets
which can not be measured. This depends on a selection argument for
classes of an easily constructed equivalence relation.

This will be discussed. Then we turn to games, games as a model for
human interaction, we have two players, Angel and Demon, playing against
each other. We describe how a game is played, and what strategies are,
in particular what constitutes winning strategies. This is done
first in the context of infinite sequences of natural numbers. The
model has the advantage of being fairly easy to grasp, it has the
additional structural advantage that we can map many applications to this
scenario. 

Actually, games become really interesting when we know that one of the
participants has actually a chance to win. Hence we postulate that our
games are of this kind, so that always either Angel or Demon has a
strategy to win the game. Unfortunately it turns out that this
postulate, called the Axiom of Determinacy, is in conflict with the
Axiom of Choice. This is of course a fairly unpleasant situation,
because both axioms seem to be reasonable statements. So we have to
see what can be done about this. We show that if we assume the latter
axiom, we can actually demonstrate that each and every subset of the
real line is measurable. This is a contradiction to the observation we
just related. 

This discussion serves two purposes. The first one is that one
sometimes wants to challenge the Axiom of Choice in favor of other
postulates, which may turn out to have more advantages (in the context
of games, the postulate that one of the players has a winning
strategy, no matter how the game is constructed, has certainly some
advantageous aspects). But the Axiom of Choice is, as we will see,
quite a fundamental postulate, so one has to find a balance between
both. This does look terribly complicated, but on the other hand does
not seem to be difficult to manage from a practical point of
view~---~and computer scientists are by definition practical people!
The second reason for introducing games and for elaborating on these
results is to demonstrate that games can actually be used as tools
for proofs. These tools are used in some branches of mathematics quite
extensively, and it appears that this may be an attractive choice for
computer scientists as well.



We work usually in what is called \emph{\index{naive set theory}naive
  set theory}, in which sets are used as a formal manner of speaking
without much thinking about it. Sets are just tools to express formal
thoughts. 

When mathematicians and logicians like G. Frege, G. Cantor or
B. Russell thought about the basic foundations of mathematics, they
found a huge pile of unposed and unanswered questions about the basic
building blocks of mathematics, e.g., the definition of a cardinal
number was usually taken for granted, without a formal foundation; a
foundation was even resisted or ridiculed\footnote{Frege's position,
  for example, was considered in the polemic by J. K. Thomae, ``Gedankenlose Denker,
  eine Ferienplauderei'' (Thinkers without a thought, a chat for the vacations). Jahresber. Deut.
  Math.Ver. 15, 1906, 434 - 438 as somewhat
  hare-brained, see Thiel's treatise~\cite{Thiel-Frege}}. 

\paragraph{The Axioms of ZFC}
\label{sec:axioms-zfc}

Nevertheless, at around the turn of the century there
seems to have been some consensus among mathematicians that the
following axioms are helpful for their work; they are called the \emph{\index{Zermelo-Fraenkel
System}Zermelo-Fraenkel
System With Choice (\index{ZFC}ZFC)} after E. Zermelo and A. A. Fraenkel. 

We will discuss them briefly and informally now. Here they are.

\begin{description}
\item[Extensionality] \emph{Two sets are equal iff they contain the same
  elements.} This requires that sets exist, and that we know which
elements are contained in them; usually these notions (set, element)
are taken for granted.
\item[Empty Set Axiom] \emph{There is a set with no elements.} This is
  of course the empty set, denoted by $\emptyset$. 
\item[Axiom of Pairs] \emph{For any two sets, there exists a set whose elements
  are precisely these sets.} From the extensionality axiom we conclude
that this set is uniquely determined. Without the axiom of pairs it
would be difficult to construct maps. Hence we can construct sets like
$\{a, b\}$ and singleton sets $\{a\}$ (because the axiom does not talk
about different elements, so we can form the set $\{a, a\}$, which, by
the axiom of extensionality, equals the set $\{a\}$). We can also
define an ordered pair through $\langle a, b\rangle := \{\{a\}, \{a,
b\}\}$. 
\item[Axiom of Separation] Let $\phi$ be a statement of the formal
  language with a free variable $z$. \emph{For any set $x$ there is a set
  containing all $z$ in $x$ for which $\phi(z)$ holds.} This permits
forming sets by describing the properties of their elements. Note the
restriction ``For any set $x$''; suppose we drop this and postulate
``There is a set containing all $z$ for which $\phi(z)$ holds.'' Let
$\phi(z)$ be the statement $z\not\in z$, then we would have postulated
 the existence of the set  $a := \{z \mid z\not\in z\}$ (is $a\in a$?). Hence we have to
 be a bit more modest. 
\item[Power Set Axiom] \emph{For any set $x$ there exists a set consisting
  of all subsets of $x$.} This set is called the power set of $x$ and
denoted by $\PowerSet{x}$. Of course, we have to define the notion of
a subset, before we can determine all subsets of a set $x$. A set $u$
is called a subset of set $x$ (in symbols $u\subseteq x$) iff every element of $u$ in a element of
$x$.
\item[Union Axiom] \emph{For any set there is a set which is the union of all
  elements of $x$.} This set is denoted by $\bigcup x$. If $x$
contains only a handful of elements like $x = \{a, b, c\}$, we
write $\bigcup x$  as $a\cup b \cup c$. The notion of a union is not
yet defined, although intuitively clear. We could rewrite this axiom a
little by stating it as: given a set $x$, there exists a set $y$ such
that $w\in y$ iff there exists a set $a$ with $a\in x$ and $w \in
a$. The intersection of two sets $a$ and $b$ can then be defined through the axiom of
separation with the predicate \emph{$\phi(z) := z\in a\text{ and }
  a\in b$}, so that we obtain
$
a\cap b := \{z \in a \cup b \mid z\in a\text{ and }
  a\in b\}.
$
\end{description}

This is the first group of axioms which are somewhat intuitive. It is
possible to build from it many mathematical notions (like maps with their
domains and ranges, finite Cartesian products). But it turns out that
they are not yet sufficient, so an extension to them is needed. 
\begin{description}
\item[Axiom of Infinity] \emph{There is an inductive set.} This means that
  there exists a set $x$ with the property that $\emptyset\in x$ and
  that $y\cup\{y\}\in x$ whenever $y\in x$. Apparently, this permits
  building infinite sets.
\item[Axiom of Replacement] Let $\phi$ be a formula with two
  arguments. \emph{If for every $a$ there exists exactly one $b$ such that
  $\phi(a, b)$ holds, then for each set $x$ there exists a set $y$
  which contains exactly those elements $b$ for which $\phi(a, b)$
  holds for some $a\in x$. } Intuitively, if we can find for  formula $\phi$
for each $a\in x$ exactly an element $b$ such that $\phi(a, b)$ is
true, then we can collect all these elements $b$ in a set. Let $\phi$
be the formula $\phi(x, y)$ iff $x$ is a set and $y = \PowerSet{x}$,
then there exists for a given family $x$ of sets the
set of all powersets $\PowerSet{a}$ with $a\in x$. 
\item[Axiom of Foundation]\label{lab-axiom-of-foundation}\emph{ Every
    set contains an $\in$-minimal element.} Sets contain other sets as
  elements, as we have seen, so there might be the danger that a
  situation like $a\in b \in c \in a$ occurs, hence that there is a
  $\in$-cycle. In some situations this might be desirable, but not in
  this very basic scenario where we try to find a fixed ground to work
  on. A formal description of this axiom reads that for each set $x$
  there exists a set $y$ such that $y\in x$ and $x\cap y =
  \emptyset$. We will have to deal with a very similar property when
  discussing ordinal numbers in Section~\ref{sec:ordinal-numbers}.
\end{description}

Now we have recorded some axioms which put the ground on our daily
work, to be used without any qualms. It permits to build up the
mathematical structures like relations, maps, injectivity,
surjectivity, what have you. We will not do this here (it gets
somewhat boring after a time if one is not after some special
effect~---~then it may become awfully hard) but rather trust that
these structures are available and familiar. 

But there still is a
catch: look at the argumentation in the following proposition which
constructs some sort of an inverse for a surjective map.

\BeginProposition{inv-exists} 
There exists for each surjective map $f:
A \to B$ a function $g: B \to A$ such that $(f\circ g)(b) = b$ for all
$b\in B$.  
\EndProposition

\BeginProof
For each $b\in B$, the set $\InvBild{f}{\{b\}}$ is not empty, because $f$ is surjective.
Thus we can pick for each $b\in B$ an element
$g(b)\in\InvBild{f}{\{b\}}$. Then
$g: B \to A$ is a map, and $f(g(b)) = b$ by construction.  
\EndProof

\textsc{Where is the catch?} The proof seems to be perfectly innocent and
straightforward. We simply have a look at all the inverse images of
elements of the image set $B$, all these inverse images are not empty,
so we pick from each of these inverse images exactly one element and
construct a map from this. 

Well, the catch lies in picking an element
from each member of this collection. The collection of axioms above
says nowhere that this selection is permitted (now you might think that
mathematicians find a sneaky way of permitting such a pick, through the
back door, so to speak; trust us --- they cannot!).

Hence we need some additional device, and this is the Axiom of Choice.
It will be discussed at length now; we take the opportunity to use
this discussion as kind of a peg onto which we hang some other objects as
well. The general approach will be that we will discuss mathematical
objects of interest, and at crucial point the discussion of {\AC} and
its equivalents will be continued (if you ever listened to a Wagner
opera, you will have encountered his \index{leitmotif}leitmotifs).  


\Subsection{The Axiom of Choice}
\label{sec:axiom-choice}

The Axiom of Choice states that
 
\AxiomBox{AC}{
Given a family $\mathcal{F}$ of non-empty subsets of some
  set $X$, there exists a function $f: \mathcal{F}\to X$ such that
  $f(F)\in F$ for all $F \in \mathcal{F}$. 
}

The function the
existence of which is postulated by this axiom is called a \emph{\index{choice
function}choice
function} on $\mathcal{F}$. 

It is at this point not quite clear why mathematicians make such a fuss about $\AC$:
\begin{description}\em
\item[W. Sierpinski] It is the great and ancient problem of existence
  that underlies the whole controversy about the axiom of choice.
\item[P. Maddy] The Axiom of Choice has easily the most tortured
  history of all set-theoretic axioms.
\item[T. Jech] There has been some controversy about the Axiom of
  Choice, indeed.  
\item[H. Herrlich] AC, the axiom of choice, because of its
  non-constructive character, is the most controversial mathematical
  axiom, shunned by some, used indiscriminately by others.  
\end{description}

In fact, let $X = \Nat$, the set of natural numbers. If $\mathcal{F}$ is a set of non-empty subsets of $\Nat$, a choice function is immediate ---~just let $f(F):= \min F$. So why bother? We will see below that $\Nat$ is a special case. B.Russell gave an interesting illustration: Suppose that you have an infinite set of pairs of shoes, and you are to select systematically one shoe from each pair. You can always take either the left or the right one. But now try the same with an infinite set of pairs of socks, where the left sock cannot be told from the right one. Now you have to have a choice function.

But we do not have to turn to socks in order to see that a choice
function is helpful, we rather prove Proposition~\ref{inv-exists} again. 


\BeginProof (of Proposition~\ref{inv-exists})\\Define 
\begin{equation*}
  \mathcal{F} := \{\InvBild{f}{\{b\}} \mid b\in B\},
\end{equation*}
then $\mathcal{F}$ is a collection of non-empty subsets of $A$, since
$f$ is onto. By assumption there exists a choice function $G:
\mathcal{F}\to A$ on $\mathcal{F}$. Put $g(b) :=
G(\InvBild{f}{\{b\}})$, then $f(g(b)) = b$.  
\EndProof

So this is a pure, simple and direct application of {\AC}, making one
wonder what application the existence of a choice function will
find. We'll see.



\Subsection{Cantor's Enumeration of $\Nat\times\Nat$}
\label{sec:schr-bernst-theor}

We will deal in this section with the comparison of sets with respect to their size. We say that two sets $A$ and $B$ have the same \index{cardinality}\emph{cardinality} iff there exists a bijection between them. This condition can sometimes be relaxed by saying that there exists an injective map $f: A \to B$ and an injective map $g: B \to A$. Intuitively, $A$ and $B$ have the same size, since the image of each set is contained in the other one. So we would expect that there exists a bijection between $A$ and $B$. This is what the famous \index{Theorem!Schröder-Bernstein}Schröder-Bernstein Theorem says.

\BeginTheorem{Schr-Bernst}
Let $f: A \to B$ and $g: B \to A$ be injective maps. Then there
exists a bijection $h: A \to B$.
\EndTheorem

\BeginProof
Define recursively
\begin{align*}
  A_0 & := A\setminus\Bild{g}{B},\\
A_{n+1} & := \Bild{g}{\Bild{f}{A_n}}
\end{align*}
and
\begin{equation*}
  B_n := \Bild{f}{A_n}.
\end{equation*}
If $a \in A$ with $a\not\in A_0$, there exists a unique $b =: g^*(a)$ such that $a =
g(b)$, because $g$ is an injection. Now define the map $h: A \to B$
through 
\begin{equation*}
h(a) := 
  \begin{cases}
    f(a), & \text{ if } a \in \bigcup_{n\geq0}A_n\\
g^*(a), & \text{ otherwise}.
  \end{cases}
\end{equation*}
Assume that $h(a) = h(a')$. If $a, a'\in\bigcup_{n\geq0}A_n$, we may
conclude that $a = a'$, since $f$ is one to one. If $a \in A_n$ for
some $n$ and $a'\not\in\bigcup_{n\geq0}A_n$, then 
$h(a) = f(a), h(a') = g^*(a')$, hence
$
a' = g(h(a')) = g(h(a)) = g(f(a)).
$
This implies 
$ 
a \in A_{n+1},
$
contrary to our assumption. Hence $h$ is an injection. If
$b \in \bigcup_{n\geq 0}B_n$, then 
$
b = f(a) = h(a).
$
Now let $b \notin\bigcup_{n\geq 0}B_n$. We claim that $g(b)\not\in
A_n$ for any $n\geq 0$. In fact, if $g(b)\in A_n$ with $n > 0$, we know that $g(b) =
g(f(a))$ for some $a \in A_{n-1}$, so $b = f(a)\in \Bild{f}{A_{n-1}}$,
contrary to our assumption. Hence $h(g(b)) = g^*(g(b)) = b$. Thus $h$
is also onto.
\EndProof

Another proof will be suggested in
Exercise~\ref{ex-schroeder-bernstein-ancora} through a fixed point
argument.

Call a set $A$ \index{countably infinite}\emph{countably infinite} there exists a
bijection $A\to\Nat$. By the Schröder-Bernstein
Theorem~\ref{Schr-Bernst} it then suffices to find an injective map
$A\to\Nat$ and an injective map $\Nat\to A$. A set is called \index{countable}\emph{countable}
iff it is either finite or countably infinite. 

We will have a closer look at countably infinite sets now and show that
the set of all finite sequences of natural numbers is countable; for
simplicity, we work with $\Natn := \Nat\cup\{0\}$. 

We start with
showing that there exists a bijection from the Cartesian product
$\Natn\times\Natn\to\Natn$. Cantor's celebrated  procedure for producing an
enumeration for $\Natn\times\Natn$ works for an initial section as
follows

{\small
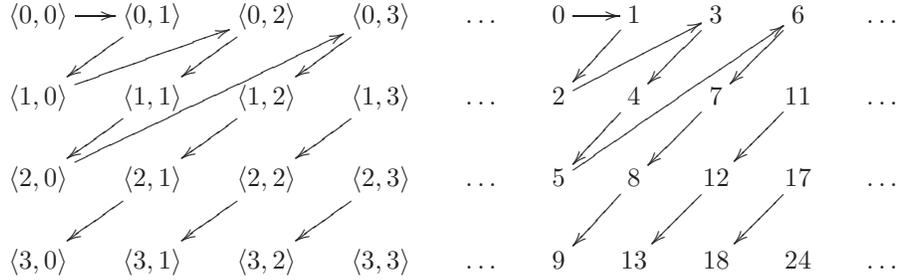
\begin{equation*}
  \xymatrix@=15pt{
\langle 0, 0\rangle\ar[r] & \langle0, 1\rangle\ar[dl] & \langle0, 2\rangle\ar[dl]&
\langle0, 3\rangle\ar[dl] & \dots & 0\ar[r] & 1\ar[dl] & 3\ar[dl] & 6\ar[dl] & \dots\\
\langle 1, 0\rangle\ar[urr] & \langle1, 1\rangle\ar[dl] & \langle1, 2\rangle\ar[dl] &
\langle1, 3\rangle\ar[dl] & \dots & 2\ar[urr] & 4\ar[dl] & 7\ar[dl] & 11\ar[dl] & \dots\\
\langle 2, 0\rangle\ar[uurrr] & \langle 2, 1\rangle\ar[dl] & \langle2, 2\rangle\ar[dl] &
\langle2, 3\rangle\ar[dl] & \dots& 5\ar[uurrr] & 8\ar[dl] & 12\ar[dl] & 17\ar[dl] & \dots\\
\langle 3, 0\rangle & \langle3, 1\rangle & \langle3, 2\rangle &
\langle3, 3\rangle & \dots& 9 & 13 & 18 & 24 & \dots
}
\end{equation*}
}

Define the function
\begin{equation*}
  J(x, y) := \binom{x+y+x}{2} + x,
\end{equation*}
then an easy computation shows that this yields just the enumeration
scheme of Cantor's procedure. We will have a closer look at $J$ now; note that the function $x \mapsto \binom{x}{2}$ increases
monotonically. 

\BeginProposition{cantor-bijects}
$J: \Natn\times\Natn\to\Natn$ is a bijection.
\EndProposition

\BeginProof
1.
$J$ is injective. We show first that $J(a, b) = J(x, y)$ implies
$a = x$. Assume that $x > a$, then $x$ can be written as $x = a + r$
for some positive $r$, so
\begin{equation*}
  \binom{a + r + y + 1}{2} + r = \binom{a + b + 1}{2},
\end{equation*}
hence $b > r + y$, so that $b$ can be written as $b = r + y + s$ with
some positive $s$. Abbreviating $c := a + r + y + 1$ we obtain 
\begin{equation*}
  \binom{c}{2} + r = \binom{c+s}{2}.
\end{equation*}
But because we have $r < c$, we get
\begin{equation*}
  \binom{c}{2} + r < \binom{c}{2} + c = \binom{c+1}{2} \geq \binom{c+s}{2}.
\end{equation*}
This is a contradiction. Hence $x\leq a$. Interchanging the r\^oles of
$x$ and $a$ one obtains $a\leq x$, so that $x = a$ may be inferred. 

Thus we obtain 
\begin{equation*}
  \binom{a+y+1}{2} = \binom{a+b+1}{2}.
\end{equation*}
This yields the quadratic equation
\begin{equation*}
  y^2 + 2ay - (b^2 + 2ab) = 0
\end{equation*}
which has the solutions $b$ and $-(2a+b)$. If $a = b = 0$ then $y = 0 =
b$, if $b > 0$, the only non-negative solution is $b$, so that $y = b$
also in this case. Hence we have shown that $J(a, b) = J(x, y)$
implies $\langle a, b\rangle = \langle x, y\rangle$.

2.
$J$ is onto. Define $Z := \Bild{J}{\Natn\times\Natn}$, then $0 = J(0,
0)\in Z$ and $1 = J(0, 1)\in Z$. Assume that $n\in Z$, so that $n =
J(x, y)$ for some $\langle x, y\rangle\in \Natn$. We consider these
cases
\begin{description}
\item[$y > 0:$] $n + 1 = J(x, y) + 1 = \binom{x+y+1}{2} + x + 1 =
  J(x+1, y-1)\in Z$.
\item[$y = 0:$] $n = \binom{x}{2}+x = \binom{x+1}{2}$, thus $n+1 =
  \binom{x+1}{2}+1$.
  \begin{description}
  \item[$x>0:$] $n + 1 = \binom{x+1}{2}+1 = \binom{1 + (x-1) + 1}{2} + 1 =
    J(1, x-1)\in Z$.
\item[$x = 0:$] Then $n = 0$, so that $n + 1 = J(0, 1)\in Z$.
  \end{description}
\end{description}

Thus we have shown that $0\in Z$, and that $n\in Z$ implies $n+1\in
Z$, from which we infer $Z=\Natn$. 
\EndProof



This construction permits us to construct an enumeration of the set of
all non-empty sequences of elements of $\Natn$. First we have a look
at sequences of fixed length. For this, define inductively
\begin{align*}
  \tau_1(x) & := x,\\
\tau_{k+1}(x_1, \dots, x_k, x_{k+1}) & := J(\tau_k(x_1, \dots, x_k), x_{k+1})
\end{align*}
($x \in \Natn$ and $k\in \Nat$, $\langle x_1, \dots, x_{k+1}\rangle\in
\Natn^{k+1}$), the idea being that an enumeration of
$\Nat^k\times\Nat$ is reduced to an enumeration of $\Nat\times\Nat$,
an enumeration of which in turn is known.

\BeginProposition{nk-to-n}
The maps $\tau_k$ are bijections $\Natn^k\to\Natn$. 
\EndProposition

\BeginProof
1.
The proof proceeds by induction on $k$. It is trivial for $k=0$.Now
assume that we have established bijectivity for $\tau_k:
\Natn^k\to\Natn$. 

2.
$\tau_{k+1}$ is injective: Assume $\tau_{k+1}(x_1, \dots, x_k, x_{k+1}) =
\tau_{k+1}(x'_1, \dots, x'_k, x'_{k+1})$, this means 
\begin{equation*}
J(\tau_k(x_1, \dots, x_k), x_{k+1}) =   J(\tau_k(x'_1, \dots, x'_k),
x'_{k+1}),
\end{equation*}
hence $\tau_k(x_1, \dots, x_k) = \tau_k(x'_1, \dots, x'_k)$ and $x_{k+1}
= x'_{k+1}$ by Proposition~\ref{cantor-bijects}. By induction
hypothesis, $\langle x_1, \dots, x_k\rangle = \langle x'_1, \dots,
x'_k\rangle$.

3. 
$\tau_{k+1}$ is onto: Given $n\in \Natn$, there exists $\langle a,
b\rangle\in\Natn\times\Natn$ with $J(a, b) = n$. Given $a\in \Natn$,
there exists $\langle x_1, \dots, x_k\rangle\in \Natn^k$ with 
$\tau_k(x_1, \dots, x_k) = a$ by induction hypothesis, so
\begin{equation*}
  n = J(a, b) = J(\tau_k(x_1, \dots, x_k), b) = \tau_{k+1}(x_1,
  \dots, x_k, b).
\end{equation*}
\EndProof

From this, we can manufacture a bijection 
\begin{equation*}
  \bigcup_{k\in\Nat} \Natn^k \to \Natn
\end{equation*}
in this way. Given a finite sequence $v$ of natural numbers, we use its
length, say, $k$, as one parameter of an enumeration of $\Nat\times\Nat$, the
other parameter for this enumeration is $\tau_k(v)$. This yields a
bijection. 

\BeginProposition{seq-to-nat-bij}
There exists a bijection $\sigma: \bigcup_{k\in\Nat} \Natn^k \to \Natn$.
\EndProposition

\BeginProof
Define 
\begin{equation*}
  \sigma(x_1, \dots, x_k) := J(k, \tau_k(x_1, \dots, x_k))
\end{equation*}
for $k\in \Nat$ and $\langle x_1, \dots, x_k\rangle\in \Natn^k$.
Because $J$ and $\tau_k$ both are injective, $\sigma$ is injective. Given
$n\in \Natn$, we can find $\langle a, b\rangle\in \Natn\times\Natn$
with $J(a, b) = n$. Given $b\in\Natn$, we can find $\langle x_1,
\dots, x_a\rangle\in \Natn^a$ with $\tau_a(x_1, \dots, x_a) = b$, so
that
\begin{equation*}
  n = J(a, b) = J(a, \tau_a(x_1, \dots, x_a)).
\end{equation*}
Hence $\sigma$ is also surjective. 
\EndProof

One wonders why we did go through this somewhat elaborate
construction. First, the construction is elegant in its simplicity,
but there is another, more subtle reason. When tracing the arguments
leading to Proposition~\ref{seq-to-nat-bij} one sees that the
argumentation is elementary, it does not require any set theoretic
assumptions like {\AC}. 

\BeginProposition{ac-implies-countable}
Let $\{A_n\mid n\in \Natn\}$ be a sequence of countably infinite sets.
Then \emph{{\AC}} implies that $\bigcup_{n\in\Natn} A_n$ is countable. 
\EndProposition

\BeginProof
We assume for simplicity that the $A_n$ are mutually disjoint. 
Given $n\in \Natn$ there exists an enumeration $\psi_n: A_n\to\Natn$.
{\AC} permits us to fix for each $n$ such an enumeration $\psi_n$,
then define
\begin{equation*}
  \psi:
  \begin{cases}
    \bigcup_{n\in\Natn} A_n & \to \Natn\\
x & \mapsto J(k, \psi_k(x)), \text{ if x}\in A_k
  \end{cases}
\end{equation*}
with $J$ as the bijection defined in Proposition~\ref{cantor-bijects}. 
\EndProof

Having {\AC}, hence Proposition~\ref{ac-implies-countable} at our
disposal, one shows by induction that 
\begin{equation*}
  \Natn^{k+1} = \bigcup_{n\in\Natn} \Natn^k\times\{n\}
\end{equation*}
is countable for every $k\in\Nat$. This establishes the countability of $
\bigcup_{k\in\Nat} \Natn^k $ immediately. On the other hand it can be
shown that Proposition~\ref{ac-implies-countable} is not valid if
{\AC} is not assumed~\cite[p. 172]{Kuratowski-Mostowski}
or~\cite[Section~3.1]{Herrlich-Choice}. This is also true if {\AC} is
weakened somewhat to postulate the existence of a choice function for
\index{countable}\emph{countable} families of non-empty sets (which in
our case would suffice).  The proof of non-validity, however, is in
either case far beyond our scope.


\Subsection{Well-Ordered Sets}
\label{sec:excurs-well-order}

A relation $R$ on a set $M$ is called an \emph{order relation} iff it
is \index{relation!reflexive}\emph{reflexive} (thus $x Rx$ holds for all $x \in M$),
\index{relation!antisymmetric}\emph{antisymmetric} (this means that $x R y$ and $y R x$ implies $x =
y$ for all $x, y \in M$) and \index{relation!transitive}\emph{transitive} (hence $x R y$ and $y R
z$ implies $x R z$ for all $x, y, z \in M$). The relation $R$ is
called \index{order!linear}\index{relation!linear}\emph{linear} iff one of the cases $x = y$, $x R y$ or $y R x$
applies for all $x, y \in M$, and it is called \emph{relation!strict \emph{strict}} iff $x R
x$ is false for each $x \in M$. If $R$ is strict and transitive, then
it is called a \emph{\index{order!strict}strict order}. 

Let $R$ be an order relation then $x\in M$ is called a \index{bound!lower}\emph{lower bound} for $\emptyset \not = A \subseteq M$ iff $x R z$ holds for
all $z \in M$ and a \index{element!smallest}\emph{smallest element} for $A$ iff it is both a
lower bound for $A$ and a member of $A$. \index{bound!upper}\emph{Upper bounds} and
\index{element!largest}\emph{largest elements} are defined similarly. An element $y$ is
called \index{element!maximal}\emph{maximal} iff there exists no element $x$ with $y R x$,
\emph{minimal} elements are defined similarly. A \index{minimal upper
  bound}\emph{ minimal upper
  bound} for a set $A \not=\emptyset$ is called the \index{supremum}\emph{supremum} of
$A$ and denoted by $\sup A$, similarly, a \index{maximal lower bound}\emph{maximal lower bound}
for $A$ is called the \index{infimum} \emph{infimum} of $A$ and denoted by $\inf A$. Neither
infimum nor supremum of a non-empty set need to exist.

\BeginExample{maximal-vs-maximum}
Look at this ordered set:

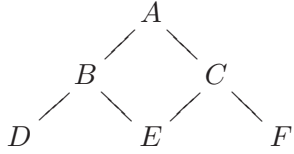
\smallBox{
\begin{equation*}
  \xymatrix@=10pt{
&&A\ar@{-}[dr]\ar@{-}[dl]&&\\
&B\ar@{-}[dr]\ar@{-}[dl]&&C\ar@{-}[dr]\ar@{-}[dl]\\
D & &E && F
}
\end{equation*}
}
\hfill\smallBox{Here $A$ is the maximum, because every element is smaller than $A$;
the minimal elements are $D$, $E$ and $F$, but there is no
minimum. The minimal elements cannot be compared to each other.} 

\EndExample

\BeginExample{ord-div}
Define $a\leq_d b$ iff $a$ divides $b$ for $a, b\in \Nat$, thus $a
\leq_d b$ iff there exists $k\in \Nat$ such that $b = k\cdot a$. Let
$g$ be the greatest common divisor of $a$ and $b$, then $g = \inf \{a,
b\}$, and if $s$ is the smallest common multiple of $a$ and $b$, then
$s = \sup \{a, b\}$. Here is why: One notes first that both $g\leq_d a$
and $g\leq_d b$ holds, because $g$ is a common divisor of $a$ and
$b$. Let $g'$ be another common divisor of $a$ and $b$, then one shows
easily that $g'$ divides $g$, so that $g'\leq_d g$ holds. Thus $g$ is
in fact the greatest common divisor. One argues similarly for the
lowest common multiple of $a$ and $b$.
\EndExample

\BeginExample{ord-p-nat}
Order $S := \PowerSet{\Nat}\setminus\{\Nat\}$ by inclusion. Then
$\Nat\setminus\{k\}$ is maximal in $S$ for every $k\in \Nat$. For we
obtain from the
definition of $S$ and its order that each element which contains
 $\Nat\setminus\{k\}$ properly would be outside the basic set $S$. The set $A := \{\{n, n+2\} \mid n\in
\Nat\}$ is unbounded in $S$. Assume that $T$ is an upper bound for $A$,  then $n\in\{n, n+2\}
\subseteq T$ and for each $n\in \Nat$, so that $T = \Nat\not\in S$.
\EndExample

Usually strict orders are written as $<$ (or $<_M$, if the basis set
is to be emphasized), and order relations as $\leq$ or $\leq_M$, resp.

Let $<_M$ be a strict order on $M$, $<_S$ be a strict order on $N$, then a
map $f: M \to N$ is called \emph{increasing} iff $x <_M y$ implies $f(x) <_S f(y)$;
$M$ and $N$ are called \emph{similar} iff $f$ is a bijection such that
$x <_M y$ is equivalent to $f(x) <_S f(y)$. An \emph{order isomorphism} is a
bijection which together with its inverse is increasing.

\BeginDefinition{d-wo-1}
The strict linear order $<$ on a set $M$ is called a \index{well-ordering}\emph{well-ordering} on $M$  iff
each non-empty subset of $M$ has a smallest element. $M$ is then
called \index{well-ordered}\emph{well-ordered} (under $<$).
\EndDefinition

These are simple examples of well-ordered sets.

\BeginExample{e-wo-1}
$\Nat$ (this shows the special r\^ole of $\Nat$ alluded to above), finite linearly ordered sets, and
$
\{1 - \frac{1}{n} \mid n \in \Nat\}
$
are well-ordered.
\EndExample

Not every ordered set, however, is well-ordered, witnessed by these
simple examples.
\BeginExample{e-wo-1a}
$\Ganz$
is not well-ordered, because it does not have a minimal element.
$\Real$ is neither, because, e.g., the open interval $]0, 1[$ does not
have a smallest element. The powerset of $\Nat$, denoted by
$\PowerSet{\Nat}$, is not well-ordered by inclusion because a
well-order is linear, and $\{1, 2\}$ and $\{3, 4\}$ are not
comparable. Finally,  
$
\{1 + \frac{1}{n} \mid n \in \Nat\}
$
is not well-ordered, because the set does not contain a smallest element.
\EndExample

\BeginExample{term-rewriting}
A \emph{\index{reduction system}reduction system} ${\cal R} = (A,
\rightarrow)$ is a set $A$ together with a relation $\rightarrow
\subseteq A \times A$; the intent is to have a set of rewrite rules,
say, $\langle a, b\rangle \in \rightarrow$ such that $a$ may be
replaced by $b$ in words over an alphabet which includes the carrier
$A$ of ${\cal R}$. Usually, one writes $a\rightarrow b$ iff $\langle
a, b\rangle\in \rightarrow$. Denote by $\stackrel{+}{\rightarrow}$ the reflexive-transitive
closure of relation $\rightarrow$, i.e., $x\stackrel{+}{\rightarrow}
y$ iff $x = y$ or there exists a chain $x = a_{0}\rightarrow \dots
\rightarrow a_{k} = y$. 

We call ${\cal R}$ \emph{\index{terminating}terminating} iff there are no
infinite chains $a_{0} \rightarrow a_{1}\rightarrow \dots \rightarrow a_{k}\rightarrow \dots$. The following proof
rule is associated with a reduction system:
\begin{equation*}
  \mathrm{(WFI)}\ \frac{
\forall x\in A (\forall y \in A: x \stackrel{+}{\rightarrow} y
\Rightarrow P(y)) \Rightarrow P(x)
}
{
\forall x \in A: P(x)
}
\end{equation*}
Here $P$ is a predicate on $A$ so that $P(x)$ is true iff $x$ has
property $P$. The the rule (WFI) says that if we can conclude for
every $x$ that $P(x)$ holds, provided the property holds for all
predecessors of $x$, the we may conclude that $P$ holds for each
element of $A$. 

This rule is equivalent to termination. In fact
\begin{itemize}
\item If $\rightarrow$ terminates, then (WFI) holds. Assume that (WFI)
  does not hold, then we find $x_{0}\in A$ such that $P(x_{0})$ does
  not hold, hence we can find some $x_{1}$ with
  $x_{0}\stackrel{+}{\rightarrow} x_{1}$ and $P(x_{1})$ does not
  hold. For $x_{1}$ we find $x_{2}$ for which $P$ does not hold with
  $x_{1}\stackrel{+}{\rightarrow} x_{2}$, etc. Hence we construct an
  infinite chain $x_{0}\stackrel{+}{\rightarrow}
  x_{1}\stackrel{+}{\rightarrow} \dots $ of elements for which $P$ doe
  not hold. But this means that $\rightarrow$ does not terminate.
\item If (WFI) holds, then $\rightarrow$ terminates. Take the
  predicate $P(x)$ iff \textsl{there is not infinite chain starting from
    $x$}. Now (WFI) says that if $y\stackrel{+}{\rightarrow} x$, and
  if $P(y)$ holds, that then $P(x)$ holds. This means that no infinite
  chain starts from $y$, and $x$ is a successor to $y$, that then no
  infinite chain starts from $x$ either. Hence, by the conclusion of
  this rule, no $x$ is the starting point of an infinite chain, consequently
  $\rightarrow$ terminates. 
\end{itemize}

Now let $(A, \rightarrow)$ be a terminating reduction system, then
each non-empty subset $B\subseteq A$ has a minimal element, because if
this is not the case, we can construct an infinite descending
chain. But $(A, \rightarrow)$ is usually not well-ordered, because
$\stackrel{+}{\rightarrow}$ is not necessarily strict. 
\EndExample

There are some helpful ways of producing a new well-order from old
ones. 

\BeginExample{e-wo-2}
Let $M$ and $N$ be well-ordered and disjoint sets, define on $M\cup N$ 
\begin{equation*}
 a < b \text{ iff }
 \begin{cases}
   a <_M b, & \text{if } a, b \in M,\\
   a <_N b, & \text{if } a, b \in N,\\
   a \in M, b \in N, & \text{otherwise}.
 \end{cases}
\end{equation*}
Then $M \cup N$ is well-ordered; this well-ordered set is usually
denoted by $M + N$. Note that $M + N$ is not the same as $N + M$. If
the sets are not disjoint, make a copy of each upon setting $M' := M
\times \{1\}, N' := N \times \{2\}$, and order these sets through,
e.g.
$
\langle m, 1 \rangle <_{M'} \langle m', 1\rangle 
$
iff 
$
m <_M m'.
$
\EndExample
\BeginExample{e-wo-3}
Define on the Cartesian product $M \times N$ 
\begin{equation*}
  \langle m, n\rangle < \langle m', n'\rangle
\text{ iff }
\begin{cases}
  m < m'\\
  n < n', & \text{if }m = m'
\end{cases}
\end{equation*}
This \index{order!lexicographic} \emph{lexicographic order} yields a well-ordering again.
\EndExample
\BeginExample{e-wo-4}
Let $Z$ be well-ordered, and assume that for each $z\in Z$ the set
$M_z$ is well-ordered so that the sets $(M_z)_{z\in Z}$ are mutually disjoint. Then
$
\bigcup_{z\in Z} M_z
$
is well-ordered.  
\EndExample

Having a look at {\AC} again, we see that it holds in a well-ordered
set: 

\BeginProposition{ac-choice-fnct}
Let $\mathcal{F}$ be a family of non-empty subsets of the
well-ordered set $M$. Then  there exists a choice function on
$\mathcal{F}$.
\EndProposition

\BeginProof
For each $F\in\mathcal{F}$ there exists a smallest element $m_F \in
\mathcal{F}$. Put $f(F) := m_F$, then $f: \mathcal{F}\to M$ is a
choice function on $\mathcal{F}$.
\EndProof

\medskip{}

Thus, if we can find a well-order on a set, then we know that we can
find choice functions. We formulate first this property.
\AxiomBox{WO}{Each set can be well-ordered.}
We will refer to this property as {\WO}. Hence we can rephrase
Proposition~\ref{ac-choice-fnct} as
\begin{equation*}
  \WO \Longrightarrow \AC
\end{equation*}

Establishing the converse will turn out to be more interesting, since
it will require the introduction of a new class of objects, viz., the
ordinal numbers. This is what we will be doing next.

We start with some preparations which deal with properties of
well-orders.

\BeginLemma{l-wo-1}
Let $M$ be well-ordered and $f: M \to M$ be an increasing map. Then $x
\leq_M f(x)$ (thus $x < f(x)$ or $x = f(x)$) holds for all $x \in M$.
\EndLemma

\BeginProof
Suppose that the set 
$
Y := \{y \in M \mid f(y) < y\}
$
is not empty, then it has a smallest element $z$. Since $f(z) < z$ we
obtain $f(f(z)) < f(z) < z$, because $f$ is increasing. This
contradicts the choice of $z$.
\EndProof

Let $M$ be well-ordered, then define for $x \in M$ the \index{initial segment}\emph{initial segment}
$O(x)$ (or $O_M(x)$) for $x$ as 
$
   O(x) := \{z \in M \mid z < x\}.
$

We obtain as a consequence
\BeginCorollary{init-segment}
No well-ordered set is order isomorphic to an initial segment of itself.
\EndCorollary

\BeginProof
An isomorphism $f: M\to O_M(x)$ for some $x\in M$ would have $f(x) <
x$, which contradicts Lemma~\ref{l-wo-1}. 
\EndProof

A surprising consequence of Lemma~\ref{l-wo-1} is that there exists at most one
isomorphism between well-ordered sets.

\BeginCorollary{unique-wo}
Let $A$ and $B$ be well-ordered sets. If $f: A\to B$ and $g: A \to B$ are order isomorphisms,
then $f = g$.
\EndCorollary

\BeginProof
Clearly both $g^{-1}\circ f$ and $f^{-1}\circ g$ are increasing,
yielding $x \leq (g^{-1}\circ f)(x)$ and $x \leq (f^{-1}\circ g)(x)$
for each $x \in A$, which means $g(x) \leq f(x)$ and $f(x) \leq g(x)$
for each $x\in A$.
\EndProof

\medskip{}

This is an important property of well-ordered sets akin to induction
in the set of natural numbers. Accordingly it is called the \index{induction!transfinite}\emph{principle of
transfinite induction}, sometimes also called \emph{Noetherian \index{induction!Noetherian}induction}
(after the eminent German mathematician Emmy Noether) or \emph{\index{induction!well-founded}well-founded
induction} (after the virtual unknown Chinese mathematician W$\bar{\text{e}}$l F$\check{\text{u}}$n Dèd). 

\BeginTheorem{t-wo-1}
Let $M$ be well-ordered, $B \subseteq M$ be a set which has for each
$x \in M$  the property that $O(x) \subseteq B$ implies $x \in
B$. Then $B = M$.
\EndTheorem

\BeginProof
Assume that $M\setminus B \not= \emptyset$, then there exists is a smallest
element $x$ in this set. Since $x$ is minimal, all elements smaller
than $x$ are elements of $B$, hence $O(x) \subseteq B$. But this
implies $x \in B$, a contradiction. 
\EndProof

Let us have a look at a proof of Lemma~\ref{l-wo-1} using the
principle of transfinite induction. Put
$
B := \{z \in M \mid z \leq f(z)\},
$
and assume $O(x) \subseteq B$. If $y \in B$ with $y \not= f(y)$, then $y
< f(y)$ and $y < x$, so that $f(y) < f(x)$, thus $y < f(x)$. 
Hence $f(x)$ is larger than any element of $O(x)$, thus
$f(x)\in M\setminus O(x)$. But $x$ is the smallest element of the
latter set, which implies $x < f(x)$, so $x \in B$. From
Theorem~\ref{t-wo-1} we see now that $B = M$.

We will show now that each set can be well ordered. In order to do
this, we construct a prototypical well order and show that each set
can be mapped bijectively to this set. This then will serve as the
basis for the construction of a well order for this set. 

Carrying out this programme requires the prototype. This will be
considered next.


\Subsection{Ordinal Numbers}
\label{sec:ordinal-numbers}

Following von Neumann~\cite[§ VII.9]{Kuratowski-Mostowski}, ordinal
numbers are defined as sets with these special properties.

\BeginDefinition{d-ord-1}
A set $\alpha$ is called an \emph{\index{ordinal}ordinal number} iff these conditions are satisfied
\begin{dingautolist}{172}
\item\label{item:1} Every element of $\alpha$ is a set.
\item\label{item:2} If $\beta\in\alpha$, then $\beta\subseteq\alpha$.
\item\label{item:3} If $\beta, \gamma\in \alpha$, then $\beta =
  \gamma$ or $\beta\in\gamma$ or $\gamma\in\beta$.
\item\label{item:4} If $\emptyset \not= B \subseteq \alpha$, then
  there exists $ \gamma\in B$ with
  $\gamma\cap B=\emptyset$. 
\end{dingautolist}
\EndDefinition

Hence in order to show that a given set is an ordinal, we have to show
that the properties \ref{item:1}, \ref{item:2}, \ref{item:3} and
\ref{item:1} hold. We will demonstrate this principle for some
examples.

\BeginExample{ord-nat}
Consider this definition of the \index{somewhat natural numbers}\emph{somewhat natural numbers} $\mathfrak{N}_0$
\begin{align*} 
  0 & := \emptyset,\\
  n + 1 & := \{0, \dots, n\},\\
\mathfrak{N}_0 & := \{0, 1, \dots\}.
\end{align*}

Then $\mathfrak{N}_0$ is an ordinal number. Each element of
$\mathfrak{N}_0$ is a set by definition. Let $\beta\in
\mathfrak{N}_0$. If $\beta = 0$,
$\beta=\emptyset\subseteq\mathfrak{N}_0$, if $\beta \not=0$, $\beta = n
= \{0, \dots, n-1\}\subseteq\mathfrak{N}_0$. One argues similarly for
property~\ref{item:3}. Finally, let  $\emptyset\not=\beta\subseteq\mathfrak{N}_0$,
and let $\gamma$ be the smallest element of $\beta$. If
$\delta\in\gamma\cap\beta$, then $\delta$ is both an element of
$\beta$ and smaller than $\gamma$, hence $\gamma\cap\delta=\emptyset$.  
\EndExample

\BeginExample{succ-ord}
Let $\alpha$ be an ordinal number, then $\beta :=
\alpha\cup\{\alpha\}$ is an ordinal. It is the smallest ordinal which
is greater than $\alpha$. Property~\ref{item:1} is evident,
so is property~\ref{item:2}. Let $\gamma, \gamma'\in\beta$ with
$\gamma \not=\gamma'$, and assume that, say, $\gamma = \alpha$, then
$\gamma'\not= \alpha$, consequently $\gamma'\in\gamma$. If neither
$\gamma$ nor $\gamma'$ are equal to $\alpha$, property~\ref{item:3}
trivially holds. Assume finally that $\emptyset\not= B \subseteq
\beta$. If $B\cap\alpha\not=\emptyset$, property~\ref{item:4} for $\beta$
follows from this property for $\alpha$, if, however, $B =
\{\alpha\}$, observe that $\alpha\in \beta$ with
$B\cap\alpha=\emptyset$. Hence this property  holds for $\beta$ as well. 
\EndExample

\BeginDefinition{d-succ-ord}
Let $\alpha$ be an ordinal, then $\alpha\cup\{\alpha\}$ is called the
\index{ordinal!successor}\emph{successor to $\alpha$} and denoted by $\alpha+1$.
\EndDefinition

It is clear from this definition that no ordinal can be squeezed in
between an ordinal $\alpha$ and it successor $\alpha+1$. 



\BeginLemma{km-9}
If $M$ is a non-empty set of ordinals, then 
\begin{enumerate}
\item\label{item:9} $\alpha_* := \bigcap M$ is an ordinal; it is the
  largest ordinal contained in all elements of $M$.
\item \label{item:10} $\alpha^* := \bigcup M$ is an ordinal; it is the
  smallest ordinal which contains all elements of $M$.
\end{enumerate}
\EndLemma

\BeginProof
We iterate over the defining properties of an ordinal number for
$\bigcap M$. Since every element $\gamma$ of $\bigcap M$ is also an
element of every $\alpha\in M$, we may conclude that $\gamma$ is a
set, and that $\gamma\subseteq\bigcap M$. If $\gamma, \delta\in\bigcap
M\subseteq\alpha$ for each $\alpha\in M$, we have either $\gamma =
\delta$, $\gamma \in \delta$ or $\delta \in \gamma$. Finally, if
$\emptyset\not= B \subseteq \bigcap M\subseteq \alpha$ for each
$\alpha\in M$, we find $\eta\in B$ such that $\eta\cap B =
\emptyset$. Thus $\alpha_* := \bigcap M$ has all the properties of an
ordinal number from Definition~\ref{d-ord-1}. It is clear that
$\alpha_*$ is the largest ordinal contained in all elements of $M$.

The proof for $\bigcup M$ works along the same lines.
\EndProof

\BeginCorollary{always-larger}
Given a non-empty set $M$ of ordinals, there is always an ordinal
which is strictly larger than all the elements of $M$.
\EndCorollary

\BeginProof
If $\alpha^* := \bigcup M \in M$, then $\alpha^* + 1$ is the desired
ordinal, otherwise $\alpha^*$ is suitable.  
\EndProof

\medskip

This is an interesting consequence.

\BeginCorollary{no-all-ordinals}
There is no set of all ordinals.
\EndCorollary

\BeginProof
If $Z$ is the set of all ordinals, then Lemma~\ref{km-9} shows that
$\alpha^* := \bigcup Z$ is an ordinal. But the successor $\alpha^*+1$
to $\alpha^*$ is an ordinal as well by Example~\ref{succ-ord}, which, however, is
not an element of $Z$. This is a contradiction. 
\EndProof

\BeginDefinition{def-lim-ord}
An ordinal $\lambda$ is called a \index{ordinal!limit}\emph{limit ordinal} iff $\alpha < \lambda$
implies $\alpha+1 < \lambda$ for all ordinals $\alpha$. 
\EndDefinition

Thus a limit ordinal is not reachable through the successor
operation. This is a convenient characterization of limit ordinals. 

\BeginProposition{char-limit}
Let $\lambda$ be an ordinal. Then
\begin{enumerate}
\item \label{item:7} If $\lambda$ is a limit ordinal, then
  $\bigcup\lambda = \lambda$.
\item \label{item:8} If $\bigcup\lambda = \lambda$, then $\lambda$ is
  a limit ordinal.
\end{enumerate}
\EndProposition

\BeginProof
1.
Let $\beta\in\bigcup\lambda$, then $\beta\in\alpha$ for some
$\alpha\in\lambda$. Since $\alpha$ is an ordinal, we conclude 
$\beta\in\alpha\subseteq\lambda$, so $\bigcup\lambda \subseteq
\lambda$. On the other hand, if $\alpha\in\lambda$, then
$\alpha+1\in\lambda$, since $\lambda$ is a limit ordinal. Thus
$\alpha\in\bigcup\lambda$, so $\bigcup\lambda \supseteq
\lambda$. This proves part~\ref{item:7}. 

2.
Let $\alpha<\lambda = \bigcup\lambda$, then $\alpha\in\beta$ for some
$\beta\in \lambda$. Then either $\alpha+1\in\beta$ or $\alpha+1 =
\beta$, in any case $\alpha+1\subseteq\beta$, so that
$\alpha+1\in\lambda$. Thus $\lambda$ is a limit ordinal. This
establishes part~\ref{item:8}.
\EndProof

Ordinals can be \index{ordinal!odd, even}\emph{odd} or \emph{even}: A limit ordinal is said to be even; if
the ordinal $\zeta$ can be written as $\zeta =\xi + 1$ and $\xi$ is
even, then $\zeta$ is odd, if $\xi$ is odd, $\zeta$ is even. This
classification is sometimes helpful, some constructions involving
ordinals depend on it, see for example
Section~\ref{sec:constr-thro-transf} on page~\pageref{sec:constr-thro-transf}. 

Several properties of ordinal numbers are established now; this is
required for carrying out the programme sketched above. The first
property states that the $\in$-relation is not cyclic, which seems to
be trivial. But since ordinal numbers have the dual face of being
elements and subsets of the same set, we will need to exclude this
property explicitly by showing that the properties of ordinals prevent
this undesired behavior.

\BeginLemma{km-5}
If $\alpha$ is an ordinal number, then there does not exist a sequence
$\beta_1, \dots, \beta_k$ of sets with $
\beta_k\in\beta_1\in\dots\beta_{k-1}\in\beta_k\in\alpha.  $
\EndLemma
\BeginProof
If there exists such sets $\beta_1, \dots, \beta_k$, put 
$
\gamma := \{\beta_1, \dots, \beta_k\}.
$
Now $\beta_k\in\alpha$ implies $\beta_k\subseteq\alpha$, thus
$\beta_{k-1}\in\alpha$, hence $\beta_{k-1}\subseteq\alpha$, so that 
$\beta_1, \dots, \beta_k\in \alpha$. But now
$\beta_{i-1}\in\beta_i\cap\gamma$ for $1 \leq i \leq k$ and $\beta_k\in\beta_1\cap\gamma$, 
so that property~\ref{item:4} in Definition~\ref{d-ord-1} is violated. 
\EndProof

\BeginLemma{km-6}
If $\alpha$ is an ordinal, then each $\beta\in\alpha$ is an ordinal
as well. 
\EndLemma

\BeginProof
1.
The properties of ordinal numbers from Definition~\ref{d-ord-1} are
inherited. This is immediate for properties~\ref{item:1},~\ref{item:3}
and \ref{item:4}, so we have to take care of property~\ref{item:2}.

2. 
Let $\gamma\in\beta$, we have to show that $\gamma\subseteq\beta$. So 
if $\eta\in\gamma$, we have by property~\ref{item:3} for $\alpha$ in the definition
of ordinals either $\eta = \gamma$ (which would imply
$\gamma\in\gamma\in\beta\in\alpha$, contradicting Lemma~\ref{km-5}), or
$\gamma\in\eta$ (which would yield
$\gamma\in\eta\in\gamma\in\beta\in\alpha$, contradicting Lemma~\ref{km-5}
again). Thus $\eta\in\beta$, so that property~\ref{item:2} also holds.
\EndProof

\BeginLemma{km-7}
Let $\alpha$ and $\beta$ be ordinals, then these properties are
equivalent
\begin{enumerate}
\item \label{item:5} $\alpha\in\beta$.
\item \label{item:6} $\alpha\subseteq\beta$ and $\alpha\not=\beta$.
\end{enumerate}
\EndLemma

\BeginProof
\labelImpl{item:5}{item:6}: 
We obtain $\alpha\subseteq\beta$ from $\alpha\in\beta$ and from
property~\ref{item:2}, and $\alpha\not=\beta$ from Lemma~\ref{km-5},
for otherwise we could conclude $\beta\in\beta$. 

\labelImpl{item:6}{item:5}:
Because $\alpha$ is a proper subset of $\beta$, thus
$\emptyset\not=\beta\setminus\alpha\subseteq\beta$, we infer from
property~\ref{item:4} for ordinals that we can find
$\gamma\in\beta\setminus\alpha$ such that
$\gamma\cap\beta\setminus\alpha=\emptyset$. We claim that
$\gamma=\alpha$. 
\begin{description}
\item[``$\subseteq$'':] Since $\gamma\in\beta$ we know that
  $\gamma\subseteq\beta$, and since
  $\gamma\cap\beta\setminus\alpha=\emptyset$, it follows
  $\gamma\subseteq\alpha$. 
\item[``$\supseteq$'':] We will show that the assumption
  $\alpha\setminus\gamma\not=\emptyset$ is contradictory.  Because
  $\emptyset\not=\alpha\setminus\gamma\subseteq\alpha$ we find
  $\eta\in\alpha\setminus\gamma$ with
  $\eta\cap\alpha\setminus\gamma=\emptyset$. Because
  $\eta\in\alpha\setminus\gamma\subseteq\alpha\subseteq\beta$ we
  conclude $\eta\in\beta$. From property~\ref{item:3} we infer that the
  cases $\eta=\gamma$, $\eta\in\gamma$ and $\gamma\in\eta$ may
  occur. Let us discuss look at these cases in turn
  \begin{itemize}
  \item $\eta=\gamma$: This is impossible, because we would have then
    $\eta\in\alpha$ and $\eta\in\beta\setminus\alpha$.
\item $\eta\in\gamma$: This is impossible because
  $\eta\in\alpha\setminus\gamma$.
\item $\gamma\in\eta$: We know that
  $\gamma\not\in\alpha\setminus\gamma$, but $\gamma\in\alpha$, which
  implies $\gamma\in\gamma\in\alpha$, contradicting  Lemma~\ref{km-5}.
  \end{itemize}
Thus we conclude that the assumption
$\alpha\setminus\gamma\not=\emptyset$ leads us to a contradiction,
from which the desired inclusion is inferred. 
\end{description}
\EndProof

Consequently, the containment relation $\in$ yields a total order on
an ordinal number.

\BeginLemma{km-10}
If $\alpha$ and $\beta$ are ordinals, then either $\alpha\subseteq
\beta$ or $\beta \subseteq \alpha$.  
\EndLemma

\BeginProof
Suppose $\alpha \not= \alpha\cap\beta \not= \beta$, then 
$\alpha\cap\beta\in\alpha$ and $\alpha\cap\beta\in\beta$ by Lemma~\ref{km-7}, hence
$\alpha\cap\beta\in\alpha\cap\beta$, contradicting Lemma~\ref{km-5}.
\EndProof

\BeginLemma{km-8}
Every ordinal is well-ordered by the inclusion relation.
\EndLemma

\BeginProof
Let $\alpha$ be an ordinal, we show first that $\alpha$ is linearly
ordered by inclusion. Take $\beta, \gamma\in\alpha$, then either
$\beta = \gamma$, $\beta \in\gamma$ or $\gamma \in \beta$. The last
two conditions translate because of property~\ref{item:2} to
$\beta\subseteq\gamma$ or $\gamma\subseteq\beta$. Now let $B$ be a
non-empty subset of $\alpha$, then we know from property~\ref{item:4}
that there exists $\gamma\in B$ with $\gamma\cap B = \emptyset$. This
is the smallest element of $B$. In fact, let $\eta\in B$ with
$\eta\not=\gamma$, then either $\gamma\in\eta$ or $\eta\in\gamma$. But
$\gamma\in\eta$ is impossible, since otherwise $ \gamma \in
B\cap\eta$. So $\eta\in\gamma$, hence $\eta\subseteq\gamma$. 
\EndProof

We can describe this strict order even a bit more precise.

\BeginLemma{km-11} 
If $\alpha$ and $\beta$ are distinct ordinals, then either $\alpha$ is
an initial segment of $\beta$ or $\beta$ is an initial segment of
$\alpha$. 
\EndLemma

\BeginProof
Because $\alpha\not=\beta$, we have either $\alpha\subseteq\beta$ or
$\beta\subseteq\alpha$ by Lemma~\ref{km-10}. Assume that
$\alpha\subseteq\beta$ holds. If $\gamma\in\alpha$,
then $\gamma\subseteq\alpha$, thus all elements of
$\alpha$ precede the element $\alpha$; conversely, if $\eta\in\beta$
with $\eta\subseteq\alpha$, then $\eta\in\alpha$. Hence $\alpha$ is a segment of
$\beta$. It cannot be similar to $\beta$ because of Corollary~\ref{init-segment}. 
\EndProof

Historically, ordinal numbers have been introduced as some sort of
equivalence classes of well-ordered sets under order isomorphisms
(note that \emph{some sort of equivalence classes} is a cautionary
expression, alluding to the fact that there is no such thing as a set
of all sets). We show now that the definition given here is not too
far away from the traditional definition. Loosely speaking, the
ordinals defined here may serve as representatives for those classes of
well-ordered sets. We want to establish 

\BeginTheorem{km-12}
If $M$ is a well-ordered set, then there exists an ordinal
$\alpha$ such that $M$ and $\alpha$ are isomorphic. 
\EndTheorem

The proof will be done in several steps. Call two well-ordered sets
$A$ and $B$ \index{similar}\emph{similar} ($A \sim B$) iff there exists an
isomorphism between them. Recall that isomorphisms preserve order
relations in both directions.

Define the set $H$ as all elements of $M$ the initial segment of which
is similar to some ordinal number, i.e.,
\begin{equation*}
  H := \{z \in M \mid \alpha_z \sim O(z)\text{ for some ordinal }\alpha_z\}.
\end{equation*}
In view of Lemma~\ref{km-11}, if $\alpha_z\sim O(z)$ and $\alpha'_z\sim
O(z)$, then $\alpha_z = \alpha_z'$, so the ordinal $\alpha_z$ is uniquely
determined, if it exists. We first show by induction that $H = M$. For this,
assume that $O(z)\subseteq H$, then we have to show that $z\in H$, so
we have to find an ordinal $\alpha_z$ with $\alpha_z\sim O(z)$. In
fact, the natural choice is 
\begin{equation*}
  \alpha_z := \{\alpha_x \in M \mid x < z\},
\end{equation*}
so we show that this is an ordinal number by going through the
properties according to Definition~\ref{d-ord-1}.  Since each element
of $\alpha_z$ is an ordinal, property~\ref{item:1} is satisfied. Let
$\alpha_x\in\alpha_z$, then $x < z$; if $\eta\in\alpha_x$, then $\eta$
is an ordinal number, hence an initial segment of $\alpha_x$ by
Lemma~\ref{km-11}, thus $\eta\sim O(t)$ for some $t$. Hence $t < x <
z$, so that $\alpha_t = \eta \in \alpha_z$. Thus property~\ref{item:2}
is satisfied. Property~\ref{item:3} follows from Lemma~\ref{km-10}:
Take $\alpha_x, \alpha_y\in\alpha_z$, then $\alpha_x$ and $\alpha_y$
are ordinals. Assume that they are different, then either
$\alpha_x\subseteq\alpha_y$ or $\alpha_y\subseteq\alpha_x$, so that by
Lemma~\ref{km-7} $\alpha_x\in\alpha_y$ or $\alpha_y\in\alpha_x$
follows. Finally, let $\emptyset\not=B\subseteq\alpha_x$. Then $B$
corresponds to a non-empty subset of $M$ with a smallest element
$y$. Then $\alpha_y\in\alpha_z$, because $y < z$, and we claim that
$\alpha_y\cap B = \emptyset$. In fact, if $\eta\in\alpha_y\cap B$,
then $\eta=\alpha_t$ for some $t \in B$, so that $y$ would not be
minimal. This shows that Property~\ref{item:4} is satisfied. Hence
$\alpha_z$ is an ordinal.  In order to establish that $z\in H$ we have
to show that $\alpha_z$ is similar to $O(z)$. But this follows from
the construction.

Consequently we know that the initial segment for each element of $M$
is similar to an ordinal. 

We are now in a position to complete the proof

\BeginProof (for Theorem~\ref{km-12})
Let 
\begin{equation*}
  \alpha := \{\alpha_z \mid \alpha_z\sim O(z) \text{ for some } z \in M\},
\end{equation*}
then one shows with exactly the arguments from above that $\alpha$ is
an ordinal. Moreover, $\alpha$ is similar to $M$: Consider the map 
$
z \mapsto \alpha_z,
$
provided $\alpha_z \sim O(z)$. It is clear that it is
one to one, since $x < y$ implies $\alpha_x \in \alpha_y$, for $O(x)$
is a (proper) initial segment of $O(y)$. It is also onto, because given
$\eta\in\alpha$, we find $z\in M$ with $\eta\sim O(z)$, so that $z
\mapsto \eta$. 
\EndProof

Let us have a brief look at all countable ordinals. They will be used
later on for a particular construction in Section~\ref{sec:games} for
the construction of a game, and in
Section~\ref{sec:constr-thro-transf} for the construction of a
$\sigma$-algebra.

\BeginProposition{omega-is-ordinal}
Let $\omega_{1} := \{\alpha \mid \alpha\text{ is a countable
  ordinal}\}$. Then $\omega_{1}$ is an ordinal.
\EndProposition

\BeginProof
Exercise~\ref{ex-omega-is-ordinal}; the proof will have to look at the
properties~\ref{item:1} through~\ref{item:4}.
\EndProof

\medskip{}

Denote by $W(\alpha) := \{\zeta \mid \zeta < \alpha\}$ all ordinals
smaller than $\alpha$, hence the initial segment of ordinals
determined by $\alpha$. Given an arbitrary non-empty set $S$, a map $f:
W(\alpha)\to S$ is called an \index{$\alpha$-sequence}\emph{$\alpha$-sequence over $S$} and
sometimes denoted by $\langle a_\zeta \mid \zeta < \alpha\rangle$, where $a_\zeta :=
f(\zeta)$. The next very general statement says that these sequences
can be defined by transfinite recursion in the following manner.

\BeginTheorem{def-transf-induction}
Let $S$ be a non-empty set, and let $\Phi$ be the set of all
$\alpha$-sequences over $S$ for some ordinal $\alpha$. Moreover, assume that
$h: \Phi\to S$ is a map of $\alpha$-sequences over $S$ to $S$. Then
there exists a uniquely determined $(\alpha+1)$-sequence
$
\langle a_\zeta \mid \zeta \leq \alpha \rangle
$
such that
\begin{equation}
\label{eq:2}
a_\zeta = h(\langle a_\xi \mid \xi < \zeta\rangle)
\end{equation}
for all $\zeta \leq \alpha$. 
\EndTheorem

\BeginProof
1.
We show uniqueness first. Assume that we have two $\alpha+1$-sequences
$\langle a_\zeta \mid \zeta \leq \alpha\rangle$ and $\langle b_\zeta
\mid \zeta \leq \alpha\rangle$ such that 
\begin{align*}
  a_\zeta & = h(\langle a_\eta \mid \eta < \zeta\rangle)\\
b_\zeta & = h(\langle b_\eta \mid \eta < \zeta\rangle)
\end{align*}
for all $\zeta\leq \alpha$. Then we show by induction on $\zeta$ that
$a_\zeta = b_\zeta$. The induction begins at the smallest ordinal
$\zeta = \emptyset$, so that $a_\emptyset = h(\emptyset) =
b_\emptyset$, and the induction step is trivial. 

2.
The sequence $\langle a_\zeta \mid \zeta \leq \alpha\rangle$ is
defined now by induction on $\zeta$. If $\langle \alpha_\eta \mid \eta \leq
\zeta\rangle$ is defined, then define 
\begin{equation*}
  \alpha_{\zeta+1} :=  h(\langle a_\eta \mid \eta \leq \zeta\rangle).
\end{equation*}
If, however, $\lambda$ is a limit ordinal such that $\langle \alpha_\eta \mid \eta \leq
\zeta\rangle$ is defined for each $\zeta<\lambda$, then one notes that
$\langle a_\xi \mid \xi < \zeta'\rangle$ is the restriction of $\langle a_\xi
\mid \xi < \zeta\rangle$ for $\zeta < \zeta' < \lambda$ by uniqueness, so
that 
\begin{equation*}
  \alpha_\lambda := h(\langle a_\zeta \mid \zeta < \lambda\rangle)
\end{equation*}
defines $\alpha_\lambda$ uniquely. 
\EndProof

\medskip{}

We are now in a position to show that the existence of a choice
function implies that each set $S$ can be well-ordered. The idea of the
proof is to find for some suitable ordinal $\alpha$ an $\alpha$-sequence $\langle a_\zeta \mid \zeta <
\alpha\rangle$ over $S$  which
exhausts $S$, hence so that $S = \{a_\zeta \mid \zeta < \alpha\}$, and
then to use the well-ordering of the ordinals by saying that 
$
a_\zeta < a_\xi
$
iff 
$
\zeta < \xi.
$ 

Constructing the sequence will use the choice function, selecting an
element in such a way that one can be sure that it has not been
selected previously. 

\BeginTheorem{ac-impl-wo}
If {\emph \AC} holds, then each set $S$ can be well-ordered.
\EndTheorem

\BeginProof
Let $f: \PowerSet{S}\setminus\{\emptyset\}\to S$ be a choice function
on the non-empty subset of $S$. Extend $f$ by putting $f(\emptyset) :=
p$, where $p \not\in S$. This element $p$ serves as an indicator that
we are done with constructing the sequence. Let $\mathcal{C}$ be the set of all ordinals
$\zeta$ such that there exists a well-order $<_B$ on a subset
$B\subseteq S$ with $(B, <_B) \sim O(\zeta)$ (cp. Theorem~\ref{km-12}). Since $\mathcal{C}$ is
a set of ordinals, there exists a smallest ordinal $\alpha$ not in
$\mathcal{C}$ by Corollary~\ref{always-larger}. 

By Theorem~\ref{def-transf-induction} there exists an $\alpha$-sequence
$\langle a_\zeta \mid \zeta < \alpha\rangle$ over $S$ such that 
$
a_\zeta
:= f(S\setminus\langle a_\eta \mid \eta < \zeta\rangle)\in S\setminus\langle a_\eta \mid \eta < \zeta\rangle
$
for all $\zeta < \alpha$.  Now if $S\setminus\langle a_\eta \mid \eta < \zeta\rangle \not=
\emptyset$, then $a_\zeta \not= p$, and $a_\zeta \notin \{a_\eta \mid
\eta < \zeta\}$, so that the $a_\zeta$ are mutually different. Suppose
that this process does not exhaust $S$, then $a_\zeta \not= p$ for all
$\zeta < \alpha$. Construct the corresponding well-order $<$ on $\{a_\zeta \mid
\zeta < \alpha\}$, then $(\{a_\zeta \mid
\zeta < \alpha\}, <)\sim O(\alpha)$. Thus $\alpha\in{\cal C}$, contradicting the choice of
$\alpha$. Hence there exists a smallest ordinal $\xi < \alpha$ with $a_\xi = p$, which
implies that $S = \{a_\zeta \mid \zeta < \xi\}$ so that elements
having different labels are in fact different. This yields a well-order on $S$.
\EndProof

Hence we have shown
\BeginTheorem{wo-equiv-ac}
The following statements are equivalent
\begin{description}
\item[\AC] The Axiom of Choice.
\item[\WO] Each set can be well-ordered.
\end{description}
\QED
\EndTheorem

{\AC} has other important and often used equivalent formulations,
which we will discuss now. 


\Subsection{Zorn's Lemma and Tuckey's Maximality Principle}
\label{sec:zorns-lemma}

Let $A$ be an ordered set, then $B\subseteq A$ is called a
\index{chain}\emph{chain} iff it is linearly ordered.  Then Zorn's Lemma states
\AxiomBox{ZL}{If $A$ is an ordered set in which every chain has an upper
  bound, then $A$ has a maximal element.} 
\BeginProposition{zl-impl-ac}
{\emph{\ZL}} implies {\emph{\AC}}.
\EndProposition

\BeginProof
Let $\mathcal{F}\not=\emptyset$ be a family of nonempty subsets of a
set $S$, we want to find a choice function on $\mathcal{F}$. Define
\begin{equation*}
  R := \{\langle F, s\rangle \mid s \in F \in \mathcal{F}\},
\end{equation*}
then $R\subseteq \mathcal{F}\times A$ is a relation. Put 
\begin{equation*}
  \mathcal{C} := \{f \mid f\text{ is a function with } f \subseteq R\}
\end{equation*}
(note that we use functions here as sets of pairs). Then
$\mathcal{C}\not=\emptyset$, because $\langle F, s\rangle\in
\mathcal{C}$ for each $\langle F, s\rangle\in R$. $\mathcal{C}$ is
ordered by inclusion, and each chain has an upper bound in
$\mathcal{C}$. In fact, if $K\subseteq\mathcal{C}$ is a chain, then
$\bigcup K$ is a map: let $\langle F, s\rangle, \langle F,
s'\rangle\in \bigcup K$, then there exists $f_1, f_2\in K$ with
$\langle F, s\rangle \in f_1, \langle F, s'\rangle\in f_2$. Because
$K$ is a chain, either $f_1\subseteq f_2$ or vice versa, let us assume
$f_1\subseteq f_2$. Thus $\langle F, s\rangle\in f_2$, and since $f_2$
is a map, we may conclude that $s = s'$. Hence $\bigcup K$ is an upper
bound to $K$ in $\mathcal{C}$.

By {\ZL}, $\mathcal{C}$ has a maximal element $f^*$. We prove that
$f^*$ is the desired choice function, hence that there exists for each
and every $F\in{\cal F}$ some $s\in F$ with $f^*(F) = s$, or,
equivalently, $\langle F, s\rangle \in f^*$. Hence the domain of $f^*$
should be all of ${\cal F}$. Assume that the domain of $f^*$ does not
contain some $F^*\in\mathcal{F}$, then the map $f^*\cup\{\langle F^*,
a\rangle\}$ contains for each $a\in F^*$ the map $f^*$ properly. This
is a contradiction. Hence $f^*: \mathcal{F}\to S$ with $f^*(F)\in F$
for all $F\in \mathcal{F}$.
\EndProof

A proof for the other direction uses a well-ordering argument for
constructing a maximal chain.

\BeginProposition{ac-impl-zl}
Assume that $A$ is an ordered set in which each chain has an upper
bound, and assume that there exists a choice function on
$\PowerSet{A}\setminus\{\emptyset\}$. Then $A$ has a maximal element.
\EndProposition

\BeginProof
1. 
As in the proof of Theorem~\ref{ac-impl-wo}, let $\mathcal{C}$ be the set of all ordinals
$\zeta$ such that there exists a well-order $<_B$ on a subset
$B\subseteq A$ with $(B, <_B) \sim O(\zeta)$, and let $\alpha$ be
the smallest ordinal not in
$\mathcal{C}$, see Corollary~\ref{always-larger}. Extend the choice
function $f$ on $\PowerSet{A}\setminus\{\emptyset\}$ upon setting
$f(\emptyset) := p$ with $p\not\in A$. This element will again serve as a marker,
indicating that the selection process is finished.

2.
Define by induction a transfinite sequence $\langle a_\zeta \mid \zeta
< \alpha\rangle$ such that $a_\emptyset\in A$ is arbitrary and 
\begin{equation*}
  a_\zeta := f(\{x\in A \mid x > a_\eta\text{ for all } \eta < \zeta\}).
\end{equation*}
Assume that $a_\zeta \not= p$, then $a_\zeta > a_\eta$ for all $\eta <
\zeta$. As in the proof of Theorem~\ref{ac-impl-wo}, there is a
smallest ordinal $\beta < \alpha$ such that $a_\beta = p$. The
selection process makes sure that $\langle a_\zeta \mid \zeta <
\beta\rangle$ is an increasing sequence, and that there does not
exists an element $x\in A$ such that $a_\zeta < x$ for all $\zeta <
\beta$. 

3.
Let $t$ be an upper bound for the chain $\langle a_\zeta \mid \zeta <
\beta\rangle$. If $t$ is not a maximal element for $A$, then there
exists $x$ with $x > t$, hence $x > a_\zeta$ for all $\zeta < \beta$,
which is a contradiction. 
\EndProof

Call a subset $\mathcal{F}\subseteq\PowerSet{A}$ of \index{finite character}\emph{finite character}
iff the following condition holds: $F$ is a member of $\mathcal{F}$ iff each finite subset of
$F$ is a member of $\mathcal{F}$. The following statement is known as
\emph{Tuckey's Lemma} or as \emph{Tuckey's Maximality Principle}: 
\AxiomBox{MP}{
Each family of finite character has a maximal element.
}

This is another equivalent to {\AC}.

\BeginProposition{mp-equiv-zl}
\emph{{\MP}} $\Leftrightarrow$ \emph{{\AC}}.
\EndProposition

\BeginProof
0.
We show {\MP} $\Rightarrow$ {\AC}, the other direction is delegated to
the Exercises.

1.
Let $\mathcal{F}\subseteq\PowerSet{S}\setminus\{\emptyset\}$ be a
family of nonempty sets. We construct a choice function for
$\mathcal{F}$. Consider
\begin{equation*}
  \mathcal{G} := \{f \mid f\text{ is a choice function for some } \mathcal{E}\subseteq\mathcal{F}\}.
\end{equation*}
Then $\mathcal{G}$ is of finite character. In fact, let $f$ be  map
from  $\mathcal{E}\subseteq\mathcal{F}$ to $S$ such that each finite
subset $f_0\subseteq f$ is a choice function for some
$\mathcal{E}_0\subseteq\mathcal{E}$, then $f$ is itself a choice
function for $\mathcal{E}$. Conversely, if $f: \mathcal{E}\to S$ is a
choice function for $\mathcal{E}\subseteq\mathcal{F}$, then each
finite subset of $f$ is a choice function for its domain. Thus there
exists by the Maximality Principle a maximal element
$f^*\in \mathcal{G}$. The domain of $f^*$ is all of $\mathcal{F}$, because
otherwise  $f^*$ could be extended as in the proof of Proposition~\ref{zl-impl-ac}, and it is clear that $f^*$ is a choice
function on $\mathcal{F}$.
\EndProof

Thus we have shown
\BeginTheorem{ac-equiv-zl-equiv-wo}
The following statements are equivalent
\begin{description}
\item[\AC] The Axiom of Choice.
\item[\WO] Each set can be well-ordered.
\item[\ZL] If $A$ is an ordered set in which every chain has an upper
  bound, then $A$ has a maximal element \index{Zorn's Lemma}(Zorn's Lemma).
\item[\MP] Each family of finite character has a maximal element
  \index{Tuckey's Maximality Principle}(Tuckey's Maximality Principle).
\end{description}
\EndTheorem

We will discuss some applications of Zorn's Lemma and the Maximality
Principle now; from
Theorem~\ref{ac-equiv-zl-equiv-wo} we know that  in each case we could use also {\AC} or
{\WO}, but an application of Zorn's Lemma appears to be
more convenient and less technical.

\Subsubsection{Compactness for Propositional Logic}
\label{sec:comp-theor-prop}

We will show that a set of \index{propositional formula}propositional
formulas is satisfiable iff each
finite subset is satisfiable. This is usually called the Compactness Theorem
for Propositional Logic. 

Fix a set $V\not=\emptyset$ of variables. A propositional formula $\phi$ is given
through this grammar
\begin{equation*}
  \phi ::= x~\mid~\phi\wedge\phi~\mid~\neg\phi
\end{equation*}
with $x\in V$. Hence a formula is either a variable, the conjunction
of two formulas, or the negation of a formula. The disjunction
$\phi\vee\psi$ is defined through $\neg(\neg\phi\wedge\neg\psi)$,
implication $\phi\to\psi$ as $\neg\phi\vee\psi$, finally
$\phi\leftrightarrow\psi$ is defined through
$(\phi\to\psi)\wedge(\psi\to\phi)$. Denote by ${\cal F}$ the set of
all \emph{propositional formulas}~---~actually, the set of all formulas depends on the set of
variables, so we ought to write ${\cal F}(V)$; since we fix $V$,
however, we use this somewhat lighter notation. 

A \emph{\index{valuation}valuation} $v$ evaluates formulas. Instead of
using \texttt{true} and \texttt{false}, we use the values $0$ and $1$, hence a valuation is a map
$V\to\{0, 1\}$ which is extended in a straightforward manner to a map ${\cal
  F}\to \{0, 1\}$, which is again denoted by $v$:
\begin{align*}
  v(\phi_{1}\wedge\phi_{2}) & := \min\{v(\phi_{1}), v(\phi_{2})\},\\
v(\neg\phi) & := 1 - v(\phi).
\end{align*}
Then we have obviously, e.g., 
\begin{align*}
  v(\phi_{1}\vee\phi_{2}) & = \max\{v(\phi_{1}), v(\phi_{2})\},\\
v(\phi_{1}\to\phi_{2}) & = 1 \text{ iff }v(\phi_{1}) \leq v(\phi_{2}),\\
v(\phi_{1} \leftrightarrow \phi_{2}) & = 1\text{ iff } v(\phi_{1}) = v(\phi_{2}).
\end{align*}

For example,
\begin{align*}
  v(\phi\to(\psi\to\gamma)) & = \max\{1-v(\phi), \max\{1-v(\psi),
  v(\gamma)\}\}\\
& = \max\{1-v(\phi), 1-v(\psi), v(\gamma)\}\\
& = \max\{1-v(\psi), \max\{1-v(\phi),
  v(\gamma)\}\}\\
& = v(\psi\to(\phi\to\gamma)).
\end{align*}
Hence 

\begin{align*}
  \bigl(\phi\to(\psi\to\gamma)\bigr) & \leftrightarrow
  \bigl(\psi\to(\phi\to\gamma)\bigr)\\
& \leftrightarrow (\phi\wedge\psi)\to\gamma.
\end{align*}

A formula is true for a valuation iff this valuation gives it the
value 1; a set ${\cal A}$ of formulas is satisfied by a valuation iff each
formula in ${\cal A}$ is true under this valuation. Formally:

\BeginDefinition{true-prop}
Let $v: {\cal F}\to\{0, 1\}$ be a valuation. Then formula $\phi$ is
\emph{true for $v$} (in symbols: $v\models \phi$) iff $v(\phi) =
1$. If ${\cal A}\subseteq{\cal F}$ is a set of propositional formulas,
then ${\cal A}$ is said to be \emph{satisfied by $v$} iff each formula
in ${\cal A}$ is true for $v$, i.e., iff $v\models \phi$ for all
$\phi\in{\cal A}$. This is written as $v\models {\cal A}$.
\EndDefinition

We are interested in the question whether or not we can find for a set
of formulas a valuation satisfying it. 

\BeginDefinition{satisf-prop}
${\cal A}\subseteq{\cal F}$ is called \emph{\index{satisfiable}satisfiable} iff there
exists a valuation $v: {\cal F}\to\{0, 1\}$ with $v\models {\cal A}$.
\EndDefinition

Depending on the size of the set of variables, the set of formulas may
be quite large. If $V$ is countable, however, ${\cal F}$ is countable
as well, so in this case the question may be easier to answer; this
will be discussed briefly after giving the proof of the Compactness
Theorem. We want to establish the general case.

Before we state and prove the result, we need a lemma which permits us
to extend the range of our knowledge of satisfiability just by one
formula.

\BeginLemma{extend-by one}
Let ${\cal A}\subseteq{\cal F}$ be satisfiable, and $\phi\not\in{\cal
  A}$ be a formula. Then one of ${\cal A}\cup\{\phi\}$ and ${\cal
  A}\cup\{\neg\phi\}$ is satisfiable.
\EndLemma

\BeginProof
If ${\cal A}\cup\{\phi\}$ is not satisfiable, but ${\cal A}$ is, let
$v$ be the valuation for which $v\models {\cal A}$ holds. Because
$v(\phi) = 0$ we conclude $v(\neg\phi) = 1$, so that $v\models {\cal
  A}\cup\{\neg\phi\}$. 
\EndProof

We establish now the \emph{\index{Theorem!Compactness}Compactness
  Theorem} for propositional logic. It permits reducing the question
of satisfiability of a set ${\cal A}$ of formulas to finite subsets of ${\cal A}$.

\BeginTheorem{completeness-propositional}
Let ${\cal A}\subseteq{\cal F}$ be a set of propositional
formulas. Then ${\cal A}$ is satisfiable iff each finite subset of
${\cal A}$ is satisfiable.
\EndTheorem

\BeginProof
We will focus on satisfiability of ${\cal A}$ provided each finite
subset of ${\cal A}$ is satisfiable, because the other half of the
assertion is trivial. 

Let
\begin{equation*}
  {\cal C} := \{\langle{\cal B}, v\rangle \mid {\cal B}\subseteq
  {\cal A}, v \models {\cal B}\}
\end{equation*}
and define 
$
\langle{\cal B}_{1}, v_{1}\rangle \leq \langle{\cal B}_{2},
v_{2}\rangle
$
iff ${\cal B}_{1}\subseteq{\cal B}_{2}$ and $v_{1}(\phi) =
v_{2}(\phi)$ for all $\phi\in{\cal B}_{1}$, so that $\langle{\cal
  B}_{1}, v_{1}\rangle \leq \langle{\cal B}_{2},v_{2}\rangle$ holds iff ${\cal
  B}_{1}$ is contained in ${\cal B}_{2}$, and if the valuations
coincide on the smaller set. This is a partial order. If ${\cal
  D}\subseteq{\cal C}$ is a chain, then put 
$
{\cal B} := \bigcup{\cal D},
$ 
and define $v(\phi) := v'(\phi)$, if $\phi\in{\cal B}'$ with
$\langle{\cal B}', v'\rangle\in{\cal D}$. Since ${\cal D}$ is a chain,
$v$ is well defined. Moreover, $v\models {\cal B}$: let $\phi\in{\cal
  B}$, then $\phi\in{\cal B}'$ for some $\langle{\cal B}',
v'\rangle\in{\cal D}$, since $v'\models {\cal B}'$, we have $v(\phi) =
v'(\phi) = 1$. Hence by Zorn's Lemma there exists a maximal element $\langle{\cal M}, w\rangle$, in
particular $w\models {\cal M}$. 

We claim that ${\cal M} = {\cal A}$. Suppose this is not the case,
then there exists $\phi\in{\cal A}$ with $\phi\not\in{\cal M}$. But
either ${\cal M}\cup\{\phi\}$ or ${\cal M}\cup\{\neg\phi\}$ is
satisfiable by Lemma~\ref{extend-by one}, hence $\langle{\cal M}, w\rangle$ is not maximal. This is
a contradiction. 

But this means that ${\cal M} = {\cal A}$, hence ${\cal A}$ is satisfiable.
\EndProof

Suppose that $V$ is countable, then we know that ${\cal F}$ is
countable as well. Then another proof for
Theorem~\ref{completeness-propositional} can be given; this will be
sketched now. Enumerate ${\cal F}$ as $\{\phi_{1},
\phi_{2}, \dots\}$.  Call ---~just temporarily~--- ${\cal
  A}\subseteq{\cal F}$ \emph{finitely satisfiable} iff each finite
subset of ${\cal A}$ is satisfiable. Let ${\cal A}$ be such a finitely satisfiable set. We construct a
sequence ${\cal M}_{0}, {\cal M}_{1}, \dots$ of finitely satisfiable
sets, starting from ${\cal M}_{0} := {\cal A}$. If ${\cal M}_{n}$ is
defined, put
\begin{equation*}
  {\cal M}_{n+1} := 
  \begin{cases}
    {\cal M}_{n}\cup \{\phi_{n+1}\}&\text{ if } {\cal M}_{n}\cup
    \{\phi_{n+1}\} \text{ is finitely satisfiable,}\\
 {\cal M}_{n}\cup \{\neg\phi_{n+1}\}&\text{ otherwise.}
  \end{cases}
\end{equation*}
This will give a finitely satisfiable set ${\cal M}^{*} :=
\bigcup_{n\geq0}{\cal M}_{n}$. Now define $v^{*}(\phi) := 1$ iff
$\phi\in{\cal M}^{*}$. We claim that $v^{*}\models \phi$ iff $\phi\in{\cal
  M}^{*}$. This is proved by a straightforward induction on
$\phi$. Because ${\cal A}\subseteq{\cal M}^{*}$, we know that
$v^{*}\models {\cal A}$. This could be modified for the general case,
well-ordering ${\cal F}$. 

The approach used for the general proof can be extended from propositional logic
to first-order logic by introducing suitable constants (they are
called \emph{Henkin constants}). We refer the reader to~\cite[Chapter
1]{Barwise-Handbook}, since we are concentrating in the present text
on applications of Zorn's Lemma.

\Subsubsection{Extending Orders}
\label{sec:extending-orders}

We will establish a generalization to the well-known fact that each
finite graph $\mathcal{G}$ can be embedded into a linear order,
provided the graph does not have any cycles. This is known as a
\index{topological sort}\emph{topological sort} of the graph~\cite[Algorithm~T,
p. 262]{Knuth-Art-1} or~\cite[Section~23.4]{CLR}. One notes first that
$\mathcal{G}$ must have a node $k$ which does not have any predecessor
(hence there is no node $\ell$ which is connected to $k$ with an edge $\ell\to
k$). If such an node $k$ would not exist, one could construct for each
node a cycle on which it lies. The algorithm proceeds recursively. If
the graph contains at most one node, it returns either the empty list
or the list containing the node. The general case constructs a list
having $k$ as its head and the list for $\mathcal{G}\setminus k$ as
its tail; here $\mathcal{G}\setminus k$ is the graph with node $k$ and
all edges emanating from $k$ removed.

Finiteness plays a special r\^ole in the argument above, because it
makes sure that we have a well-order among the nodes, which in turn is
needed for making sure that the algorithm terminates. Let us turn to
the general case. Given a partial order $\leq$ on a set $S$, we show
that $\leq$ can be extended to a strict order $\leq_s$ (hence $a\leq
b$ implies $a\leq_s b$ for all $a, b\in S$).

This will be shown through Zorn's Lemma. Put
\begin{equation*}
  \mathcal{G} := \{R \mid R\text{ is a partial order on $S$ with } \leq~\subseteq~R\}
\end{equation*}
and order $\mathcal{G}$ by inclusion. Let
$\mathcal{C}\subseteq\mathcal{G}$ be a chain, then we claim that $R_0
:= \bigcup \mathcal{C}$ is a partial order. It is obvious that $R_0$
is reflexive; if $a R_0 b$ and $b R_0 a$, then there exists relations
$R_1, R_2\in\mathcal{C}$ with $a R_1 b$ and $b R_2 a$. Since
$\mathcal{C}$ is a chain, we know that $R_1\subseteq R_2$ or
$R_2\subseteq R_1$ holds. Assume that the former holds, then $a R_2 b$
follows, so that we may conclude $a = b$. Hence $R_0$ is
antisymmetric. Transitivity is proved along the same lines, using that
$\mathcal{C}$ is a chain. By Zorn's Lemma, $\mathcal{G}$ has a maximal
element $M$; since $M\in\mathcal{G}$, $M$ is a partial order which
contains the given partial order $\leq$.

We have to show that $M$ is linear. Assume that it is not, so that
there exists $a, b\in S$ such that both $a M b$ and $b M a$ are
false. Put
\begin{equation*}
  M' := M \cup \{\langle x, y\rangle \mid x M a \text{ and } b M y\}.
\end{equation*}
Then $M'$ contains $M$ properly. If we can show that $M'$ is a partial order, we have shown that $M$ is not maximal, which is a contradiction. Let's see:
\begin{itemize}
\item $M'$ is reflexive: Since $M\subseteq M'$ and $M$ is reflexive, $x M x$ holds for all $x \in S$.
\item $M'$ is transitive: Let $x M' y$ and $y M' z$, then these cases are possible
  \begin{enumerate}
  \item \label{item:13} $x M y$ and $y M z$, hence $x M z$, thus $x M' z$.
  \item \label{item:14} $x M y$ and $y M a$ and $b M z$, thus $x M a$ and $b M z$, so that $x M' z$.
  \item \label{item:15} $x M a$ and $b M y$ and $y M z$, hence $x M' z$.
  \item \label{item:16} $x M a$ and $b M y$ and $y M a$ and $b M z$, but then $b M a$ contrary to our assumption. Hence this case cannot occur.
  \end{enumerate}
Thus we may conclude that $M'$ is transitive.
\item $M'$ is antisymmetric. Assume that $x M' y$ and $y M' x$, and look at the cases above with $z = x$. Case~\ref{item:14} would imply $x M a$ and $b M x$, so is not possible, case~\ref{item:15} is excluded for the same reason, so only case~\ref{item:13} is left, which implies $x = y$. 
\end{itemize}
Thus the assumption that there exists  $a, b\in S$ such that both $a M
b$ and $b M a$ are false leads to the conclusion that $M$ is not
maximal in $\mathcal{G}$, which is a contradiction. 

Then $a <_s b$ iff $\langle a, b\rangle\in M$ defines the desired
total order, and by construction it extends the given order. 

Hence we have shown:
\BeginProposition{extend-partial-orders}
Each partial order on a set can be extended to a total order.
\QED
\EndProposition

\Subsubsection{Bases in Vector Spaces}
\label{sec:bases-vector-spaces}

Fix a vector space $V$ over field $K$. A set  $B\subseteq V$
is called \emph{linear independent} iff 
$
\sum_{b\in B_0} a_b\cdot b = 0 
$
implies $a_b =0$ for all $b\in B_0$, whenever $B_0$ is a finite
non-empty subset of $B$. Hence, e.g., a single vector $v$ with $v
\not= 0$ is linear independent. 

\BeginExample{sqrt-is-indep}
The reals $\Real$ form a vector space over the rationals
$\Rational$. Then $\sqrt{2}$ and $\sqrt{3}$ are independent. In fact,
assume that $q_1\sqrt{2} + q_2\sqrt{3} = 0$ with rational numbers $q_1
= r_1/s_1$ and $q_2 = r_2/s_2$. Then we can find integers $t_1, t_2$
such that $t_1\sqrt{2} = t_2\sqrt{3}$ so that $t_1$ and $t_2$ have no
common divisors. But $2t_1^2 = 3t_2^2$ implies that $2$ and $3$ are
both common divisors to $t_1$ and to $t_2$. 
\EndExample

The linear independent set $B$ is called a \index{base}\emph{base}
for $V$ iff $B$ is linear independent, and if each element $v\in V$
can be represented as
\begin{equation*}
  v = \sum_{i = 1}^n a_i\cdot b_i
\end{equation*}
for some $a_1, \dots, a_n\in K$ and $b_1, \dots, b_n \in B$. 

\BeginProposition{v-has-basis}
Each vector space $V$ has a base.
\EndProposition

\BeginProof
0. 
We first find a maximal independent set, and then we show that this
set is a base.

1.
Let 
\begin{equation*}
  \mathcal{V} := \{B\subseteq V \mid B\text{ is linear independent}\}.
\end{equation*}
Then $\mathcal{V}$ contains all singletons with non-null vectors,
hence it is not empty. Order $\mathcal{V}$ by inclusion, and let
$\mathcal{B}$ be a chain in $\mathcal{V}$. Then $B_0 := \bigcup
\mathcal{B}$ is independent. In fact, if 
$
\sum_{i=1}^n a_i\cdot b_i = 0, 
$
let $b_i\in B_i\in \mathcal{B}$ for $1 \leq i \leq n$. Since
$\mathcal{C}$ is linearly ordered, we find some $k$ such that $b_i\in
B_k$, since $B_k$ is independent, we may conclude $b_1 = \dots = b_n =
0$. By Zorn's Lemma there exists a maximal independent set
$B^*\in\mathcal{V}$. 

2.
If $B^*$ is not a basis, then we find a vector $x$ which cannot be
represented as a finite linear combination of elements of
$B^*$. Clearly $x\not\in B^*$. But
then $B^*\cup\{x\}$ is linear independent, for it could otherwise be
represented by elements from $B^*$. This contradicts the
maximality of $B^*$. 
\EndProof

One notes that part 1. of the proof could as well argue with the
Maximality Principle, because a set is linear independent iff each finite
subset is linear independent. Hence the set $\mathcal{V}$ constructed
in the proof is of finite character, hence contains by {\MP} a maximal
element. Then one argue exactly as in part 2. of the proof. This shows
that {\ZL} and {\MP} are close relatives. 

The proofs above not constructive, since they do not tell us how to
construct a base for a given vector space, not even in the finite
dimensional case. 

\Subsubsection{Extending Linear Functionals}
\label{sec:hahn-banach}

Sometimes one is given a linear map from a sub-vector space to the
reals, and one wants to extend this map to a linear map on the whole
vector space. Usually there is the constraint that both the given map
and the extension should be dominated by a sub linear map.

Let $V$ be a vector space over the reals. A map $f: V\to \Real$ is
said to be a \emph{\index{functional!linear}linear functional} (or a
\emph{linear map}) on $V$ iff $f(\alpha\cdot x+\beta y) = \alpha\cdot
f(x)+\beta\cdot f(y)$ holds for all $x, y\in V$ and $\alpha,
\beta\in\Real$. Thus a linear functional is compatible with the vector
space structure of $V$. Call $p: V\to \Real$ \emph{sub linear} iff
$p(x+y) \leq p(x) + p(y)$, and $p(\alpha\cdot x) = \alpha\cdot p(x)$
for all $x, y\in V$ and $\alpha\geq 0$.

We have a look at the situation in the finite dimensional case first.

\BeginProposition{kid-hahn-banach}
Let $V$ be a finite dimensional real vector space with a sub linear
functional $p: V\to \Real$. Given a subspace $V_{0}$, and a linear map
$f_{0}: V_{0}\to \Real$ such that $f_{0}(x)\leq p(x)$ for all $x\in
V_{0}$. Then there exists a linear functional $f: V\to \Real$ which
extends $f_{0}$ such that $f(x)\leq p(x)$ for all $x\in V$.
\EndProposition

\BeginProof
1.
It is enough to show that $f_{0}$ can be extended to a linear functional dominated by $p$ to the vector space generated by $V_{0}\cup\{z\}$ with $z\not\in V_{0}$. In fact, we can then repeat this procedure a finite number of times, in each step adding a new basis vector nit contained in the previous subspace. Since $V$ is finite dimensional, this will eventually give us $V$ as the domain for the linear functional.

2.
Let $z\not\in V_{0}$, then $\{v+\alpha\cdot z\mid v\in V_{0}, \alpha\in\Real\}$ is the vector space generated by $V_{0}$ and $z$. This is clearly a vector space containing $V_{0}\cup\{z\}$, and each vector space containing $V_{0}\cup\{z\}$
must also contain linear combinations of the form $v+\alpha\cdot z$ with $v\in V_{0}$ and $\alpha\in\Real$. Moreover, the representation of an element in this vector space is unique: assume $v +\alpha\cdot z = v'+\alpha'\cdot z$, then $v-v' = (\alpha-\alpha')\cdot z$, and because $z\not\in V_{0}$, this implies $v-v'=0$, hence also $\alpha=\alpha'$. 

3.
Now set 
\begin{equation*}
f(v+\alpha\cdot z) := f_{0}(v)+\alpha\cdot c
\end{equation*}
with a value $c$ which will have to be determined. Consider $v, v'\in V_{0}$, then we have
\begin{equation*}
  f_{0}(v) - f_{0}(v') = f_{0}(v-v') \leq p(v-v') \leq p(v+z) + p(-z-v')
\end{equation*}
for an arbitrary $v_{1}\in V$. Thus we obtain
$
-p(z-v') - f_{0}(v') \leq p(v+z)-f_{0}(v).
$
Note that the left hand side of this inequality is independent of $v$, and that the right hand side is independent of $v'$, which means that we can find $c$ with
\begin{itemize}
\item\label{lab:hb-1} $c \leq p(v+z)-f_{0}(v)$ for all $v\in V_{0}$,
\item\label{lab:hb-2} $c \geq -p(-z-v')-f_{0}(v')$ for all $v\in V_{0}$. 
\end{itemize}
Now let's see what happens. Fix $\alpha$. If $\alpha=0$, we have $f(v + 0\cdot z) = f_{0}(v) \leq p(v+0\cdot z)$. If $\alpha>0$, we have 
\begin{equation*}
f(v+\alpha\cdot z) = \alpha\cdot f(v/\alpha + z) = \alpha\cdot (f_{0})(v/\alpha) + c) \leq \alpha\cdot (f_{0}(v) + p(v/\alpha+z)-f_{0}(v/\alpha) = p(v+\alpha\cdot z)
\end{equation*}
by~(\ref{lab:hb-1}) and sub linearity. If, however, $\alpha<0$, we use the inequality~(\ref{lab:hb-2}) and sub linearity of $p$; note that the coefficient $-z/\alpha$ of $z$ is positive in this case.

Summarizing, we have $f(v+\alpha\cdot z) \leq p(v+\alpha\cdot z)$ for all $v\in V_{0}$ and $\alpha\in\Real$. 
\EndProof 

When having a closer look at the proof, we see that the assumption on
working in a finite dimensional vector space is only important for
making sure that the extension process ends in a finite number of
steps. The core of this proof, however, consists in the observation that we can extend a linear functional from a vector space $V_{0}$ to a vector space $\{v+\alpha\cdot z\mid v\in V_{0}, \alpha\in\Real\}$ with $z\not\in V_{0}$ without loosing domination by the sub linear functional $p$. Let us record this important intermediate result.

\BeginCorollary{kid-hahn-banach-cor}
Let $V_{0}$ be a vector space, $V_{0}\subseteq V$, $p: V\to \Real$ be a sub linear functional, and $z\not\in V_{0}$. Then each linear functional $f: V_{0}\to \Real$ which is dominated by $p$ can be extended to a linear functional $f$ on the vector space generated by $V_{0}$ and $z$ such that $f$ is also dominated by $p$. \QED
\EndCorollary

Now we are in a position to formulate and prove the Hahn-Banach
Theorem. We will use Zorn's Lemma for the proof by setting up a partial
order such that each chain has an upper bound. We may conclude then
that there exists a maximal element. By the ``dimension free'' version
of the extension just stated we will then show that the assumption
that we did not capture the whole vector space through our maximal
element will yield a contradiction.

\BeginTheorem{hahn-banach}
Let $V$ be a real vector space with a sub linear
functional $p: V\to \Real$. Given a subspace $V_{0}$, and a linear map
$f_{0}: V_{0}\to \Real$ such that $f_{0}(x)\leq p(x)$ for all $x\in
V_{0}$. Then there exists a linear functional $f: V\to \Real$ which
extends $f_{0}$ such that $f(x)\leq p(x)$ for all $x\in V$.
\EndTheorem

\BeginProof
1.  Define $\langle V', f'\rangle \in{\cal W}$ iff $V'$ is a vector
space with $V_{0}\subseteq V'\subseteq V$, and $f': V'\to \Real$
extends $f_{0}$ and is dominated by $p$. Define $\langle V', f'\rangle
\leq \langle V", f"\rangle$ iff $V'$ is a subspace of $V"$ and $f"$ is
an extension to $f'$ for $\langle V', f'\rangle, \langle V",
f"\rangle\in{\cal W}$. Then $\leq$ is a partial order on ${\cal
  W}$. Let $\bigl(\langle V_{i}, f_{i}\rangle\bigr)_{i\in I}$ be a
chain in ${\cal W}$, then $V' := \bigcup_{i\in I}V_{i}$ is a subspace
of $V$. In fact, let $x, x'\in V'$, then $x\in V_{i}$ and $x'\in
V_{i'}$. Then either $V_{i}\subseteq V_{i'}$ or $V_{i'}\subseteq
V_{i}$. Assume the former, hence $x, x'\in V_{i'}$, thus $\alpha\cdot
x+\beta\cdot x'\in V_{i'}\subseteq V'$ for all $\alpha,
\beta\in\Real$. Put $f'(x) := f_{i}(x)$, if $x\in V_{i}$ for some
$i\in I$, then $f': V'\to \Real$ is well defined, linear and dominated
by $p$, moreover, $f'$ extends every $f_{i}$, hence, by transitivity,
$f_{0}.$ This implies $\langle V', f'\rangle\in{\cal W}$, and this is
obviously an upper bound for the chain.

2.  Hence each chain has an upper bound in ${\cal W}$, so that Zorn's
Lemma implies the existence of a maximal element $\langle V^{+},
f^{+}\rangle\in{\cal W}$. Assume that $V^{+}\not= V$, then there
exists $z\in V$ with $z\not\in V^{+}$. Then the vector space $V^{*}$
generated by $V^{+}\cup\{z\}$ contains $V^{+}$ properly, and $f^{+}$
has a linear extension $f^{*}$ to $V^{*}$ which is dominated by $p$ by
Corollary~\ref{kid-hahn-banach-cor}. But this means $\langle V^{+},
f^{+}\rangle$ is strictly smaller than $\langle V^{*},
f^{*}\rangle\in{\cal W}$, a contradiction. Hence $V^{+}= V$, and
$f^{+}$ is the desired extension.
\EndProof

\Subsubsection{Maximal Filters}
\label{sec:maximal-filters}

Fix a set $S$. The power set $\PowerSet{S}$ is ordered by
inclusion, exhibiting some interesting properties.

\BeginDefinition{set-filter}
A non-empty subset $\mathcal{F}\subseteq\PowerSet{S}$ is called a
\index{filter}\emph{filter} iff
\begin{enumerate}
\item $\emptyset\not\in\mathcal{F}$,
\item if $F_1, F_2\in\mathcal{F}$, then $F_1\cap F_2\in\mathcal{F}$,
\item if $F\in\mathcal{F}$ and $F\subseteq F'$, then $F'\in\mathcal{F}$.
\end{enumerate}
\EndDefinition
Thus a filter is closed under finite intersections and closed with
respect to super sets, and it must not contain the empty set. 

\BeginExample{pointed-filter-0}
Given $s\in S$, the set  $\mathcal{F}_s :=\{A\subseteq S \mid s \in A\}$ is a filter. 
\EndExample

\BeginExample{cofinite-filter}
Let $M$ be an infinite set. Then ${\cal F} := \{A\subseteq M\mid
M\setminus A \text{ is finite}\}$ is a filter, the filter of cofinite sets. 
\EndExample

The filter from Example~\ref{pointed-filter-0} is special because it
is maximal, we cannot find a filter $\mathcal{G}$ which properly
contains $\mathcal{F}_s$. Let's try: Take $G\in \mathcal{G}$ with
$G\not\in \mathcal{F}_s$, then $s\not\in G$, hence $s\in S\setminus
G$, so that both $G\in\mathcal{G}$ and $S\setminus G\in \mathcal{G}$,
the latter one via ${\cal F}$. This implies $\emptyset\in\mathcal{G}$,
since a filter is closed under finite intersections. We have arrived
at a contradiction, giving rise to the definition of a maximal filter
(Definition~\ref{max-filter-0}).

Before stating it, we will introduce filter bases. Sometimes we are not presented with a filter but rather with a family
of sets which generates one.

\BeginDefinition{filter-base-1}
A subset $\mathcal{B}\subseteq\PowerSet{S}$ is called a \index{filter base}\emph{filter base}
iff no intersection of a finite collection of elements of
$\mathcal{B}$ is empty, thus iff
$
  \emptyset\not\in \{B_1\cap\dots\cap B_n \mid B_1, \dots, B_n\in \mathcal{B}\}.
$
\EndDefinition

\BeginExample{nbhd-filter}
Fix $x\in\Real$, then the set ${\cal B} := \bigl\{]a, b[ \mid a < x < b\bigr\}$
of all open intervals  containing $x$ is a
filter base. Let $\Folge{a}$ be a sequence in $\Real$, then the set ${\cal E}
:= \bigl\{\{a_{k}\mid k\geq n\}\mid n \in\Nat\bigr\}$ of infinite tails of the
sequence is a filter base as well.
\EndExample

Clearly, if $\mathcal{B}$ is to be contained in a filter $\mathcal{F}$, then it must
not have the empty sets among its finite intersections, because all
these finite intersections are elements of $\mathcal{F}$. It is easy
to characterize the filter generated by a base.

\BeginLemma{gen-filter-base}
Let $\mathcal{B}$ be a filter base, then 
\begin{equation*}
  \mathcal{F} := \{B\subseteq S \mid B\supseteq B_1\cap\dots\cap B_n
  \text{ for some }  B_1, \dots, B_n\in \mathcal{B}\}
\end{equation*}
is the smallest filter containing $\mathcal{B}$.
\EndLemma

\BeginProof
It is clear that $\mathcal{F}$ is a filter, because it cannot contain
the empty set, it is closed under finite intersections, and it is
closed under super sets. Let $\mathcal{G}$ be a filter containing
$\mathcal{B}$, and let $B\supseteq B_1\cap\dots\cap B_n$
for some $B_1, \dots, B_n\in \mathcal{B}\subseteq\mathcal{G}$, hence
$B\in\mathcal{G}$. Thus $\mathcal{F}\subseteq\mathcal{G}$, so that
$\mathcal{F}$ is in fact the smallest filter containing $\mathcal{B}$.
\EndProof

Let us return to the properties of the filter defined in
Example~\ref{pointed-filter-0}. 

\BeginDefinition{max-filter-0}
A filter is called \emph{maximal} iff it is not properly
contained in another filter. Maximal filters are also called \index{ultrafilter}\emph{ultrafilters}.
\EndDefinition

This is an easy characterization of maximal filters.

\BeginLemma{l-max-filter}
These conditions are equivalent for a filter $\mathcal{F}$
\begin{enumerate}
\item \label{item:11} $\mathcal{F}$ is maximal.
\item \label{item:12} For each subset $A\subseteq S$, either $A\in\mathcal{F}$ or $S\setminus
  A\in\mathcal{F}$. 
\end{enumerate}
\EndLemma

\BeginProof
\labelImpl{item:11}{item:12}: Assume there is a set $A\subseteq S$
such that both $A\not\in\mathcal{F}$ and $S\setminus
A\not\in\mathcal{F}$ holds. Then 
\begin{equation*}
  \mathcal{G}_0 := \{F\cap A \mid F\in \mathcal{F}\}
\end{equation*}
is a filter base, because $F\cap A=\emptyset$ for some $F\in
\mathcal{F}$ would imply $F\subseteq S\setminus A$, thus $S\setminus
A\in \mathcal{F}$. Because $F\cap A \not\in \mathcal{F}$ for all
$F\in\mathcal{F}$ we conclude that the filter $\mathcal{G}$ generated
by $\mathcal{G}_0$ contains $\mathcal{F}$ properly. This ${\cal F}$ is
not maximal.
 
\labelImpl{item:12}{item:11}: A filter $\mathcal{G}$ which contains
$\mathcal{F}$ properly will contain a set $A\not\in\mathcal{F}$. By
assumption, $S\setminus A\in\mathcal{F}\subseteq\mathcal{G}$, so that $\emptyset\in\mathcal{G}$.
\EndProof

\BeginExample{ex-cofinite-not-max}
The filter ${\cal F}$ of cofinite sets from Example~\ref{cofinite-filter} for an
infinite set $M$ is not an ultrafilter. In fact, decompose $M =
M_{0}\cup M_{1}$ into disjoint sets $M_{0}$ and $M_{1}$ which are both
infinite. Then neither $M_{0}$ nor its complement are contained in
${\cal F}$. 
\EndExample

The existence of  ultrafilters is trivial by
Example~\ref{pointed-filter-0}, but we do not know whether each filter
is actually contained in an ultrafilter. The answer is in the
affirmative.

\BeginTheorem{ultra-filter-1}
Each filter can be extended to a maximal filter.
\EndTheorem

\BeginProof
Let $\mathcal{F}$ be a filter on $S$, and define
\begin{equation*}
  \mathcal{V} := \{\mathcal{G} \mid \mathcal{G}\text{ is a filter with
    } \mathcal{F}\subseteq\mathcal{G}\}.
\end{equation*}
Order $\mathcal{V}$ by inclusion. Then each chain ${\cal C}$ in $\mathcal{V}$ has
an upper bound in $\mathcal{V}$. In fact, let ${\cal H} := \bigcup
{\cal C}$. If $A\in {\cal H}$ and $A\subseteq B$, there exists a
filter ${\cal G}\in {\cal C}$ with $A\in {\cal G}$, hence $B\in {\cal
  G}$, so that $B\in {\cal H}$. If $A, B\in {\cal H}$, we find ${\cal
  G}_A, {\cal G}_B\in {\cal H}$ with either ${\cal G}_A\subseteq{\cal
  G}_B$ or ${\cal G}_B\subseteq{\cal G}_A$, because ${\cal C}$ is
linearly ordered. Assume the former, hence $A, B\in {\cal C}_B$, hence
$A\cap B\in {\cal G}_B\subseteq{\cal H}$. So ${\cal H}$ is a filter in
${\cal V}$. 

Thus there
exists a maximal element $\mathcal{F}^*$ which is a maximal
filter (just repeat the argument in the proof
of \labelImpl{item:12}{item:11} for
Lemma~\ref{l-max-filter}). $\mathcal{F}^*$ contains $\mathcal{F}$.  
\EndProof

\BeginCorollary{exists-uf-a}
Let $\emptyset\not= A \subseteq X$ be a non-empty subset of a set
$X$. Then there exists an ultrafilter containing $A$. 
\EndCorollary

\BeginProof
Using Theorem~\ref{ultra-filter-1}, extend the filter $\{B\subseteq X \mid A\subseteq B\}$ to an
ultrafilter. 
\EndProof

\Subsubsection{Ideals and Filters}
\label{sec:prime-ideals}

Recall that a \index{lattice}\emph{lattice} $(L, \leq)$ is an set $L$ with an order relation $\leq$ such
that each non-empty finite subset has a lower bound and an upper
bound. Put 
\begin{align*}
  a \wedge b & := \inf \{a, b\},\\
  a \vee b   & := \sup \{a, b\}.
\end{align*}

We note these properties ($a, b\in L$):
\begin{description}
\item[Impotency] $a \wedge a = a \vee a = a$.
\item[Commutativity] $a\wedge b = b \wedge a$ and $a\vee b = b \vee
  a$.
\item[Absorption] $a\wedge(a\vee b) = a$ and $a\vee (a\wedge b) = a$.
  In fact, $a \leq a\vee b$, thus $a =  a \wedge a \leq a\wedge (a\vee
  b)$, on the other hand $a\wedge (a\vee b) \leq a$. The second
  equality is proved similarly.
\end{description}
For simplicity we assume that the lattice is \emph{\index{lattice!bounded}bounded}, i.e., that it has a smallest element $\bot$ and a largest element $\top$, so that we can put $\bot := \sup~\emptyset$ and $\top := \inf~\emptyset$, resp.

\BeginExample{power-set-lattice}
The power set $\PowerSet{S}$ of a set $S$ is a lattice, where $A\leq
B$ iff $A \subseteq B$, so that 
\begin{align*}
  A\cap B & = \inf\{A, B\},\\
A\cup B & = \sup\{A, B\}.
\end{align*}
\EndExample

\BeginExample{ex-from-d-and-pr}
Look at this example

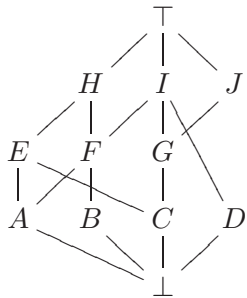
\smallBox{
\begin{equation*}
  \xymatrix@=12pt{
&&\top\ar@{-}[dl]\ar@{-}[d]\ar@{-}[dr]\\
&H\ar@{-}[d]\ar@{-}[dl]&I\ar@{-}[d]\ar@{-}[dl]\ar@{-}[ddr]&J\ar@{-}[dl]\\
E\ar@{-}[d]\ar@{-}[drr]&F\ar@{-}[d]\ar@{-}[dl]&G\ar@{-}[d]\\
A\ar@{-}[drr]&B\ar@{-}[dr]&C\ar@{-}[d]&D\ar@{-}[dl]\\
&&\bot
}
\end{equation*}
}
\hfill\smallBox{
Then $\{B, C\}$ has these upper bounds: $\{\top, H, I\}$, thus has no
smallest upper bound, so that ---~probably contrary to the first
view~--- $B \vee C$ does not exist. trying to determine $A \vee B$, we
see that the set of upper bounds to $\{A, B\}$ is just $\{\top, F, H,
I\}$, hence $A \vee B = F$. 
}

\EndExample

\BeginExample{lattice-real-line}
Consider the set ${\cal J}$ of all open intervals $]a, b[$ with $a,
b\in\Real$, and take the order inherited from $\PowerSet{\Real}$, then
${\cal I}$ is closed under taking the infimum of two elements (since
the intersection of two open intervals is again an open interval), but
${\cal J}$ is not closed under taking the supremum of two elements in
$\PowerSet{\Real}$, since the union of two open intervals is not
necessarily an open interval. Nevertheless, ${\cal J}$ is a lattice in
its own right, because we have 
\begin{equation*}
  ]a_{1}, b_{1}[ ~\vee~ ]a_{2}, b_{2}[~=~]\min\{a_{1}, a_{2},
  \max\{b_{1}, b_{2}\}[ 
\end{equation*}
in ${\cal J}$. Hence we have to be careful where to look for the
supremum. 
\EndExample
\BeginExample{rectangles-in-the-plane}
Similarly, consider the set ${\cal R}$ of all closed rectangles in the
plane $\Real\times\Real$, again with the order inherited from
$\PowerSet{\Real\times\Real}$. The intersection $R_{1}\cap R_{2}$ of two closed
rectangles $R_{1}, R_{2}\in {\cal R}$ is an element of ${\cal R}$ and
is indeed the infimum of $R_{1}$ and $R_{2}$. But what do we take as
the supremum in ${\cal R}$, if it exists at all? From the definition of the supremum we have
\begin{equation*}
  R_{1}\vee R_{2} = \bigcap \{R\in {\cal R} \mid R_{1}\subseteq
  R\text{ and } R_{2}\subseteq R\},
\end{equation*}
in plain words, the smallest closed rectangle which encloses both
$R_{1}$ and $R_{2}$. Hence, e.g., 
\begin{equation*}
  [0, 1]\times[0, 1]~\vee~[5, 6]\times[8, 9] = [0, 6]\times[0, 9].
\end{equation*}
This renders ${\cal R}$ a lattice indeed. 
\EndExample

A lattice is called \index{lattice!distributive}\emph{distributive} iff 
\begin{align*}
  a \wedge (b \vee c) & = (a \wedge b) \vee (a \wedge c)\\
a \vee (b \wedge c) & = (a \vee b) \wedge (a \vee c)
\end{align*}
holds (both equations are actually equivalent, see Exercise~\ref{lattice-equiv-distr}). 

\BeginExample{powerset-is-distrib}
The powerset lattice $\PowerSet{S}$ is a distributive lattice, because
unions and intersections are distributive. 

But \textsc{beware!} Distributivity is not necessarily
inherited. Consider the lattice ${\cal J}$ of closed intervals of the
real line, as in Example~\ref{lattice-real-line}, then
\begin{align*}
  I_{1}\wedge I_{2} & = I_{1}\cap I_{2},\\
I_{1}\vee I_{2} & = [\min I_{1}\cup I_{2}, \max I_{1}\cup I_{2}],
\end{align*}
as above. Put $A := [-3, -2], B:= [-1, 1], C := [2, 3]$, then 
\begin{align*}
  (A\wedge B)\vee(B\wedge C) & = \emptyset,\\
B\wedge(A\vee C) & = [-1, 1].
\end{align*}
Thus ${\cal J}$ is not distributive, although the order has been
inherited from the powerset. 
\EndExample

\BeginExample{down-sets}
Let $P$ be a set with a partial order $\leq$. A set $D\subseteq P$ is
called a \emph{\index{down set}down set} iff $t\in D$ and $s\leq t$
imply $s\in D$. Hence a down set is downward closed in the sense that
all elements below an element of the set belong to the set as well. A
generic example for a down set is $\{s \in P \mid s \leq t\}$ with
$t\in P$. Down sets of this shape are called \emph{\index{principal
    down sets}principal down sets}. The intersection and the union of
two down sets are down sets again. For example, let $D_{1}$ and
$D_{2}$ be down sets, let $t\in D_{1}\cup D_{2}$, and assume $s\leq
t$. Because $t\in D_{1}$ or $t\in D_{2}$ we may conclude that $s\in
D_{1}$ or $s\in D_{2}$, hence $s\in D_{1}\cup D_{2}$. Let ${\cal
  D}(P)$ be the set of all down sets of $P$, then ${\cal D}(P)$ is a
distributive lattice; this is so because the infimum and the supremum
of two elements in ${\cal D}(P)$ are the same as in $\PowerSet{P}$.

Define 
\begin{equation*}
  \Psi: 
  \begin{cases}
    P & \to {\cal D}(P)\\
    t & \mapsto \{s\in P \mid s \leq t\}
  \end{cases}
\end{equation*}
Then $t_{1}\leq t_{2}$ implies $\Psi(t_{1})\subseteq \Psi(t_{2})$,
hence the order structure carries over from $P$ to ${\cal D}(P)$.
Moreover $\Psi(t_{1}) = \Psi(t_{2})$ implies $t_{1} = t_{2}$, so that
$\Psi$ is injective. Hence we have embedded the partially ordered set
$P$ into a distributive lattice. 
\EndExample

A \index{algebra!Boolean}\emph{Boolean algebra} $B$ is a distributive lattice such that there exists
a unary operation $-: B \to B$ such that 
\begin{align*}
  a \vee -a & = \top\\
a \wedge -a & = \bot
\end{align*}
$-a$ is called the \index{complement}complement of $a$. We assume that $\wedge$ binds
stronger than $\vee$, and that complementation binds stronger than the
binary operations.  

Filters and and ideals are important structures in a lattice.

\BeginDefinition{ideal-lattice-filter}
Let $L$ be a lattice.

\smallBox{%
 $J\subseteq L$ is called an \index{ideal}\emph{ideal} iff
  \begin{itemize}
  \item \label{sec:prime-inde} $\emptyset \not= J \not= L$.
  \item \label{sec:prime-inde-1} If $a, b \in J$, then $a \vee b
    \in J$.
  \item \label{sec:prime-inde-2} If $a\in J$ and $b\in L$ with $b
    \leq a$, then $b \in J$.
  \end{itemize}
  The ideal $J$ is called \index{prime}\emph{prime} iff $a \wedge b \in J$ implies
  $a\in J$ or $b \in J$, it is called \emph{maximal} iff it is not
  properly contained in another ideal.
}
\hfill\smallBox{%
 $F\subseteq L$ is called a \index{filter}\emph{filter} iff
  \begin{itemize}
  \item \label{sec:prime-filt} $\emptyset \not= F \not= L$.
  \item \label{sec:prime-filt-1} If $a, b \in F$, then $a \wedge b
    \in F$.
  \item \label{sec:prime-filt-2} If $a\in F$ and $b\in L$ with $b
    \geq a$, then $b \in F$.
  \end{itemize}
  The filter $F$ is called \index{prime}\emph{prime} iff $a \vee b \in F$ implies
  $a\in F$ or $b \in F$, it is called \emph{maximal} iff it is not
  properly contained in another filter. 
}
\EndDefinition

Maximal filters are also called \emph{\index{ultrafilter}ultrafilters}. Recall the definition of an ultrafilter in
Definition~\ref{max-filter-0}; we have defined already
ultrafilters for the special case that the
underlying Boolean algebra is the power set of a given set.  The notion of a filter base fortunately
carries over directly from Definition~\ref{filter-base-1}, so that we
may use Lemma~\ref{gen-filter-base} in the present context as well. We will be in
this section a bit more general, but first some simple examples.

\BeginExample{ideal-bsp}
$I := \{F\subseteq \Nat \mid F\text{ is
  finite }\}$ is an ideal in $\PowerSet{\Nat}$ with set inclusion as the
partial order. This is so since the intersection of two finite sets is
finite again, and because subsets of finite sets are finite
again. Also $\emptyset\not= I \not= \PowerSet{Nat}$. This ideal is not
prime. 
\EndExample

\BeginExample{divis-ex}
Consider all divisors of $24$.
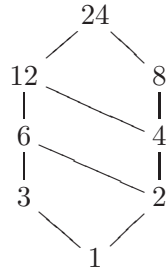
\begin{equation*}
  \xymatrix@=10pt{
& 24\ar@{-}[dl]\ar@{-}[dr] &\\
12\ar@{-}[d]\ar@{-}[drr] && 8\ar@{-}[d]\\
6\ar@{-}[d]\ar@{-}[drr]  && 4\ar@{-}[d]\\
3\ar@{-}[dr] && 2\ar@{-}[dl]\\
& 1 \\
}
\end{equation*}
$\{1, 2, 3, 6\}$ is an ideal, $\{1, 2, 3, 4, 6\}$ is not.
\EndExample

\BeginExample{senza-m}
Let $S \not=\emptyset$ be a set, $a\in S$. Then
$\PowerSet{S\setminus\{a\}}$ is a prime ideal in $\PowerSet{S}$ (with
  set inclusion as the partial order). In fact, $\emptyset \not=
  \PowerSet{S\setminus\{a\}} \not= \PowerSet{S}$, and if $a\not\in A$
  and $a\not\in B$, then $a\not\in A\cup B$. On the other hand, if
  $a\not\in A\cap B$, then $a\not\in A$ or $a\not\in B$. 
\EndExample

\BeginLemma{prop-filter}
Let $L$ be a lattice, $\emptyset \not= F \not= L$ be a proper
non-empty subset of $L$. 
\begin{itemize}
\item  These conditions are equivalent
  \begin{enumerate}
  \item \label{item:19} $F$ is a filter.
  \item \label{item:20} $\top \in F$ and ($a\wedge b\in F
    \Leftrightarrow a\in F\text{ and } b\in F$).
  \end{enumerate}
\item If filter $F$ is maximal and $L$ is
  distributive, then $F$ is a prime filter
\end{itemize}
\EndLemma

\BeginProof
1.
The implication \labelImpl{item:19}{item:20} in the first part is
trivial, for \labelImpl{item:20}{item:19} one notes that $a\leq b$ is
equivalent to $a\wedge b = a$.

2. In order to show that the maximal filter $F$ is prime, we show that
$a\vee b\in F$ implies $ a\in F$ or $b\in F$. Assume that $a\vee b\in
F$ with $a\not\in F$. Consider $B := \{f\wedge b \mid f\in F\}$, then
$\bot\not\in B$. In fact, assume that $f\vee b = \bot$ for some $f\in
F$, then we could write $ a = (f\wedge b)\vee a = (f\vee
a)\wedge(b\vee a) $ by distributivity. Since $f\in F$ and $F$ is a
filter, $f\vee a\in F$ follows, and since $b\vee a\in F$, we obtain $a\in F$, contradicting the
assumption. Thus $B$ is a filter base, and because $F$ is maximal we
may conclude that $B\subseteq F$, which in turn implies $b\in F$.
\EndProof

Hence maximal filters are prime in a distributive
lattice. If the lattice is not distributive, this may not be
true. Look at this example

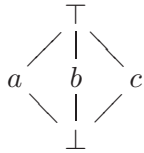
\smallBox{\begin{equation*}
  \xymatrix@=10pt{
& \top\ar@{-}[dl]\ar@{-}[d]\ar@{-}[dr] & \\
a\ar@{-}[dr] & b\ar@{-}[d]& c\ar@{-}[dl]\\
&\bot    
}
\end{equation*}
}
\hfill\smallBox{
The lattice is  not distributive, because $(a\wedge b)\vee c = c \not=
\top = (a\vee c)\wedge (b\vee c)$. Then  $\{\top\}$, $\{\top, a\}$, $\{\top, b\}$, $\{\top, c\}$ and
$\{\top, d\}$ are filters,  $\{\top, a\}$, $\{\top, b\}$, $\{\top,
c\}$ are maximal, but none of them is prime. }

Prime ideals and prime filters are not only dual notions, they are
also complementary concepts.

\BeginLemma{compl-prime-id}
In a lattice $L$ a subset $F$ is a prime filter iff its complement
$L\setminus F$ is a prime ideal.
\EndLemma

\BeginProof
Exercise~\ref{ex-compl-prime-id}.
\EndProof

We can say more in a Boolean algebra:
\BeginLemma{prime-equiv-max}
Let $B$ be a Boolean algebra. Then an ideal is maximal iff it is
prime, and a filter is maximal iff it is prime.
\EndLemma

\BeginProof
Note that a Boolean algebra is a distributive lattice with more than
one element (viz., $\bot$ and $\top$). We prove the assertion only for
filters. That a maximal filter is prime has been shown in
Lemma~\ref{prop-filter}. If
$\mathcal{F}$ is not maximal, there exists $a$ with $a\not\in {\cal
  F}$ and $-a\not\in {\cal F}$ by Lemma~\ref{l-max-filter}. But $\top
= a \vee -a\in {\cal F}$, hence ${\cal F}$ is not prime. 
\EndProof

This is another and probably surprising equivalent to {\AC}.

\AxiomBox{MI}{Each lattice with more than one element contains a maximal ideal.}

\BeginTheorem{pi-equiv-ac}
\emph{{\MI}} is equivalent to \emph{{\AC}}.
\EndTheorem

\BeginProof
1.
{\MI}~$\Rightarrow$~{\AC}:
We show actually that {\MI} implies {\MP}, an application of
Theorem~\ref{ac-equiv-zl-equiv-wo} will then establish the claim. Let
$\mathcal{F}\subseteq\PowerSet{S}$ be a family of finite character. In
order to apply {\MI}, we need a lattice $L$, which we will define
now. Define $\mathcal{L} := \mathcal{F}\cup\{S\}$, and put
for $X, Y\in\mathcal{F}$
\begin{align*}
  X\wedge Y & := X\cap Y,\\
X\vee Y & := 
\begin{cases}
  X\cup Y,& \text{ if } X\cup Y\in\mathcal{F},\\
S, &\text{otherwise}
\end{cases}
\end{align*}
Then $\mathcal{L}$ is a lattice with top element $S$ and bottom element
$\emptyset$. Let $\mathcal{M}$ be a maximal ideal in $\mathcal{L}$, then we assert that
$M^* := \bigcup \mathcal{M}$ is a maximal element of
$\mathcal{F}$. Then $M^* \not= S$. 

First we show that $M^*\in\mathcal{F}$. If 
$\{a_1, \dots, a_k\}\in M^*,$
then we can find $M_i\in\mathcal{M}$ such that $m_i\in M_i$ for $1
\leq i \leq n$. Since $\mathcal{M}$ is an ideal in $\mathcal{L}$, we
know that $M_1\vee\dots\vee M_n\in \mathcal{M}$, so that $\{a_1,
\dots, a_k\}\in\mathcal{F}$, hence $M^*\in \mathcal{F}$. 

Now assume that $M^*$ is not maximal, then we can find $N\in
\mathcal{F}$ such that $M^*$ is a proper subset of $N$, hence there
exists $t\in N$ such that $t\not\in M^*$. Because $N\in
\mathcal{F}$ and $\mathcal{F}$ is of finite character,
$\{t\}\in\mathcal{F}$. Now put
$
\mathcal{M}' := \mathcal{M}\cup\{M\vee\{t\}\mid M\in \mathcal{M}\} = \mathcal{M}\cup\{M\cup\{t\}\mid M\in \mathcal{M}\},
$
then $\mathcal{M}'$ is an ideal in $\mathcal{L}$ which properly
contains $\mathcal{M}$. This is a contradiction, hence we have found a
maximal element of $\mathcal{F}$.

2.
{\AC}~$\Rightarrow$~{\MI}: Again, we use the equivalences in
Theorem~\ref{ac-equiv-zl-equiv-wo}, because we actually show
{\ZL}~$\Rightarrow$~{\MI}. Let $L$ be a lattice with at
least two elements, and order
\begin{equation*}
  \mathcal{I} := \{I\subseteq L \mid I\text{ is an ideal in } L\}
\end{equation*}
by inclusion. Because $\{b\in L \mid b \leq a\}\in \mathcal{I}$ for
$a\in L, a \not=\top$ (by assumption, such an element exists), we know
that $\mathcal{I}\not=\emptyset$. If $\mathcal{C}\subseteq\mathcal{I}$
is a chain, then $I := \bigcup\mathcal{C}\in\mathcal{I}$. In fact,
$\emptyset\not=I\not=L$, because $\top\not\in I$, and if $a, b\in I$,
we find $I_1, I_2$ with $a\in I_1, b \in I_2$, because $\mathcal{C}$
is a chain, we may assume that $I_1\subseteq I_2$, hence $a, b\in
I_2$, so that $a\vee b\in I_2\subseteq I$. If $a \leq b$ and $b\in I$
then $a\in I$, because $b\in I_1$ for some $I_1\in \mathcal{I}$. Hence each
chain has an upper bound in $\mathcal{I}$. {\ZL} implies the existence
of a maximal element $M\in \mathcal{I}$. 
\EndProof

Since each Boolean algebra is a lattice with more than the top
element, the following corollary is a consequence of
Theorem~\ref{pi-equiv-ac}. It is known under the name \index{Theorem!Prime Ideal}\emph{Prime
  Ideal Theorem}. We know from Lemma~\ref{prime-equiv-max} that prime
ideals and maximal ideals are really the same.

\BeginCorollary{prime-ideal-thm}
\emph{{\AC}} implies the existence of a prime ideal in a Boolean algebra. 
\QED
\EndCorollary

\Subsubsection{The Stone Representation Theorem}
\label{sec:stone-repr-theor}

Let us stick for a moment to Boolean algebras and discuss the famous
Stone Representation Theorem, which requires the Prime Ideal
Theorem at a crucial point.

Fix a Boolean algebra $B$ and define for two elements $a, b\in B$
their \index{symmetric difference}\emph{symmetric difference} $a\ominus b$ through
\begin{equation*}
  a \ominus b := (a\wedge -b)\vee(-a\wedge b)
\end{equation*}
If $B =\PowerSet{S}$ for some set $S$, and if $\wedge, \vee, -$ are
the respective set operations $\cap, \cup, S\setminus \cdot$, then $A\ominus B$
is in fact equal to the symmetric difference 
$
(A\setminus B)\cup(B\setminus A) = (A\cup B)\setminus(B\cap A).
$

Fix an ideal $I$ of $B$, and define
\begin{equation*}
  a \sim_I b \Leftrightarrow a\ominus b \in I
\end{equation*}

Then $\sim_I$ is a \index{congruence}congruence, i.e., an equivalence
relation which is compatible with the operations on the Boolean
algebra. This will be shown now through a sequence of statements. 

We state some helpful properties.
\BeginLemma{helpful-symm-diff}
Let $B$ be a Boolean algebra, then
\begin{enumerate}
\item \label{item:2-z} $a \ominus a = \bot$, $a \ominus b = b \ominus
  a$ and $a \ominus b = (-a) \ominus (-b)$.
\item \label{item:3-z} $a \ominus b = (a\vee b)\wedge -(a\wedge b)$.
\item\label{item:1-z} $(a\ominus b)\wedge c = (a\wedge
  c)\ominus(b\wedge c)$ and $c\wedge(a\ominus b) = (c\wedge a)\ominus
  (c \wedge b)$.
\end{enumerate}
\EndLemma

\BeginProof
The properties under~\ref{item:2-z}. are fairly obvious, \ref{item:3-z}. is calculated
directly using distributivity, finally the first part
of~\ref{item:1-z}. follows thus
\begin{align*}
  (a\wedge c)\ominus(b\wedge c) & = (a\wedge c \wedge -(b \wedge
  c))\vee (b\wedge c \wedge -(a \wedge c))\\
& = (a \wedge c \wedge (-b \vee -c)) \vee (b \wedge c \wedge (-a \vee
-c)) \\
& = (a\wedge -b \wedge c) \vee (b\wedge -a \wedge c)\\
& = (a \ominus b)\wedge b, 
\end{align*}
because $a \wedge c \wedge -c = \bot = b \wedge c \wedge -c$. 
\EndProof

\BeginLemma{symmdiff-is-equiv}
$\sim_I$ is an equivalence relation on $B$ with these properties
\begin{enumerate}
\item \label{item:17}  $\isEquiv{a}{a'}{\sim_I}$ and
  $\isEquiv{b}{b'}{\sim_I}$ imply $\isEquiv{a\wedge b}{a'\wedge
    b'}{\sim_I}$ and $\isEquiv{a\vee b}{a'\vee b'}{\sim_I}$.
\item \label{item:18} $\isEquiv{a}{a'}{\sim_I}$  implies  $\isEquiv{-a}{-a'}{\sim_I}$.
\end{enumerate}
\EndLemma

\BeginProof
Because $a \ominus a'\in I$ and $b \ominus b'\in I$ we conclude that
$(a \ominus a')\vee(b \ominus b')\in I$, thus
\begin{align*}
  (a\vee a')\ominus(b\vee b') & \leq ((a\vee b)\wedge-(a\wedge b)) \vee
  ((a'\vee b')\wedge-(a'\wedge b'))\\ &= (a \ominus a')\vee(b \ominus b')\in I.
\end{align*}
Since $I$ is an ideal, we conclude $(a\vee a')\ominus(b\vee b')\in
I$. 

From Lemma~\ref{helpful-symm-diff} we conclude that $a\wedge b\sim_I
a'\wedge b \sim_I a'\wedge b'$. The assertion about complementation
follows from Lemma~\ref{helpful-symm-diff} as well. 
\EndProof

Denote by $\Klasse{x}{\sim_I}$ the equivalence class of $x \in B$, and
let $\fMap{\sim_I}: x \mapsto \Klasse{x}{\sim_I}$ be the associated
factor map. Define on the factor space $\Faktor{B}{I} :=
\{\Klasse{x}{\sim_I} \mid x \in B\}$ the operations
\begin{align*}
\Klasse{a}{\sim_I}\wedge\Klasse{b}{\sim_I} & := \Klasse{a\wedge
  b}{\sim_I},\\
\Klasse{a}{\sim_I}\vee\Klasse{b}{\sim_I} & := \Klasse{a\vee b}{\sim_I},\\
-\Klasse{a}{\sim_I} & := \Klasse{-a}{\sim_I}.
\end{align*}

We have  also
\begin{align*}
  \Klasse{a}{\sim_I} \leq \Klasse{b}{\sim_I} & \Leftrightarrow a\ominus
  (a\wedge b)\in I \Leftrightarrow b\ominus (a\vee b) \in I,\\
a \in I & \Leftrightarrow a \sim_I \bot.
\end{align*}
\medskip

The following statement is now fairly easy to prove. Recall that a
\index{homomorphism}homomorphism $f: (B, \wedge, \vee, -)\to (B', \wedge', \vee', -')$ is
a map $f: B \to B'$ such that $f(a\wedge b) = f(a)\wedge'f(b), f(a\vee
b) = f(a)\vee'f(b)$ and $f(-a) = -'f(a)$ for all $a, b\in B$ are
valid. 

\BeginProposition{factor-is-ba}
The \index{factor algebra}factor space $\Faktor{B}{I}$ is a Boolean algebra, and
$\fMap{\sim_I}$ is a homomorphism of Boolean algebras.
\EndProposition

\BeginProof
The operations on $\Faktor{B}{I}$ are well defined by Lemma~\ref{symmdiff-is-equiv}
and yield a lattice with $\Klasse{\top}{\sim_I}$ as the largest and
and $\Klasse{\bot}{\sim_I}$ as the smallest element, resp. Hence $-$ is a
complementation operator on $\Faktor{B}{I}$ because
\begin{align*}
  \Klasse{a}{\sim_I}\wedge\Klasse{-a}{\sim_I} & =
  \Klasse{\bot}{\sim_I},\\
\Klasse{a}{\sim_I}\vee\Klasse{-a}{\sim_I} & =
  \Klasse{\top}{\sim_I}.
\end{align*}
It is evident from the construction that $\fMap{\sim_I}$ is a homomorphism. 
\EndProof

\medskip{}

The Prime Ideal Theorem implies that the Boolean algebra $\Faktor{B}{I}$
has a prime ideal $J$ by Corollary~\ref{prime-ideal-thm}. This observation leads to a stronger version of this
theorem for the given Boolean algebra.

\BeginTheorem{extended-prime-ideal}
Let $I$ be an ideal in a Boolean algebra. Then \emph{{\AC}} implies that
there exists a prime ideal $K$ which contains $I$. 
\EndTheorem

\BeginProof
1.
Construct the factor algebra $\Faktor{B}{I}$, then {\AC} implies that
this Boolean algebra has a prime ideal $J$. We claim that 
\begin{equation*}
  K := \{x \in B \mid \Klasse{x}{\sim_I}\in J\}
\end{equation*}
is the desired prime ideal. Since $I = \Klasse{\bot}{\sim_I}\in J$, we
see that $I \subseteq K$ holds, thus $K\not=\emptyset$. 

2.
$K$ is an ideal. If $K = B$, then $\top\in K$ which would mean
$\Klasse{\top}{\sim_I}\in J$, but this is impossible. Let $a \leq b$
with $b\in K$, hence $a = a\wedge b$, so that $\Klasse{a}{\sim_I} =
\Klasse{a\wedge b}{\sim_I}$. Because $b\in K$, we infer
$\Klasse{a\wedge b}{\sim_I}\in J$, hence $\Klasse{a}{\sim_I}\in J$, so that
$a\in K$. If $a, b\in K$, then $a\vee b\in K$, because $J$ is an ideal.

3.
$K$ is prime. In fact, we have 
\begin{align*}
  a\wedge b \in K
&\Leftrightarrow 
\Klasse{a\wedge b}{\sim_I}\in J
\Leftrightarrow 
\Klasse{a}{\sim_I}\wedge\Klasse{b}{\sim_I}\in J\\
& \Rightarrow
\Klasse{a}{\sim_I}\in J\text{ or }\Klasse{b}{\sim_I}\in J
 \Leftrightarrow
a \in K \text{ or }  b \in K
\end{align*}
\EndProof

As a consequence, we can find  in a Boolean
algebra for any given element $a \not= \top$ a prime ideal which does contain it.

\BeginCorollary{prime-separate-elements}
Let $B$ be a Boolean algebra and assume that \emph{{\AC}} holds. 
\begin{enumerate}
\item Given $a\not= \top$, there exists a prime ideal which contains
  $a$.
\item Given $a, b \in B$ with $a\not= b$, there exists a prime
  ideal which contains $a$ but not $b$.
\item Given $a, b \in B$ with $a\not= b$, there exists an
  ultrafilter which contains $a$ but not $b$.
\end{enumerate}
\EndCorollary

\BeginProof
We find a prime ideal $K$ which extends the ideal $\{x \in B \mid x
\leq a\}$. This establishes the first part.

If $a \not= b$, we have $a \ominus b \not= \bot$, so
$a\wedge -b \not= \bot$ or $-a\wedge b \not= \bot$. Assume the former,
then there exists a prime ideal $K$ with $-(a\wedge-b)\in K$, so that
both $b\in K$ and $-a\in K$ holds. Since $-a\in K$ implies $a\not\in
K$, we are done with the second part. The third part follows through Lemma~\ref{compl-prime-id}.
\EndProof

This yields one of the true classics, the \index{Theorem!Stone Representation}Stone Representation
Theorem. It states that each Boolean algebra is essentially a set
algebra, i.e., a Boolean algebra comprised of sets. 

\BeginTheorem{stone-repr}
Let $B$ be a Boolean algebra, and assume that \emph{{\AC}} holds. Then there
exists a Boolean set algebra $S$ such that $B$ is isomorphic to $S$. 
\EndTheorem

\BeginProof
Define 
\begin{align*}
  {\cal S}_0 & := \{U \mid U\text{ is an ultrafilter on } B\},\\
\psi(b) & := \{U\in {\cal S}_0 \mid b \in U\}.
\end{align*}
Then 
\begin{align*}
  \psi(b_1\wedge b_2) & = \psi(b_1)\cap\psi(b_2),\\
\psi(b_1\vee b_2) & = \psi(b_1)\cup\psi(b_2),\\
\psi(-b) & = {\cal S}_0\setminus\psi(b).
\end{align*}

For example, we obtain from Lemma~\ref{prop-filter} that
\begin{align*}
  U \in \psi(b_1\wedge b_2) 
& \Leftrightarrow b_1\wedge b_2\in U\\
& \Leftrightarrow b_1\in U \text{ and } b_2\in U\\
& \Leftrightarrow U\in \psi(b_1) \text{ and }  U\in \psi(b_2)\\
& \Leftrightarrow U\in \psi(b_1)\cap\psi(b_2).
\end{align*}
Similarly, $U\in \psi(-b) \Leftrightarrow -b\in U \Leftrightarrow
b\not\in U \Leftrightarrow U\not\in \psi(b)$ by Lemma~\ref{l-max-filter}, because $U$ is an
ultrafilter. 

Because we can find for $b_1\not=b_2$ an ultrafilter which contains
$b_1$ but not $b_2$ by Corollary~\ref{prime-separate-elements}, we
conclude that $\psi$ is injective (this is actually the place where
{\AC} is used). Thus the
Boolean algebras $B$ and $\Bild{\psi}{B}$ are isomorphic, and the
latter one is comprised of sets.
\EndProof

\Subsubsection{Compactness and Alexander's Subbase Theorem}
\label{sec:alexander-subbase}

The closed interval $[u, v]$ with $-\infty < u < v < +\infty$ is an important example of a compact space. It has the following property: each cover through a countable number of open intervals contains a finite subcover which already cover the interval. This is what the famous Heine-Borel Theorem states. We give below Borel's proof~\cite[vol. I, p. 163]{Fichtenholz}.

This section assumes that {\AC} holds.

We will prove in this section Alexander's Subbase Theorem as an
application for Zorn's Lemma. The Theorem states that when proving a
space compact one may restrict one's attention to a particular
subclass of open sets, a class which is usually easier to handle than
the full family of open sets. This application of Zorn's Lemma is
interesting because it shows in which way a maximality argument can be
used for establishing a property through a subclass (rather than
extending a property until maximality puts a stop to it, as we did in
showing that each vector space has a basis). Alexander's Theorem is
also a very practical tool, as we will see later.

This is the Heine-Borel Theorem.

\BeginTheorem{heine-borel}
Let an interval $[u, v]$ with $-\infty < u < v < +\infty$ be
given. Then each cover $\bigl\{]x_n, y_n[ \mid n\in\Nat\bigr\}$ of
$[u, v]$ through a countable number of open intervals contains a
finite cover $]x_{n_{1}}, y_{n_{1}}[, \dots, ]x_{n_{k}}, y_{n_{k}}[$.
\EndTheorem

\BeginProof
Suppose the assertion is
false, then either $[u, 1/2(u+v)]$ or $[1/2(u+v), v]$ is not covered
by finitely many of those intervals; select the corresponding one, call it
$[a_1, b_1]$. This interval can be halved, let $[a_2, b_2]$ be the half which
cannot be covered by finitely many intervals. Repeating this process,
one obtains a sequence $\{[a_n, b_n]\mid n\in \Nat\}$ of intervals,
each having half of the length of its predecessor, and each one not
being covered by an finite number of intervals from $\{]x_n, y_n[ \mid
n\in\Nat\}$. Because the lengths of the intervals shrink to zero,
there exists $c\in[u, v]$ with $\lim_{n\to\infty} a_n = c = \lim_{n\to\infty} b_n$, hence
$c\in ]x_m, y_m[$ for some $m$. But there is some $n_0\in\Nat$ with $[a_n,
b_n]\subseteq]x_m, y_m[$ for $n\geq n_0$, contradicting the assumption
that $[a_n, b_n]$ cannot be covered by a finite number of those intervals.
\EndProof

Although the proof is given for a countable cover, its analysis
shows that it goes through for an arbitrary cover of open
intervals (this is so because each cover induces a partition of the
interval considered into two parts, so a sequence of intervals will
result in any case). This section will discuss compact spaces which
have the property that an arbitrary cover contains a finite one.
To be on firm ground, we first introduce topological spaces as the kind of
objects to be discussed here.

\BeginDefinition{top-space}
Given a set $X$, a subset $\tau\subseteq\PowerSet{X}$ is called a \emph{\index{topology}topology} iff these conditions are satisfied:
\begin{itemize}
\item $\emptyset, X\in \tau$.
\item If $G_{1}, \dots, G_{k}\in \tau$, then $G_{1}\cap\dots\cap G_{k}\in \tau$, thus $\tau$ is closed under finite intersections.
\item If $\tau_{0}\subseteq \tau$, then $\bigcup\tau_{0}\in\tau$, thus $\tau$ is closed under arbitrary unions. 
\end{itemize}
The pair $(X, \tau)$ is then called a \emph{\index{space!topological}topological space}, the elements of $\tau$ are called \emph{open sets}. An \emph{open neighborhood} $U$ of an element $x\in X$ is an open set $U$ with $x\in U$. 
\EndDefinition

These are the topologies one can always find on a set $X$.

\BeginExample{top-spaces-ex1}
$\PowerSet{X}$ and $\{\emptyset, X\}$ are always topologies; the former one is called the \emph{discrete topology}, the latter one is called \emph{indiscrete}. 
\EndExample

The topology one deals with usually on the reals is given by intervals, and the plane is topologically described by open balls (well, they really are circles, but they are given through measuring a distance, and in this case the name ``ball'' sticks). 

\BeginExample{top-spaces-ex2}
Call a set $G\subseteq \Real$ open iff for each $x\in G$ there exists $a, b\in \Real$ with $a < b$ such that $x\in ]a, b[\subseteq G$; note that $\emptyset$ is open. Then the open sets form a topology on the reals, which is also called the \emph{interval topology}. Clearly, $G$ is open iff, given $x\in G$, there exists $\epsilon>0$ with $]x-\epsilon, x+\epsilon[\subseteq G$. Call a subset $G\subseteq R^{2}$ of the Euclidean plane open iff, given $x\in G$, there exists $\epsilon>0$ such that $B_{\epsilon}(x) \subseteq G$, where 
$
B_{r}(x_{1}, x_{2}):= \{\langle y_{1}, y_{2}\rangle \mid \sqrt{(y_{1}-x_{1})^{2}+(y_{2}-x_{2})^{2}} < r\}
$
is the open ball centered at $\langle x_{1}, x_{2}\rangle$ with radius $r$. 
\EndExample

Let $(X, \tau)$ be a topological space. If $Y\subseteq X$, then the
trace of $\tau$ on $Y$ gives a topology $\tau_{Y}$ on $Y$, formally,
$\tau_{Y} := \{G\cap Y\mid G\in \tau\}$, the \emph{subspace
  \index{topology!subspace}topology}. This permits sometimes to
transfer a property from the space to its subsets. 

A set $F\subseteq X$ is called \emph{closed} iff its complement
$X\setminus F$ is open. Then both $\emptyset$ and $X$ are closed, and
the closed sets are closed (no pun intended) under arbitrary
intersections and finite unions. We associate with each set an open
set and a closed set:

\BeginDefinition{interior-and-closure}
Let $M\subseteq X$, then 
\begin{itemize}
\item $\Interior{M} := \bigcup\{G\in \tau\mid G\subseteq M\}$ is called the \emph{\index{interior}interior} of $M$.
\item $\Closure{M} := \bigcap\{F\subseteq X\mid M\subseteq F\text{ and $F$ is closed}\}$ is called the \emph{\index{closure}closure} of $M$.
\item $\partial M := \Closure{M}\setminus\Interior{M}$ is called the \emph{\index{boundary}boundary} of $M$. 
\end{itemize}
\EndDefinition

We have always $\Interior{M}\subseteq M \subseteq \Closure{M}$; this
is apparent from the definition.  Clearly, $\Interior{M}$ is an open
set, and it is the largest open set which is contained in $M$, so that
$M$ is open iff $M = \Interior{M}$. Similarly, $\Closure{M}$ is a
closed set, and it is the smallest closed set which contains $M$. We
also have $M$ is closed iff $M = \Closure{M}$. The boundary $\partial
M$ is also a closed set, because it is the intersection of two closed
sets, and we have $\partial M = \partial{X\setminus M}$. $M$ is closed
iff $\partial M \subseteq M$. All this is easily established through
the definitions.

Look at the indiscrete topology: Here we have $\Interior{\{x\}} = \emptyset$ and
$\Closure{\{x\}} = X$ for each $x\in X$. For the discrete topology one sees
$\Interior{A} = \Closure{A} = A$ for each $A\subseteq X$.

\BeginExample{boundary-euclidean}
In the Euclidean topology on $\Real^{2}$ of Example~\ref{top-spaces-ex2}, we have 
\begin{align*}
  \Closure{B_{r}(x_{1}, x_{2})} & = \{\langle y_{1}, y_{2}\rangle \mid
  \sqrt{(y_{1}-x_{1})^{2}+(y_{2}-x_{2})^{2}}\leq r\},\\
  \partial{B_{r}(x_{1}, x_{2})} & = \{\langle y_{1}, y_{2}\rangle \mid
  \sqrt{(y_{1}-x_{1})^{2}+(y_{2}-x_{2})^{2}}= r\}.
\end{align*}
\EndExample

Just to get familiar with boundaries:

\BeginLemma{char-boundary}
Let $(X, \tau)$ be a topological space, $A\subseteq X$. Then
$x\in\partial A$ iff each neighborhood of $x$ has a non-empty
intersection with $A$ and with $X\setminus A$. In particular $\partial
A = \partial (X\setminus A)$ and $\partial (A\cup B) \subseteq
(\partial A) \cup (\partial B)$.
\EndLemma

\BeginProof
Let $x\in \partial A$, and $U$ an open neighborhood of $x$. If $A\cap
U = \emptyset$, then $A\subseteq X\setminus U$, so $x\not\in
\Closure{A}$, if $U\cap X\setminus A = \emptyset$, it follows $x \in
\Interior{A}$. So we arrive at a contradiction. Assume that
$x\in\bigcap\{U\mid x\in U, U\cap A \not= \emptyset, U\cap X\setminus
A \not= \emptyset\}$, then $x\not\in \Interior{A}$, similarly,
$\not\in \Interior{X\setminus A} = X\setminus(\Closure{A})$.
\EndProof

A set without a boundary is both closed and open, so it is called
\emph{\index{clopen}clopen}. The clopen sets of a topological space
form a Boolean algebra.

Sometimes it is sufficient to describe the topology in terms of some
special sets, like the open balls for the Euclidean topology.

\BeginDefinition{def-top-base}
A subset ${\cal B}\subseteq \tau$ of the open sets is called a
\emph{\index{topology!base}base for the topology} iff for each open
set $G\in\tau$ and for each $x\in G$ there exists $B\in {\cal B}$ such
that $x\in B\subseteq G$, thus each open set is the union of all base
elements contained in it. ${\cal S}\subseteq \tau$ is called a
\emph{\index{topology!subbase}subbase} for $\tau$ iff the set of
finite intersections of elements of ${\cal S}$ forms a base for
$\tau$.
\EndDefinition

Then the open intervals are a base for the interval topology, and the
open balls are a base for the Euclidean topology (actually, we did
introduce the respective topologies through their bases). A subbase
for the interval topology is given by the sets $\{]-\infty, a[\mid
a\in\Real\}$, because the set of finite intersections includes all
open intervals, which in turn form a base. Bases and subbases are not
uniquely determined, for example is $\{]r, s[\mid r<s, r, s
\in\Rational\}$ a base for the interval topology.

Let us return to the problem discussed in the opening of this
section. We have seen that finite open intervals have the remarkable
property that, whenever we cover them by an arbitrary number of open
intervals, we can find a finite collection among these intervals which
already cover the interval. This property can be generalized to
arbitrary topological spaces; subsets with this properties are called
compact, formally:

\BeginDefinition{compact-subset}
The topological space $(X,\tau)$  is called \emph{compact} iff each cover of $X$ by open sets contains a finite subcover. 
\EndDefinition

Thus $X$ is compact iff, whenever $(G_{i})_{i\in I}$ is a collection of open sets with $X = \bigcup_{i\in I}G_{i}$, there exists $I_{0}\subseteq I$ finite such that $C\subseteq \bigcup_{i\in I_{0}}G_{i}$. It is apparent that compactness is a generalization of finiteness, so that compact sets are somewhat small, measured in terms of open sets. Consider as a trivial example the discrete topology. Then $X$ is compact precisely when $X$ is finite. 

This is an easy consequence of the definition.

\BeginLemma{closed-is-compact}
Let $(X, \tau)$ be a compact topological space, and $F\subseteq X$. Then $F$ is compact (thus $(F, \tau_{F})$ is a compact topological space). \QED
\EndLemma

The following example shows a close connection of Boolean algebras to compact topological spaces; this is the famous \emph{Stone Duality \index{Theorem!Stone Duality}Theorem}.

\BeginExample{stone-space-compact}
Let $B$ be a Boolean algebra with $\wp_{B}$ as the set of all prime ideals of $B$. Define 
\begin{equation*}
  X_{a} := \{I\in \wp_{B}\mid a\not\in I\}.
\end{equation*}
Then we have these properties
\begin{itemize}
\item $X_{\top} = \wp_{B}$, since an ideal does not contain $\top$. 
\item $X_{-a} = \wp_{B}\setminus X_{a}$. To see this, let $I$ be a prime filter, then $I\in X_{-a}$ iff $-a\not\in I$, this is the case iff $-a\in B\setminus I$, hence iff $a\not\in B\setminus I$, since $B\setminus I$ is a maximal filter by Lemma~\ref{compl-prime-id} and Lemma~\ref{prime-equiv-max}; the latter condition is equivalent to $a\in I$, hence to $I\not\in X_{a}$.
\item $X_{a\wedge b} = X_{a}\cap X_{b}$ and $X_{a\vee b} = X_{a}\cup X_{b}$ This follows similarly from Lemma~\ref{prop-filter}. 
\end{itemize}
Define a topology $\tau$ on $\wp_{B}$ by taking the sets $X_{a}$ as a base, formally
\begin{equation*}
  {\cal B} := \{X_{a}\mid a\in B\}.
\end{equation*}
We claim that $(\wp_{B}, \tau)$ is compact. In fact, let ${\cal U}$ be
a cover of $\wp_{B}$ with open sets. Because each $U\in{\cal U}$ can
be written as a union of elements of ${\cal B}$, we may and do assume
that ${\cal U}\subseteq{\cal B}$, so that ${\cal U} = \{X_{a}\mid a\in
A\}$ for some $A\subseteq B$. Now let $J$ be the ideal generated by $A$, so that $J$ can be written as
\begin{equation*}
  J := \{b\in B\mid b\leq a_{1}\vee \dots\vee a_{k}\text{ for some }a_{1}, \dots, a_{k}\in A\}.
\end{equation*}
We distinguish these cases:
\begin{description}
\item[$\top\in J$:] In this case we have $\top = a_{1}\vee\dots\vee a_{k}$ for some $a_{1}, \dots, a_{k}\in A$, which means 
  \begin{equation*}
\wp_{B} = X_{\top} = X_{a_{1}\vee\dots\vee a_{k}} = X_{a_{1}}\cup \dots\cup X_{a_{k}}
\end{equation*}
with $X_{a_{1}}, \dots, X_{a_{k}}\in {\cal U}$, so we have found a finite subcover in ${\cal U}$.
\item[$\top\not\in J$:] Then $J$ is a proper ideal, so by Corollary~\ref{prime-ideal-thm} there exists a prime ideal $K$ with $J\subseteq K$. But we cannot find $a\in A$ such that $K$ equals $X_{a}$, so $K\in\wp_{B}$, but $K$ fails to be covered by ${\cal U}$, which is a contradiction. 
\end{description}
Thus $(\wp_{B}, \tau)$ is a compact space, which is sometimes called the \emph{prime ideal space} of the Boolean algebra. 

We conclude that the sets $X_{a}$ are clopen, since $X_{-a} =
\wp_{B}\setminus X_{a}$. Moreover, each clopen set in this space can
be represented in this way. In fact, let $U$ be clopen, thus $U =
\bigcup\{X_{a}\mid a\in A\}$ for some $A\subseteq B$. Since $U$ is closed, it is compact by
Lemma~\ref{closed-is-compact}, so there exist $a_{1}, \dots, a_{n}\in
A$ such that $U = X_{a_{1}}\cup\dots\cup X_{a_{n}} =
X_{a_{1}\vee\dots\vee a_{n}}$.
\EndExample

Compactness is formulated in terms of a cover through arbitrary open sets. Alexander's Theorem states that it is sufficient to consider covers which come from a subbase for the topology. This is usually quite a considerable help, since subbases are sometimes easier to handle than the connection of all open sets. The proof comes as an application of Zorn's Lemma. The proof follows essentially the one given for~\cite[Theorem 6.40]{Hewitt-Stromberg}.

\BeginTheorem{alexander-subbase}
Let $(X, \tau)$ be a topological space with a subbase ${\cal S}$. Then the following statements are equivalent.
\begin{enumerate}
\item\label{alexander-subbase:1} $X$ is compact.
\item\label{alexander-subbase:2} Each cover of $X$ by elements of ${\cal S}$ contains a finite subcover.
\end{enumerate}\index{Theorem!Alexander's Subbase}
\EndTheorem

\BeginProof
Because the elements of a subbase are open, the
implication \labelImpl{alexander-subbase:1}{alexander-subbase:2} is
trivial, hence we have to
show \labelImpl{alexander-subbase:2}{alexander-subbase:1}. Assume that
the assertion is false, and define
\begin{equation*}
  {\mathfrak Z} := \{{\cal C}\mid {\cal C} \text{ is on open cover of $X$ without a finite subcover}\}.
\end{equation*}
Order ${\mathfrak Z}$ by inclusion, and let ${\mathfrak
  Z}_{0}\subseteq {\mathfrak Z}$ be a chain, then ${\cal C} := \bigcup
{\mathfrak Z}_{0}\in {\mathfrak Z}$. In fact, it is clear that ${\cal
  C}$ is a cover, and assume that ${\cal C}$ has a finite subcover,
say $\{E_{1}, \dots, E_{k}\}$. Then $E_{j}\in {\cal
  C}_{j}\in{\mathfrak Z}_{0}$, and since ${\mathfrak Z}_{0}$ is a
chain with respect to inclusion, we find some ${\cal
  C}_{i}\in{\mathfrak Z}_{0}$ with $\{E_{1}, \dots, E_{k}\}\subseteq
{\cal C}_{i}$, which is a contradiction. By Zorn's Lemma, ${\mathfrak
  Z}$ has a maximal element ${\cal V}$. This means
that 
\begin{itemize}
\item ${\cal V}$ is an open cover of $X$.
\item ${\cal V}$ does not contain a finite subcover.
\item If $U\in \tau$ is open with $U\not\in{\cal V}$, then ${\cal V}\cup\{U\}$ contains a finite subcover. 
\end{itemize}
Let ${\cal W} := {\cal V}\cap{\cal S}$, hence all elements of ${\cal
  V}$ which are taken from the subbase. By assumption, no finite
subfamily of ${\cal W}$ covers $X$, hence ${\cal W}$ is not a cover
for $X$, which implies that $R := X\setminus \bigcap {\cal
  W}\not=\emptyset$. Let $x\in R$, then there exists $V\in {\cal V}$
such that $x\in V$. Since $V$ is open and ${\cal S}$ is a subbase, we
find $S_{1}, \dots, S_{k}\in{\cal S}$ with $x\in S_{1}\cap\dots\cap
S_{n}\subseteq V$. Because $x\not\in\bigcup{\cal W}$, we conclude that
no $S_{j}$ is an element of ${\cal V}$ (otherwise $S_{j}\in {\cal
  V}\cap{\cal S}={\cal W}$, a contradiction). ${\cal V}$ is maximal,
each $S_{j}$ is open, thus ${\cal V}\cup\{S_{j}\}$ contains a finite
cover of $X$. Hence we can find for each $j$ some open set $A_{j}$
which is a finite union of elements in ${\cal V}$ such that $A_{j}\cup
S_{j} = X$. But this means
\begin{equation*}
  V\cup \bigcup_{j=1}^{k}A_{j} \supseteq (\bigcap_{j=1}^{k}S_{j})\cup(\bigcup_{j=1}^{k}A_{j}) = X.
\end{equation*}
Hence $X$ can be covered through a finite number of elements in ${\cal V}$; this is a contradiction to the maximality of ${\cal V}$. 
\EndProof

The Priestley topology as discussed by
Goldblatt~\cite{Goldblatt_TopProofs} provides a first example for the
use of Alexander's Theorem.

\BeginExample{pristeley-topology}
Given $x\in X$, define
\begin{align*}
  \|x\| & := \{A \subseteq X \mid  x \in A\},\\
-\|x\| & := \{A\subseteq X \mid x\not\in A\}.
\end{align*}
The \emph{Priestley \index{topology!Priestley}topology} on $\PowerSet{X}$ is defined as the topology which is generated by the subbase
\begin{equation*}
  {\cal S} := \{\|x\|\mid x\in X\}\cup\{-\|x\|\mid  x \in X\}.
\end{equation*}
Hence the basic sets of this topology have the form
\begin{equation*}
  \|x_{1}\|\cap\dots\cap\|x_{k}\|\cap-\|y_{1}\|\cap\dots\cap-\|y_{n}\|
\end{equation*}
for $x_{1}, \dots, x_{k}, y_{1}, \dots, y_{n}\in X$ and some $n, k\in\Nat$. 

We claim that $\PowerSet{X}$ is compact in the Priestley topology. In
fact, let ${\cal C}$ be a cover of $\PowerSet{X}$ with elements from
the subbase ${\cal S}$. Put $P := \{x\in X\mid -\|x\|\in{\cal
  C}\}$. Then $P\in\PowerSet{X}$, so we must find some element from
${\cal C}$ which contains $P$. If $P\in -\|x\|\in{\cal C}$ for some
$x\in X$, this means $x\not\in P$, so by definition $-\|x\|\not\in
{\cal C}$, which is a contradiction. Thus there exists $x\in X$ such
that $P\in\|x\|\in{\cal C}$. But this means $x\in P$, hence
$-\|x\|\in{\cal C}$, so $\{\|x\|, -\|x\|\}\subseteq{\cal C}$ is a
cover of $\PowerSet{X}$. This $\PowerSet{X}$ is compact by Alexander's Theorem~\ref{alexander-subbase}.
\EndExample 


\Subsection{Boolean $\sigma$-Algebras}
\label{sec:bool-sigma-algebr}

We generalize the notion of a Boolean algebra by introducing countable
operations, leading to Boolean $\sigma$-algebras. This extension
becomes important e.g., when working with probabilities or, more
general, with measures. For example, one of the fundamental
probability laws states that the probability of a disjoint union of
countable events equals the infinite sum of the events'
probabilities. In order to express this adequately, the domain of the
probability must be closed under countable unions.

\medskip{}
We assume in this section that {\AC} holds.
\medskip{}

Given a Boolean algebra $B$, we associate with the lattice operations
on $B$ an order relation $\leq$ by 
\begin{equation*}
  a \leq b \Longleftrightarrow a\wedge b = a~(\Longleftrightarrow a \vee b = b).
\end{equation*}
We will switch in the discussion below between the order and the
use of the algebraic operations.

\BeginDefinition{boolean-sigma-algebra}
A Boolean algebra $B$ is called a \index{Boolean \index{$\sigma$-algebra}$\sigma$-algebra}Boolean $\sigma$-algebra iff it is
closed under countable suprema and infima.
\EndDefinition

\BeginExample{exmpl-bool-algs}
The power set of each set is a Boolean $\sigma$-algebra. Consider
\begin{equation*}
  {\cal A} := \{A\subseteq\Real \mid A\text{ is countable or
  }\Real\setminus A \text{ is countable}\}.
\end{equation*}
Then ${\cal A}$ is a Boolean $\sigma$-algebra (we use here that the
countable union of countable set is countable again, hence
{\AC}). This is sometimes called the \emph{\index{$\sigma$-algebra!countable-cocountable}countable-cocountable
$\sigma$-algebra}. On
the other hand, her little sister, 
\begin{equation*}
  {\cal D} := \{A\subseteq\Real \mid A\text{ is finite or
  }\Real\setminus A \text{ is finite}\},
\end{equation*}
the \index{algebra!finite-cofinite}finite-cofinite algebra, is a Boolean algebra, but evidently no $\sigma$-algebra. 
\EndExample

If $\Folge{a}$ is an at most countable subset of the Boolean
$\sigma$-algebra $B$, then we define 
\begin{align*}
  \bigwedge_{n\in \Nat} a_n & := \inf \{a_n \mid n\in \Nat\},\\
\bigvee_{n\in \Nat} a_n & := \sup \{a_n \mid n\in \Nat\}.
\end{align*}
In addition, we note that
\begin{align*}
  \inf \emptyset & = \top,\\
\sup \emptyset & = \bot.
\end{align*}
We know that a Boolean algebra is a distributive lattice, for a Boolean $\sigma$-algebra a
stronger infinite distributive law holds.

\BeginLemma{sigma-bool-is-distrib}
Let $B$ be a Boolean $\sigma$-algebra, $\Folge{a}$ be a sequence of
elements in $B$, then
\begin{align*}
  b\wedge\bigvee_{n\in\Nat} a_n & = \bigvee_{n\in\Nat} (b\wedge a_n),\\
b\vee\bigwedge_{n\in\Nat} a_n & = \bigwedge_{n\in\Nat} (b\vee a_n)
\end{align*}
holds for any $b\in B$.
\EndLemma

\BeginProof
We establish the first equality, the second one follows by
duality. Since $b\wedge a_n \leq b$ and $b\wedge a_n\leq a_n$, we see that 
$ \bigvee_{n\in\Nat} (b\wedge a_n) \geq b\wedge\bigvee_{n\in\Nat}
a_n.$
For establishing the reverse inequality, assume that $s$ is an upper
bound to $\{b\wedge a_n \mid n\in\Nat\}$, hence  $b\wedge a_n\leq
s$ for all $n\in\Nat$, consequently, 
$
a_n = (b\wedge a_n)\vee(-b\wedge a_n) \leq s \vee (-b\wedge a_n)\leq
s\vee -b.
$ 
Thus
\begin{equation*}
b\wedge\bigvee_{n\in \Nat} a_n \leq b \wedge (s\vee -b) = (b\wedge
s)\vee(b\wedge-b) \leq b\wedge s \leq s.
\end{equation*}
Hence $s$ is an upper bound to $b\wedge\bigvee_{n\in \Nat} a_n$ as
well. Now apply this to the upper bound $s := \bigvee_{n\in\Nat} (b\wedge a_n)$.
\EndProof

Let $A$ be a non-empty subset of a Boolean $\sigma$-algebra $B$, then
there exists a smallest $\sigma$-algebra $C$ which contains $A$. In
fact, this must be
\begin{equation*}
  C = \bigcap\{D \subseteq B\mid D\text{ is a $\sigma$-algebra with }
  A\subseteq D\}.
\end{equation*}
We first note that the intersection of a set of $\sigma$-algebras is s
$\sigma$-algebra again. Moreover, there exists always a
$\sigma$-algebra which contains $A$, viz., the superset
$B$. Consequently, $C$ the the object of our desire, it is denoted by
$\sigma(A)$, so that $\sigma(A)$ denotes the smallest $\sigma$-algebra
containing $A$. $\sigma$ is an example for a \emph{closure operator}:
We have $A\subseteq \sigma(A)$, and $A_1\subseteq A_2$ implies
$\sigma(A_1)\subseteq\sigma(A_2)$, moreover, applying the operator
twice does not yield anything new: $\sigma(\sigma(A)) = \sigma(A)$. 

\BeginExample{borel-in-unit-line}
Let $A := \{[a, b] \mid a, b\in [0, 1]\}$ be the set of all closed
intervals $[a, b] := \{x \in \Real \mid a \leq x \leq b\}$ of the unit
interval $[0, 1]$. Denote by $B := \sigma(A)$ the
$\sigma$-algebra generated by $A$; the elements of $B$ are sometimes
called the\emph{ \index{Borel sets}Borel sets} of $[0, 1]$. Then the half open intervals $[a,
b[$ and $]a, b]$ are members of $B$. We can write, e.g.,  
$[a, b[ = \bigcup_{n\in \Nat} [a, b-1/n]$. Since $[a, b-1/n]\in A
\subseteq B$ for all $n\in \Nat$, and since $B$ is closed under
countable unions, the claim follows.

A more complicated Borel set is constructed in this way: define
$C_0 := [0, 1]$ and assume that $C_n$ is defined already as a 
union of $2^n$ mutually disjoint intervals of length $1/3^n$ each, say
$C_n = \bigcup_{1 \leq j \leq 2^n}I_j$. Obtain $C_{n+1}$ by removing
the open middle third of each interval $I_j$. For example
\begin{align*}
  C_0 = & [0, 1],\\
C_1   = & [0, 1/3]\cup[2/3, 1],\\
C_2  = & [0, 1/9]\cup[2/9, 1/3]\cup[2/3, 7/9]\cup[8/9, 1],\\
C_3  = & [0, 1/27]\cup[2/27, 1/9]\cup[2/9, 7/27]\cup[8/27]\cup[2/3,
19/27]\\&\cup[20/27, 7/9]\cup[8/9, 25/27]\cup[26/27, 1]
\end{align*}
and so on. Clearly $C_n\in B$, because this set is the finite union
of closed intervals. Now put
\begin{equation*}
  C := \bigcap_{n\in \Nat} C_n,
\end{equation*}
then $C\in B$, because it is the countable intersection of sets in
$B$. This set is known as the \emph{\index{Cantor ternary set}Cantor ternary set}. 
\EndExample

The next two examples deal with $\sigma$-algebras of sets, each defined
on the infinite product $\{0, 1\}^{\Nat}$. It may be used as a model
for an infinite sequence of flipping coins --- $0$ denoting head, $1$
denoting tail. But we can only observe a finite numbers of these
events, probably as long as we want. So we cater for that by having a
look at the $\sigma$-algebra which is defined by these finite
observations. 

\BeginExample{coin-flip-1}
Let $X := \{0, 1\}^\Nat$ be the set of all infinite binary sequences, and put
$
{\cal B} := \sigma(\{A_{k, i} \mid k \in \Nat, i = 0, 1\})
$
with $A_{k, i} := \{\langle x_1, x_2, \dots\rangle \mid x_k = i\}$ as
the set of all sequences the $k$-th component of which is $i$. 

We claim that that for $r\in\Natn$ both
$
S_{k, r} := \{\langle x_1, x_2, \dots\rangle \in X\mid x_1 + \dots + x_k = r\}
$
and 
$
T_r := \{\langle x_1, x_2, \dots\rangle  \in X\mid \sum_{i=0}^\infty x_i  = r\}
$
are elements of ${\cal B}$.

In fact, given a finite binary sequence $v := \langle
  v_1, \dots, v_k\rangle$, the set 
\begin{equation*}
Q_v := \{x\in X \mid \langle x_1, \dots x_k\rangle =v\} =
\bigcap_{i=1}^k A_{i, v_i}
\end{equation*}
is a member of ${\cal B}$, the set $L_{k, r}$ of binary sequences of
length $k$ which sum up to $r$ is finite. Thus
\begin{equation*}
S_{k, r} = \bigcup_{v\in L_{k, r}} Q_v \in{\cal B}.
\end{equation*} 
\EndExample

We continue the example by looking at all sequences for which the
average result of flipping a coin $n$ times will converge as $n$ tends
to infinity. 
\BeginExample{coin-flip-2} 
Let $X := \{0, 1\}^\Nat$ be the set of all infinite binary sequences
as in Example~\ref{coin-flip-1}, and put
\begin{equation*}
W := \{\langle x_1, x_2, \dots\rangle \in X \mid
\frac{1}{n}\sum_{i=1}^n x_i\text{ converges}\}.
\end{equation*}
We claim that $W\in {\cal B}$, noting that  a real sequence $\Folge{y}$
converges iff it is a Cauchy sequence, i.e., iff given
$0<\epsilon\in\Rational$ there exists $n_0\in\Nat$ such that $|y_m -
y_n| < \epsilon$ for all $n, m\geq n_0$. 

Given $F\subseteq\Nat$ finite, the set 
\begin{align*}
  H_F & := \{x\in X \mid x_j=1 \text{ for all } j\in F\text{ and }x_i =
  0 \text{ for all }i\not\in F\}\\
&  = \bigcap_{j\in F} A_{j, 1} \cap
  \bigcap_{i\not\in F} A_{i, 0}
\end{align*}
is a member of ${\cal B}$; since there are countably many finite
subsets of $\Nat$ which have exactly $r$ elements, we obtain
$
T = \bigcup \{H_F \mid F\subseteq\Nat\text{ with } |F|=r\},
$ 
which is a countable union of elements of ${\cal B}$, hence an
element of ${\cal B}$.

The sequence $\bigl(\frac{1}{n}\sum_{i=1}^n x_i\bigr)_{n\in\Nat}$
converges iff
\begin{equation*}
\forall \epsilon>0, \epsilon\in\Rational\exists
n_0\in\Nat\forall n\geq n_0\forall m \geq n_0: \bigl|\frac{1}{n}\sum_{i=1}^n
x_i - \frac{1}{m}\sum_{i=1}^m x_i\bigr| < \epsilon,
\end{equation*}
thus iff
\begin{equation*}
\langle x_1, x_2, \dots\rangle\in 
\bigcap_{\epsilon>0,
  \epsilon\in\Rational}\bigcup_{n_0\in\Nat}\bigcap_{\Nat\ni n\geq
  n_0}\bigcap_{\Nat\ni m\geq n_0} W_{n, m, \epsilon}.
\end{equation*}
with
\begin{equation*}
W_{n, m, \epsilon} := \{\langle x_1, x_2, \dots \rangle \mid \bigl|\frac{1}{n}\sum_{i=1}^n
x_i - \frac{1}{m}\sum_{i=1}^m x_i\bigr| < \epsilon\}.
\end{equation*}
Now $\langle x_1, x_2, \dots \rangle\in W_{n, m, \epsilon}$ iff 
$
\bigl|m\cdot \sum_{i=1}^n x_i - n\cdot \sum_{j=1}^m x_j\bigr| < n\cdot
m\cdot \epsilon.
$
If $n < m$, this is equivalent to 
\begin{equation*}
  -n\cdot m\cdot \epsilon < (m-n)\cdot\sum_{i=1}^n x_i - n\cdot
  \sum_{j=n+1}^m x_j < n\cdot m\cdot \epsilon
\end{equation*}
hence 
$ -n\cdot m\cdot \epsilon < (m-n)\cdot a - n\cdot b < n\cdot m\cdot
\epsilon$
for $a = \sum_{i=1}^n x_i$ and $b = \sum_{j=n+1}^m x_j$; the same
applies to the case $m < n$. Since there are only finitely many
combinations of $\langle a, b\rangle$ satisfying these constraints, we
conclude that $ W_{n, m, \epsilon}\in{\cal B}$, so that the set $W$ of all
sequences for which the average sum converges is a member of ${\cal
  B}$ as well.
\EndExample

\Subsubsection{Construction Through Transfinite Induction}
\label{sec:constr-thro-transf}

The description of $\sigma(A)$ given above is  non-constructive; it is
done through a
closure operation, from the outside, so to speak. Transfinite
induction permits us to construct $\sigma(A)$. In order to describe
it, we introduce two operators on the subsets of $B$ as follows. Let
$H\subseteq B$, then
\begin{align*}
  H_\sigma & := \{\bigvee_{n\in\Nat} a_n\mid a_n\in H\text{ for all }
  n\in \Nat\},\\
H_\delta & := \{\bigwedge_{n\in\Nat} a_n\mid a_n\in H\text{ for all }
  n\in \Nat\}.
\end{align*}
Thus $H_\sigma$ contains all countable suprema of elements of $H$, and
$H_\delta$ contains all countable infima. Hence $A$ is a Boolean
sub $\sigma$-algebra of $B$ iff
\begin{equation*}
  A_\sigma \subseteq A, A_\delta \subseteq A, \text{ and }\{-a \mid a\in
  A\}\subseteq A
\end{equation*}
hold. 

So couldn't we, when constructing $\sigma(A)$, just take all
complements, then all countable infima and suprema of elements in $A$,
and then their countable suprema and infima, and so on? This is the
basic idea for the construction. But since the process indicated above
is not guaranteed to terminate after a finite number of applications
of the $\sigma$-and the $\delta$-operations, we do a transfinite
construction.

So fix $A\subseteq B$, and define by transfinite
induction\label{transfinite-def-xxx}
\begin{align*}
  A_0 & := A\cup\{-a \mid a\in A\},\\
A_\zeta & := \bigcup_{\eta < \zeta} A_\eta, \text{ if $\zeta$ is a
  limit ordinal}\\
A_{\zeta+1} & := (A_\zeta)_\sigma, \text{ if $\zeta$ is odd},\\
A_{\zeta+1} & := (A_\zeta)_\delta, \text{ if $\zeta$ is even},\\
A_\omega & := \bigcup_{\zeta<\omega} A_\zeta.
\end{align*}
It is clear that $A_\zeta\subseteq C$ holds for each $\sigma$-algebra
$C$ which contains $A$, so that $A_\omega\subseteq\sigma(A)$ is
inferred. 

Let us work on the other inclusion. It is sufficient to show that
$A_\omega$ is a $\sigma$-algebra. This is so because $A\subseteq
A_\omega$, so that is this case $A_\omega$ would contribute to the
intersections defining $\sigma(A)$, hence we could infer
$A_\omega\subseteq\sigma(A)$. We proof the assertion through a series
of auxiliary statements, noting that $\langle A_\zeta \mid
\zeta<\omega\rangle$ forms a chain with respect to set inclusion.

\BeginLemma{transf-sigma-compl}
For each $\zeta<\omega$, if $a\in A_\zeta$, then $-a\in A_{\zeta+1}$.
\EndLemma

\BeginProof
The proof proceeds by transfinite induction. The assertion is true for
$\zeta = 0$; assume that it is true for all $\eta < \zeta$. 

If $\zeta$ is a limit ordinal, we know that we can find for $a\in
A_\zeta$ an ordinal $\eta < \zeta$ with $a\in A_\eta$, hence by
induction hypothesis $-a\in A_{\eta+1}\subseteq A_\zeta$, because
$\eta+1<\zeta$ by the definition of a limit ordinal (see
Definition~\ref{def-lim-ord} on page~\pageref{def-lim-ord}).

If $\zeta$ is even, but not a limit ordinal, we can write $\zeta$ as
$\zeta = \xi+1$. Then $A_\zeta = (A_\xi)_\sigma$, hence 
$a = \bigvee_{n\in \Nat} a_n$ for some $a_n\in A_\xi\subseteq
A_\zeta$, so that $-a = \bigwedge_{n\in \Nat} (-a_n)\in (A_\zeta)_\delta
= A_{\zeta+1}$. The argumentation for $\zeta$ odd is exactly the
same. 
\EndProof

Thus $A_\omega$ is closed under complementation. Closure under
countable infima and suprema is shown similarly, but we have to cater
somehow for a countable sequence of countable ordinals.

\BeginLemma{transf-sigma-countable}
$A_\omega$ is closed under countable infima and countable suprema.
\EndLemma

\BeginProof
We focus on countable suprema, the proof for infima works exactly in
the same way. Let $\Folge{a}$ be a sequence of elements in $A_\omega$,
then we find ordinal numbers $\zeta_n<\omega$ such that $a_n\in
A_{\zeta_n}$. Because $\zeta_n$ is countable for each $n\in \Nat$, we
conclude from Proposition~\ref{omega-is-ordinal} that $\zeta^* :=
\bigcup_{n\in \Nat} \zeta_n$ is a countable ordinal, so that
$\zeta^*<\omega$. Because  $\langle A_\zeta\mid
\zeta<\omega\rangle$ forms a chain, we infer that $a_n\in A_{\zeta^*}$
for all $n\in \Nat$. Consequently, $\bigvee_{n\in \Nat} a_n\in
(A_{\zeta^*})_\sigma\subseteq A_\omega$.  
\EndProof

Thus we have shown

\BeginProposition{transf-constr-sigma-alg}
$A_\omega = \sigma(A)$.
\QED
\EndProposition

\Subsubsection{Factoring Through $\sigma$-Ideals}
\label{sec:sigma-ideals}

Factoring a Boolean $\sigma$-algebra through an ideal works as for
general Boolean algebras, resulting in a Boolean algebra again. There
is no reason why the factor algebra should be a $\sigma$-algebra,
however, so if we want to obtain a $\sigma$-algebra we have to make
stronger assumptions on the object used for factoring. 

\BeginDefinition{sigma-ideal}
Let $B$ be a Boolean algebra, $I\subseteq B$ an ideal. $I$ is called a
\index{$\sigma$-ideal}$\sigma$-ideal iff $\sup_{n\in \Nat} a_n\in I$, provided $a_n\in I$
for all $n\in \Nat$. 
\EndDefinition

Not every ideal is a $\sigma$-ideal: $\{F\subseteq\Nat \mid F\text{ is
  finite}\}$ is an ideal but certainly not a $\sigma$-ideal in
$\PowerSet{\Nat}$, even if $\PowerSet{\Nat}$ is a Boolean
$\sigma$-algebra.

The following statement is the $\sigma$-variant of
Proposition~\ref{factor-is-ba}.

\BeginProposition{factoring-sigma-ideal}
Let $B$ be a Boolean $\sigma$-algebra and $I\subseteq B$ be a
$\sigma$-ideal. Then $\Faktor{B}{I}$ is a Boolean $\sigma$-algebra.
\EndProposition

\BeginProof\Rand{\cite[p. 79]{Aumann}}
1.
Because each Boolean $\sigma$-algebra is a Boolean algebra, and each
$\sigma$-ideal is an ideal, we may conclude from
Proposition~\ref{factor-is-ba} that $\Faktor{B}{I}$ is a Boolean
algebra. Hence it remains to be shown that this Boolean algebra is
closed under countable suprema; since $\Faktor{B}{I}$ is closed under
complementation,  closedness under countable infima will follow.

2.
Let $a_n\in B$, then $a := \bigvee_{n\in \Nat} a_n\in B$. We claim
that $\Klasse{a}{\sim_I} = \bigvee_{n\in \Nat}
\Klasse{a_n}{\sim_I}$. Because $a_n \leq a$ for all $n\in \Nat$, we
conclude that $\Klasse{a_n}{\sim_I}\leq\Klasse{a}{\sim_I}$ for all
$n\in\Nat$, hence $\bigvee_{n\in\Nat}
\Klasse{a_n}{\sim_I}\leq\Klasse{a}{\sim_I}$. Now let
$\Klasse{a_n}{\sim_I}\leq \Klasse{b}{\sim_I}$ for all $n\in \Nat$,
then we show that $\Klasse{a}{\sim_I}\leq \Klasse{b}{\sim_I}$. In
fact, because $\Klasse{a_n}{\sim_I}\leq \Klasse{b}{\sim_I}$ we
conclude that $c_n := a_n\ominus (a_n\wedge b)\in I$ and $b\wedge
c_n=\bot$ for all $n\in \Nat$ (since $c_n = a_n\wedge-(a_n\wedge
b)$). Thus $a_n = c_n\vee (a_n\wedge c_n)$, so that we have $\bigvee_{n\in \Nat} a_n = (a\wedge b) \vee \bigvee_{n\in \Nat}
c_n$ by the
infinite distributive law from Lemma~\ref{sigma-bool-is-distrib}. This implies $a\ominus (a\wedge b) = \bigvee_{n\in \Nat} c_n\in
I$, or, equivalently, $\Klasse{a}{\sim_I}\leq
\Klasse{b}{\sim_I}$. Consequently, $\Klasse{a}{\sim_I}$ is the
smallest upper bound to $\{\Klasse{a_n}{\sim_I} \mid n\in \Nat\}$. 
\EndProof

\Subsubsection{Measures}
\label{sec:measures}

Boolean $\sigma$-algebras model events. The top element $\top$ is
interpreted as an event which can happen unconditionally and always,
the bottom element $\bot$ is the impossible event. The complement of
an event is an event, and if we have a countable sequence of events,
then their supremum is an event, viz., the event that at least one of
the event in the sequence happens. 

To illustrate, suppose that we have a set $T$ of
traders which may form unions or coalitions, then $T$ as well as
$\emptyset$ are coalitions; if $A$ is a coalition, then $T\setminus A$
is a coalition as well, and if $A_n$ is a coalition for each $n\in
\Nat$, then we want to be able to form the ``big'' coalition $\bigcup_{n\in
  \Nat} A_n$. Hence the set of all coalitions forms a
$\sigma$-algebra.

We deal in the sequel with set based $\sigma$-algebras, so we fix a
set $S$ of events. 

\BeginDefinition{superadd-set-fnct}
Let ${\cal C}\subseteq\PowerSet{S}$ be a family of sets with
$\emptyset\in{\cal C}$. A map $\mu: ${\cal C}$\to[0, \infty]$ with $\mu(\emptyset) = 0$ is called
\begin{enumerate}
\item \emph{monotone} iff $\mu(A) \leq \mu(B)$ for $A, B\in {\cal C}, A\subseteq B$,
\item \emph{\index{additive}additive} iff $\mu(A\cup B) = \mu(A) + \mu(B)$ for
  all $A, B\in{\cal C}$, with $A\cup B\in {\cal C}$ and $A\cap B = \emptyset$,
\item \emph{\index{countably subadditive}countably subadditive} iff 
$
\mu(\bigcup_{n\in \Nat} A_n) \leq  \sum_{n\in \Nat} \mu(A_n),
$ 
whenever $\Folge{A}$ is a sequence of sets in ${\cal C}$ with
$\bigcup_{n\in\Nat} A_n\in{\cal C}$,
\item \emph{\index{countably additive}countably additive} iff 
$
\mu(\bigcup_{n\in \Nat} A_n) =  \sum_{n\in \Nat} \mu(A_n),
$ 
whenever $\Folge{A}$ is a mutually disjoint  sequence of sets in ${\cal C}$ with
$\bigcup_{n\in\Nat} A_n\in{\cal C}$,
\end{enumerate}
If ${\cal C}$ is a $\sigma$-algebra, then a map $\mu: ${\cal C}$\to[0,
\infty]$ with $\mu(\emptyset) = 0$ is called a \emph{\index{measure}measure} iff $\mu$ is
monotone and countably additive.  
\EndDefinition

Note that we permit that $\mu$ assumes the value $+\infty$. Clearly, a
countably additive set function is additive, and it is countably
subadditive, provided it is monotone. 

\BeginExample{dirac-measure}
Let $S$ be a set, and define for $a\in S, A\subseteq S$
\begin{equation*}
  \delta_a(A) :=
  \begin{cases}
    1, & \text{ if } a \in A\\
    0, & \text{otherwise}.
  \end{cases}
\end{equation*}
Then $\delta_a$ is a measure on the power set of $S$. It is usually
referred to a the \emph{\index{measure!Dirac}Dirac measure} on $a$. 
\EndExample

A slightly more complicated example indicates the connection to ultrafilters.

\BeginExample{meas-through-ultra}
Let $\mu: \PowerSet{S}\to\{0, 1\}$ be a binary valued measure. Define
\begin{equation*}
  {\cal F} := \{A\subseteq S \mid \mu(A) = 1\}
\end{equation*}
Then ${\cal F}$ is an ultrafilter on $\PowerSet{S}$. First, we check
that ${\cal F}$ is a filter: $\emptyset\not\in{\cal F}$ is obvious,
and if $A\in{\cal F}$ with $ A\subseteq B$, then certainly $B\in{\cal
  F}$. Let $A, B\in{\cal F}$, then $2 = \mu(A) + \mu(B) = \mu(A\cup B) +
\mu(A\cap B)$, hence $\mu(A\cap B)  = 1$, thus $A\cap B\in {\cal
  F}$. Thus ${\cal F}$ is indeed a filter. It is also an ultrafilter
by Lemma~\ref{l-max-filter}, because $A\not\in{\cal F}$ implies $S\setminus A\in{\cal F}$. 

The converse construction, viz., to generate a binary valued measure
from a filter, would require $\bigcup_{n\in\Nat} A_n \in{\cal F}$ if
and only if there exists $n\in\Nat$ with $A_n\in{\cal F}$ for any
disjoint family $\Folge{A}$. This, however, leads to very deep
questions on Set Theory, see~\cite[Chapter 10]{Jech} for a
discussion. 
\EndExample

Let us have a look at an important example.

\BeginExample{length-of-intervals}
Let ${\cal C} := \{ ]a, b] \mid a, b\in [0, 1]\}$ be all left open,
right closed intervals of the unit interval. Put $\ell(]a, b]) := b -
a$, hence $\ell(I)$ is the length of interval $I$. Note that
$\ell(\emptyset) =  \ell(]a, a]) = 0$. Certainly $\ell: {\cal
  C}\to\pReal$ is monotone and additive. 
\begin{enumerate}
\item If $\bigcup_{i=1}^k ]a_i, b_i]\subseteq ]a, b]$, and the
  intervals are disjoint, then $\sum_{i=1}^k \ell(]a_i, b_i])  \leq
  \ell(]a, b])$. The proof proceeds by induction on the number $k$ of
  intervals. For the induction step we have mutually disjoint intervals
  with $\bigcup_{i=1}^{k+1} ]a_k, b_k]\subseteq ]a, b]$. Renumbering,
  if necessary, we may assume that $a_1\leq b_1 \leq  a_2\leq b_2 \dots
  a_k\leq b_k\leq a_{k+1}\leq b_{k+1}$. Then $\sum_{i=1}^k \ell(]a_i, b_i]) +
  \ell(]a_{k+1}, b_{k+1}]) \leq \ell(]a_1, b_k]) + \ell(]a_{k+1},
  b_{k+1}]) \leq \ell(]a, b])$, because $\ell$ is monotone and
  additive.
\item If $\bigcup_{i=1}^\infty ]a_i, b_i]\subseteq ]a, b]$, and the
  intervals are disjoint, then $\sum_{i=1}^k \ell(]a_i, b_i])  \leq
  \ell(]a, b])$ for all $k$, hence
  \begin{equation*}
    \sum_{i=1}^\infty \ell(]a_i, b_i]) = \sup_{k\in\Nat}\sum_{i=1}^k \ell(]a_i, b_i])  \leq
  \ell(]a, b]).
  \end{equation*}
\item If $]a, b] \subseteq \bigcup_{i=1}^k ]a_i, b_k]$ then $\ell(]a,
  b]) \leq \sum_{i=1}^k \ell(]a_i, b_i])$ with not necessarily disjoint
  intervals. This is established by induction on $k$. If $k = 1$, the
  assertion is obvious. The induction step proceeds as follows: Assume
  that $]a, b] \subseteq \bigcup_{i=1}^{k+1} ]a_i, b_k]$. By
  renumbering, if necessary, we can assume that $a_{k+1} < b \leq
  b_{k+1}$. If $a_{k+1} \leq a$, the assertion follows, so let us
  assume that $a < a_{k+1}$. Then 
$
]a, a_{k+1}] \subseteq \bigcup_{i=1}^k ]a_i, b_i],
$
so that by the induction hypothesis 
$
a_{k+1}-a = \ell(]a, a_{k+1}])  \geq \sum_{i=1}^k \ell(]a_i, b_i]).
$
Thus 
\begin{equation*}
\ell(]b, a]) = b - a \leq (a_{k+1} - a) + (b_{k+1} - a_{k+1})
\leq \sum_{i=1}^{k+1} \ell(]a_i, b_i]).
\end{equation*}
\item Now assume that 
$
]a, b] \subseteq \bigcup_{i=1}^\infty ]a_i, b_i].
$
This is a little bit more complicated since we do not know whether
the interval $]a, b]$ is covered already by a finite number of
intervals, so we have to resort to a little trick. The interval
$[a+\epsilon, b]$ is closed and bounded, hence compact, for every fixed
$\epsilon > 0$; we also know that for each $i\in \Nat$ the semi-open
interval $]a_i, b_i]$ is contained in the open interval $]a_i,
b_i+\epsilon/2^{i}[$, so that we have
\begin{equation*}
[a+\epsilon, b] \subseteq \bigcup_{i=1}^\infty]a_i, b_i+\epsilon/2^{i}[  
\end{equation*}
By the Heine-Borel Theorem~\ref{heine-borel}
we can find a finite subset of these intervals which cover
$[a+\epsilon, b]$, say
$
[a+\epsilon, b] \subseteq \bigcup_{i\in K} ]a_i, b_i+\epsilon/2^{i}[,
$
with $K\subseteq\Nat$ finite. Hence
\begin{equation*}
]a+\epsilon, b] \subseteq \bigcup_{i\in K} ]a_i, b_i+\epsilon/2^{i}],
\end{equation*}
and we conclude from the finite case that 
\begin{equation*}
  b - (a+\epsilon) = \ell([a+\epsilon, b]) \leq \sum_{i\in K} \ell(]a_i,
  b_i+\epsilon/2^{i}]) = \sum_{i=1}^\infty (b_i+\epsilon/2^i-a_i)< \sum_{i=1}^\infty \ell(]a_i, b_i]) + \epsilon.
\end{equation*}
Since $\epsilon>0$ was arbitrary, we have established
\begin{equation*}
  \ell(]a, b]) \leq \sum_{i=1}^\infty \ell(]a_i, b_i]).
\end{equation*}
\end{enumerate}
\EndExample

Thus we have shown
\BeginProposition{interval-length}
Let ${\cal C}$ be the set of all left open, right closed intervals of
the unit interval, and denote by $\ell(]a, b]) := b - a$ the length of
interval $]a, b]\in {\cal C}$. Then $\ell: {\cal C}\to\pReal$ is
monotone and countably additive. \QED
\EndProposition

When having a look at ${\cal C}$ we note that this family is not
closed under complementation, but the complement of a set in ${\cal
  C}$ can be represented through elements of ${\cal C}$, e.g., 
$]0, 1]\setminus]1/3, 1/2] = ]0, 1/3]\cup]1/2, 1]$. This is captured
through the following definition.

\BeginDefinition{semi-ring}
${\cal R}\subseteq\PowerSet{S}$ is called a \emph{\index{semiring}semiring} iff
\begin{enumerate}
\item $\emptyset\in{\cal R}$,
\item ${\cal R}$ is closed under finite intersections,
\item If $B\in{\cal R}$, then there exists a finite family $C_1,
  \dots, C_k\in{\cal R}$ with
$
S\setminus B = C_1\cup\dots\cup C_k.
$
\end{enumerate}
\EndDefinition

Thus the complement of a set in ${\cal R}$ can be represented through
a finite disjoint union of elements of ${\cal R}$.

We want to extend $\ell: {\cal C}\to\pReal$ from the semiring of left
open, right closed intervals to a measure $\lambda$ on the
$\sigma$-algebra $\sigma({\cal C})$. This measure is fairly important,
it is called the \emph{\index{measure!Lebesgue}Lebesgue measure} on the unit interval. 

A first step towards an extension of $\ell$ to the $\sigma$-algebra generated by
the intervals is the extension to the algebra generated by them. This
can be accomplished easily once this algebra has been identified.

\BeginLemma{alg-gen-interv}
Let ${\cal C}$ be the set of all left open, right closed intervals in
$]0, 1]$. Then the algebra generated by ${\cal C}$ consists of all
disjoint unions of elements of ${\cal C}$.
\EndLemma

\BeginProof
Denote by 
\begin{equation*}
  {\cal D} := \{\bigcup_{1 \leq i \leq n} ]a_i, b_i] \mid n\in\Nat,
    a_1\leq b_1\leq a_2 \leq b_2 \dots \leq a_n\leq b_n\}
\end{equation*}
Then all elements of ${\cal D}$ are certainly contained in the algebra
generated by ${\cal C}$. If we can show that ${\cal D}$ is an algebra itself,
we are done, because then ${\cal D}$ is the smallest algebra
containing ${\cal C}$. 

${\cal D}$ is certainly closed under finite unions and
finite intersections, and $\emptyset\in{\cal D}$. The complement of
$\bigcup_{1 \leq i \leq n} ]a_i, b_i]$ is $]0, a_1]\cup]b_1,
a_2]\cup\dots\cup]b_n, 1]$, which is a member of ${\cal D}$ as well.
Thus ${\cal D}$ is also closed under complementation, hence is an
algebra. 
\EndProof

This permit us to extend $\ell$ to the algebra generated by the
intervals:

\BeginCorollary{extend-to-algebra-lebesgue}
$\ell$ extends uniquely to the algebra generated by ${\cal C}$ such
that the extension is monotone and countably additive.
\EndCorollary

\BeginProof
Put 
\begin{equation*}
  \ell(\bigcup_{1 \leq i \leq n} ]a_i, b_i]) := \sum_{i=1}^n
  \ell(]a_i, b_i]),
\end{equation*}
whenever $]a_i, b_i]\in{\cal C}$. This is well defined. Assume
\begin{equation*}
  \bigcup_{1 \leq i \leq n} ]a_i, b_i] = \bigcup_{1 \leq j \leq m} ]c_j, d_j],
\end{equation*}
then $]a_i, b_i] $ can be represented as a disjoint union of those
intervals $]c_j, d_j]$ which it contains, so that we have
\begin{align*}
  \sum_{i=1}^n \ell(]a_i, b_i]) & = \sum_{i=1}^n\sum_{j=1}^m
  \ell(]a_i, b_i]\cap]c_j, d_j])\\
& = \sum_{j=1}^m\sum_{i=1}^m
  \ell(]c_j, d_j]\cap]a_i, b_i])\\
& = \sum_{j=1}^m \ell(]c_j, d_j])
\end{align*}
We may conclude from Example~\ref{length-of-intervals} that $\ell$ is
countably additive on the algebra. 
\EndProof

For the sake of illustration, let us assume that we have Lebesgue
measure constructed already, and let us compute $\lambda(C)$ where $C$
is the Cantor ternary set constructed in
Example~\ref{borel-in-unit-line} on page~\pageref{borel-in-unit-line}.
The construction of the ternary set is done through sets $C_n$, each
of which which is the union of $2^n$ mutually disjoint intervals of
length $1/3^n$. If $I$ is an interval of length $3^{-n}$, we know that
$\lambda(I) = 3^{-n}$, so that $ \lambda(C_n) = (2/3)^n.  $ We also
know that $C_1\supseteq C_2 \supseteq \dots $, so that we have a
descending chain of sets with $ C = \bigcap_{n\in \Nat} C_n.  $

In order to compute $\lambda(C)$, we need so know something about the
behavior of measures when monotone limits of sets are encountered. 

\BeginLemma{meas-monotone}
Let $\mu: {\cal A}\to [0, \infty]$ be a measure on the
$\sigma$-algebra ${\cal A}$. 
\begin{enumerate}
\item \label{meas-monotone-item:1} If $A_n\in {\cal A}$ is a monotone
  increasing sequence of sets in ${\cal A}$, and $A = \bigcup_{n\in
    \Nat} A_n$, then 
$
\mu(A) = \sup_{n\in\Nat} \mu(A_n).
$
\item \label{meas-monotone-item:2}  If $A_n\in {\cal A}$ is a monotone
  decreasing sequence of sets in ${\cal A}$, and $A = \bigcap_{n\in
    \Nat} A_n$, then 
$
\mu(A) = \inf_{n\in\Nat} \mu(A_n),
$
provided $\mu(A_k)<\infty$ for some $k\in\Nat$.
\end{enumerate}
\EndLemma

\BeginProof

1.
We can write $A_n = \bigcup_{i=1}^n B_i$ with $B_1 := A_1$ and $B_i :=
A_i\setminus A_{i-1}$. Because the $A_n$ form an increasing sequence,
the $b_n$ are mutually disjoint. Assume without loss of generality
that $\mu(A_n) < \infty$ for all $n\in \Nat$ (otherwise the assertion
is trivial), then by countable additivity and through telescoping
\begin{equation*}
\mu(A) = \sum_{i=1}^\infty \mu(B_i) = \mu(A_1) + \sum_{i=1}^\infty (\mu(A_{i+1})
- \mu(A_i)) = \lim_{n\to\infty} \mu(A_n) = \sup_{n\in\Nat}\mu(A_n).
\end{equation*}

2.
Assume $\mu(A_1) < \infty$, then the sequence $A_1\setminus A_n$ is
increasing towards $A_1\setminus A$, hence
\begin{equation*}
  \mu(A) = \mu(A_1) - \mu(A_1\setminus A) = \mu(A_1) -
  \sup_{n\in\Nat}\mu(A_1\setminus A_n) = \inf_{n\in\Nat}\mu(A_n).
\end{equation*}
\EndProof

Ok, so let us return to the discussion of Cantor's set. We know that
$\lambda(C_n) = (2/3)^n$, and that $C_1 \supseteq C_2 \supseteq C_3\dots$,
so we conclude
\begin{equation*}
  \lambda(C) = \inf_{n\in \Nat} \lambda(C_n) = 0.
\end{equation*}

So we have identified a geometrically fairly complicated set which has
measure zero. This set is geometrically not easy to visualize, since
it does not contain an interval of positive length. 

Now fix a semiring ${\cal C}\subseteq \PowerSet{S}$ and $\mu: {\cal
  C}\to [0, \infty]$ with $\mu(\emptyset) = 0$, which is monotone and
countably subadditive. We will first compute an outer approximation
for each subset of $S$ by elements of ${\cal C}$. But since the
subsets of $S$ may be as a whole somewhat inaccessible, and since
${\cal C}$ may be somewhat small, we try to cover the subsets of $S$
by countable unions of elements of ${\cal C}$ and take the best
approximation we can, i.e., we take the infimum. Define
\begin{equation*}
  \mu^*(A) := \inf\{\sum_{n\in\Nat}\mu(C_n) \mid A\subseteq
  \bigcup_{n\in\Nat} C_n, C_n\in{\cal C}\}
\end{equation*}
for $A\subseteq S$. This is the \emph{outer measure} of $A$ associated with
$\mu$.

These are some interesting (for us, that is) properties of $\mu^*$.

\BeginLemma{prop-outer}
$\mu^*: \PowerSet{S}\to [0, \infty]$ is monotone and countably subadditive, $\mu^*(\emptyset) =
0$. If $A\in{\cal C}$, then $\mu^*(A) = \mu(A)$.
\EndLemma
 
\BeginProof
1.
Let $\Folge{A}$ be a sequence of subset of $S$, put $A:=
\bigcup_{n\in\Nat} A_n$. If $\sum_{n\in\Nat}\mu^*(A_n) < \infty$, we
find for $A_n$ a cover $\{C_{n, m}\mid m\in\Nat\}\subseteq {\cal C}$
with
$
\mu(A_n) \leq \sum_{m\in\Nat}\mu(C_{n, m}) \leq \mu^*(A_n) + \epsilon/2^n,
$
thus $\{C_{n, m} \mid n, m\in\Nat\}\subseteq{\cal C}$ is a cover of
$A$ with
$
\mu(A) \leq \sum_{n, m\in\Nat}\mu(C_{n, m}) \leq
\sum_{n\in\Nat}\mu(A_n) + \epsilon.
$
Since $\epsilon>0$ was arbitrary, we conclude
$
\mu^*(A) \leq \sum_{n\in\Nat}\mu^*(A_n).
$
If, however, $\sum_{n\in\Nat}\mu^*(A_n) = \infty$, the assertion is
immediate. 

2.
The other properties are readily seen. 
\EndProof

The next step is somewhat mysterious~---~it has been suggested by
Carathéodory around 1914 for the construction of a measure extension. It splits a
set $A = (A\cap X) \cup (A\cap S\setminus X)$ along an arbitrary
other set $X$, and looks what happens to the outer
measure. If 
$
\mu^*(A) = \mu^*(A\cap X) + \mu(A\cap S\setminus X),
$
then $A$ is considered well behaved. Those sets which are well behaved no matter
what set $X$ we use for splitting are considered next. 

\BeginDefinition{cara-measurable-sets}
A set $A\subseteq S$ is called \emph{\index{$\mu$-measurable}$\mu$-measurable} iff 
$
\mu^*(X) = \mu^*(X\cap A) + \mu^*(X\cap S\setminus A)
$
holds for all $X\subseteq S$. The set of all $\mu$-measurable sets is
denoted by ${\cal C}_\mu$
\EndDefinition

So take a $\mu$-measurable set $A$ and an arbitrary subset $X\subseteq
S$, then $X$ splits into a part $X\cap A$ which belongs to $A$ and
another one $X\cap S\setminus A$ which does not belong to
$A$. Measuring these pieces through $\mu^*$, we demand that they add up
to $\mu^*(X)$ again.

These properties are immediate:
\BeginLemma{prop-mu-outer-star}
The outer measure has these properties
\begin{enumerate}
\item \label{prop-mu-outer-star_item:1} $\mu^*(\emptyset) = 0$.
\item \label{prop-mu-outer-star_item:2} $\mu^*(A) \geq 0$ for all
  $A\subseteq S$.
\item \label{prop-mu-outer-star_item:3} $\mu^*$ is monotone.
\item \label{prop-mu-outer-star_item:4} $\mu^*$ is countably subadditive.
\end{enumerate}
\EndLemma

\BeginProof
We establish only the last property. Here we have to show that 
$
\mu^*(\bigcup_{n\in\Nat} A_n) \leq \sum_{n\in \Nat} \mu^*(A_n).
$ We may and do assume that all $\mu^*(A_n)$ are finite. Given
$\epsilon>0$ we find for each $n\in\Nat$ a sequence $B_{n, k}\in {\cal C}$ for
$A_n$ such that 
$
A_n\subseteq\bigcup_{k\in\Nat} B_{n, k} 
$
and 
$
\sum_{k\in\Nat}\mu(B_{n, k}) \leq \mu^*(A_n) + \epsilon/2^n.
$
Thus 
$
\sum_{n, k\in\Nat}\mu(B_{n, k}) \leq \sum_{n\in\Nat}(\mu^*(A_n) +
\epsilon/2^n) < \sum_{n\in\Nat}\mu^*(A_n) + \epsilon,
$
which implies
$
\sum_{n\in\Nat}\mu^*(A_n) \leq \mu^*(\bigcup_{n\in \Nat} A_n),
$
because 
$
\bigcup_{n\in\Nat}A_n \subseteq \bigcup_{{n, k\in\Nat}} B_{n, k},
$
and because $\epsilon>0$ was arbitrary.
\EndProof

Because countably subadditivity, we conclude 

\BeginCorollary{prop-mu-outer-star-cor}
$A\in{\cal C}_\mu$ iff $\mu^*(X\cap A) + \mu^*(X\cap S\setminus A)
\leq \mu^*(X)$ for all $X\subseteq S$.
\QED
\EndCorollary

Let us have a look at the set of all $\mu$-measurable sets. It turns
out  that the originally given sets are all $\mu$-measurable, and
that ${\cal C}_\mu$ is an algebra. 

\BeginProposition{meas-is-algebra}
${\cal C}_\mu$ is an algebra. Also if $\mu$ is additive,  ${\cal C}\subseteq{\cal C}_\mu$ and
$\mu(A) = \mu^*(A)$ for all $A\in{\cal C}$. 
\EndProposition

\BeginProof
1.
${\cal C}_\mu$ is closed under complementation; this is obvious from
its definition, and $S\in{\cal C}_\mu$ is also clear. So we have only
to show that ${\cal C}_\mu$ is closed under finite intersections. For
simplicity, denote complementation by $\cdot^c$.

Now let $A, B\in{\cal C}_\mu$, we want to show 
\begin{equation*}
 \mu^*(X) \geq \mu^*((A\cap B)\cap X) + \mu^*((A\cap B)^c\cap X), 
\end{equation*}
for each $X\subseteq S$; from Corollary~\ref{prop-mu-outer-star-cor} we infer that this implies
$A\cap B\in {\cal C}_\mu$. Since $B\in{\cal C}_\mu$ and then $A\in {\cal C}_\mu$, we know
\begin{align*}
 \mu^*(X) & = \mu^*(X\cap B) + \mu^*(X \cap B^c)\\
& = \mu^*(X \cap (A\cap B)) + \mu^*(X \cap (A^c\cap B)) + \mu^*(X\cap
(A\cap B^c)) + \mu^*(X\cap (A^c\cap B^c))\\
& \geq  \mu^*(X \cap (A\cap B)) + \mu^*(X\cap((A^c\cap B)\cup (A\cap B^c)
\cup (A^c\cap B^c)))\\
& \stackrel{(\ddag)}{=} \mu^*(X\cap (A\cap B)) + \mu^*(X \cap (A\cap B)^c).
\end{align*}
Equality $(\ddag)$ uses
\begin{align*}
  (A^c\cap B)\cup(A\cap B^c)\cup(A^c\cap B^c) & = \bigl(A^c\cap(B\cup
  B^c)\bigr)\cup(A\cap B^c)\\
& = A^c\cup (A\cap B^c) \\
& = A^c\cup B^c.
\end{align*}
Hence we see that $A\cap B$ satisfies the defining inequality. 

2.
We still have to show that ${\cal C}\subseteq{\cal C}_\mu$, and that
$\mu^*$ extends $\mu$. Let $A\in{\cal C}$, then $S\setminus A =
D_1\cup\dots\cup D_k$ for some disjoint $D_1, \dots, D_k\in{\cal C}$,
because ${\cal C}$ is a semiring. Fix $X\subseteq S$, and assume that
$\mu^*(S) < \infty$ (otherwise, the assertion is trivial). Given
$\epsilon>0$ there exists in ${\cal C}$ a cover
$\Folge{A}$   of $X$ with 
$
\mu^*(X) < \sum_{n\in\Nat} \mu(A_n) + \epsilon.
$
Now put $B_n := A\cap A_n$ and $C_{i, n} := A_n\cap D_i$. Then $X\cap
A \subseteq \bigcup_{n\in\Nat} B_n$ with $B_n\in {\cal
  C}$ and $X\cap A^c \subseteq \bigcup_{n\in\Nat, 1\leq i \leq k}
C_{i, n}$ with $C_{i, n}\in {\cal C}$. Hence
\begin{align*}
  \mu^*(X\cap A) + \mu^* (X\cap A^C) 
& \leq \sum_{n\in\Nat} \mu(B_n) + \sum_{n\in\Nat, 1\leq i \leq k}
\mu(C_{i, n})\\
& \leq \sum_{n\in\Nat} \mu(A_n) \\
& < \mu^*(X) - \epsilon,
\end{align*}
because $\mu$ is (finitely) additive. Hence $A\in{\cal C}_\mu$. $\mu^*$ is an
extension to $\mu$ by Lemma~\ref{prop-outer}. 
\EndProof

But we can in fact say more on the behavior of $\mu^*$ on ${\cal
  C}_\mu$: It turns out to be additive on the splitting parts.

\BeginLemma{outer-splits}
Let ${\cal D}\subseteq{\cal C}_\mu$ be a finite or infinite family of
mutually disjoint sets in ${\cal C}_\mu$, then
\begin{equation*}
\mu^*(X\cap\bigcup_{D\in{\cal D}} D) = \sum_{D\in{\cal D}}\mu^*(X\cap D)
\end{equation*}
holds for all $X\subseteq S$. 
\EndLemma

\BeginProof
1.
We establish the equality above for finite ${\cal D}$, say, ${\cal D}
= \{A_1, \dots, A_n\}$ with $A_n\in{\cal C}_\mu$ for $1 \leq j \leq
n$. From this we obtain that the equality holds in the countable case
as well, because then
\begin{equation*}
  \mu^*(X\cap\bigcup_{i=1}^\infty A_i) \geq 
 \mu^*(X\cap\bigcup_{i=1}^n A_i) = \sum_{i=1}^n\mu^*(X\cap A_i), 
\end{equation*}

for all $n\in\Nat$, so that 
$
\mu^*(X\cap\bigcup_{i=1}^\infty A_i) \geq \sum_{i=1}^\infty\mu^*(X\cap
A_i),
$
which together with countable subadditivity gives the desired result.

2.
The proof for 
$
\mu^*(X\cap\bigcup_{i=1}^n A_i) = \sum_{i=1}^n\mu^*(X\cap A_i)
$
proceeds by induction on $n$, starting with $n = 2$. If $A_1\cup A_2 =
S$, this is just the definition that $A_1$ (or $A_2$) is
$\mu$-measurable, so the equality holds. If $A_1\cup A_2\not= S$, we
note that 
$
\mu^*(X) = \mu^*((X\cap(A_1\cup A_2))\cap A_1) + \mu^*((X\cap(A_1\cup
A_2))\cap S\setminus A_1).
$
Evaluating the pieces, we see that
\begin{align*}
  (X\cap(A_1\cup A_2))\cap A_1 & = X\cap A_1,\\
(X\cap(A_1\cup A_2))\cap S\setminus A_1) & = X\cap A_2,
\end{align*}
because $A_1\cap A_2=\emptyset$. 
The induction step is straightforward:
\begin{align*}
\mu^*(X\cap\bigcup_{i=1}^{n+1} A_i) & =
\mu^*((X\cap\bigcup_{i=1}^n A_n)\cup(X\cap A_{n+1}))\\
& = \sum_{i=1}^n\mu^*(X\cap A_i)  + \mu^*(X\cap A_{n+1})\\
& = \sum_{i=1}^{n+1}\mu^*(X\cap A_i)
\end{align*}
\EndProof

We can relax the condition on a set being a member of ${\cal C}_{\mu}$
if we know that the domain ${\cal C}$ from which we started is an
algebra, and that $\mu$ is additive on ${\cal C}$. Then we do not have
to test whether a $\mu$-measurable set splits all the subsets of $S$,
but it is rather sufficient that $A$ splits $X$, to be specific:

\BeginProposition{outer-splits-all}
Let ${\cal C}$ be an algebra, and $\mu: {\cal C}\to [0, +\infty]$ be additive. Then $A\in {\cal C}_{\mu}$ iff $\mu^{*}(A) + \mu^{*}(X\setminus A) = \mu^{*}(X)$.
\EndProposition

\BeginProof
This is a somewhat lengthy and laborious computation similarly to the one above, see~\cite[1.11.7, 1.11.8]{Bogachev}.
\EndProof

Returning to the general discussion, we have:
\BeginProposition{outer-splits-is-sigma-algebra}
${\cal C}_\mu$ is a $\sigma$-algebra, and $\mu^*$ is countably
additive on ${\cal C}_\mu$. 
\EndProposition

\BeginProof
0.
Let $\Folge{A}$ be a countable family of mutually disjoint sets in
${\cal C}_\mu$, then we have to show that $A := \bigcup_{n\in\Nat}
A_n\in {\cal C}_\mu$, thus we have to show that 
\begin{equation*}
  \mu^*(X\cap A) + \mu^*(X\cap A^c) \leq \mu^*(X)
\end{equation*}
for each $X\subseteq S$ (here $\cdot^c$ is complementation again). Fix
$X$. 

1.
We know that ${\cal C}_\mu$ is closed under finite unions, so we have
for each $n\in\Nat$
\begin{align*}
  \mu^*(X) & \geq \sum_{i=1}^n\mu^*(X\cap A_i) +
  \mu^*(X\cap\bigcap_{i=1}^n A_i^c)\\
& \geq \sum_{i=1}^n\mu^*(X\cap A_i) +
  \mu^*(X\cap A),
\end{align*}
because $\bigcap_{i=1}^n A_i^c\supseteq A^c$. Letting $n\to\infty$ we
obtain the desired inequality. 

3.  Thus ${\cal C}_\mu$ is closed under disjoint countable
unions. Using the \index{first entrance trick}first entrance trick
(Exercise~\ref{ex-subadditive-additive}) and the observation that
${\cal C}_\mu$ is an algebra by Proposition~\ref{meas-is-algebra}, we
convert each countable union into a disjoint countable union, so we
have shown that ${\cal C}_\mu$ is a $\sigma$-algebra. Countable
additivity of $\mu^*$ on ${\cal C}_\mu$ follows from
Lemma~\ref{outer-splits} when putting $X := S$.
\EndProof

Summarizing, we have demonstrated this Extension Theorem.

\BeginTheorem{caratheodory}
Let ${\cal C}$ is an algebra over a set $S$ and $\mu: {\cal C}\to [0, \infty]$ 
monotone and countably additive.
\begin{enumerate}
\item There exists an extension of $\mu$ to a measure on the
  $\sigma$-algebra $\sigma({\cal C})$ generated by ${\cal C}$. 
\item If $\mu$ is \emph{$\sigma$-finite}, i.e., if $S$ can be written as $S = \bigcup_{n\in\Nat} S_n$ with
  $S_n\in{\cal C}$ and $\mu(S_n)<\infty$ for all $n\in\Nat$, then the
  extension is uniquely determined.
\end{enumerate}
\EndTheorem

\BeginProof
1. 
Proposition~\ref{outer-splits-is-sigma-algebra} shows that ${\cal
  C}_\mu$ is a $\sigma$-algebra containing ${\cal C}$, and that
$\mu^*$ is a measure on ${\cal C}_\mu$. Hence $\sigma({\cal
  C})\subseteq{\cal C}_\mu$, and we can restrict $\mu^*$ to
$\sigma({\cal C})$. Denote this restriction also by $\mu$, then $\mu$ is a
measure on $\sigma({\cal C})$. 

2.
In order to establish uniqueness, assume first that
$\mu(S)<\infty$. Let $\nu$ be a measure which extends $\mu$ to
$\sigma({\cal C})$. Recall the construction of $\sigma({\cal C})$
through transfinite induction on page~\pageref{transfinite-def-xxx}. We claim that 
\begin{equation*}
  \mu(A) = \nu(A)\text{ for all } A\in{\cal C}_\zeta
\end{equation*}
holds for all ordinals $\zeta<\omega$. Because ${\cal C}$ is an algebra, it is easy to see that for odd
ordinals $\zeta$ a set $A\in
({\cal C}_\zeta)_\delta$ iff there exists a decreasing sequence
$\Folge{A}\subseteq {\cal C}_\zeta$ with $A = \bigcap_{n\in\Nat} A_n$;
similarly, each element of $({\cal C}_\zeta)_\sigma$ can be represented as
the union of an increasing sequence of elements of $A_\zeta$ if
$\zeta$ is even. Assume for the induction step that $\zeta$ is odd,
and let $A\in {\cal C}_{\zeta+1}$, thus $A = \bigcap_{n\in\Nat} A_n$
with $A_1 \supseteq A_2 \supseteq \dots$ and $A_n\in{\cal
  C}_\zeta$. Hence by Lemma~\ref{meas-monotone}
\begin{equation*}
  \mu(A) = \mu(\bigcap_{n\in\Nat} A_n) = \inf_{n\in\Nat}\mu(A_n) = \inf_{n\in\Nat}
  \nu(A_n) = \nu(A).
\end{equation*}
Thus $\mu$ and $\nu$ coincide on ${\cal C}_{\zeta+1}$, if $\zeta$ is
odd. One argues similarly, but with a monotone increasing sequence in
the case that $\zeta$ is even. If $\mu$ and $\nu$ coincide on all
${\cal C}_\eta$ for all $\eta$ with $\eta < \zeta$ for a limit number
$\zeta$, then it is clear that they also coincide on ${\cal C}_\zeta$
as well. 

3. 
Assume that $\mu(S) = \infty$, but that there exists a sequence
$\Folge{S}$ in ${\cal C}$ with $\mu(S_n) < \infty$. Because 
$
\mu(S_1\cup\dots\cup S_n) \leq \mu(S_1) + \dots + \mu(S_n) < \infty,
$
we may and do assume that the sequence is monotonically
increasing. Let $\mu_n(A) := \mu(A\cap S_n)$ be the localization of
$\mu$ to $S_n$. $\mu_n$ has a unique extension to $\sigma({\cal C})$,
and since we have 
$
\mu(A) = \sum_{n\in\Nat} \mu_n(A)
$
for all $A\in\sigma({\cal C})$, the assertion follows. 
\EndProof

But we are not quite done yet, witnessed by a glance at Lebesgue
measure. There we started from the semiring of intervals, but our
uniqueness theorem states only what happens when we carry out our
extension process starting from an
algebra. 
\setboolean{isBook}{true}
\ifthenelse{\boolean{isBook}}{
It turns out to be
most convenient to have a closer look at the construction of
$\sigma$-algebras when the family of sets we start from has already
some structure. This gives the occasion to introduce Dynkin's
\index{$\pi$-$\lambda$-Theorem}\index{Theorem!$\pi$-$\lambda$}$\pi$-$\lambda$-Theorem. This is an important tool, which makes it
sometimes simpler to identify the $\sigma$-algebra generated from
some family of sets.

\BeginTheorem{Pi-Lambda}{\textbf{($\pi$-$\lambda$-Theorem)}}
Let $\mathcal{P}$ be a family of
subsets of $S$ that is closed under finite intersections (this is
called a
\emph{$\pi$-class}). Then $\sigma(\mathcal{P})$ is the smallest
$\lambda$-class containing $\mathcal{P}$, where a family
$\mathcal{L}$ of subsets of $S$ is called a \emph{$\lambda$-class}
iff it is closed under complements and countable disjoint unions.
\EndTheorem

\BeginProof
1.
Let $\mathcal{L}$ be the smallest $\lambda$-class containing $P$, then
we show that $\mathcal{L}$ is a $\sigma$-algebra.

2.
We show first that
it is an algebra. Being a $\lambda$-class, $\mathcal{L}$ is closed under
complementation. Let
$A \subseteq S$, then
$
\mathcal{L}_A := \{B \subseteq S \mid A \cap B \in \mathcal{L}\}
$
is a $\lambda$-class again: if $A \cap B \in \mathcal{L}$, then
$$
A \cap \left(S\setminus B\right) = A\setminus B = S\setminus ((A \cap B) \cup (S\setminus A)),
$$
which is in $\mathcal{L}$, since $(A \cap B) \cap S\setminus A = \emptyset$, and
since $\mathcal{L}$ is closed under disjoint unions.

If $A \in \mathcal{P}$, then $\mathcal{P} \subseteq \mathcal{L}_A$, because
$\mathcal{P}$ is closed under intersections. Because $\mathcal{L}_A$ is a
$\lambda$-system, this implies
$
\mathcal{L} \subseteq\mathcal{L}_A
$
for all $A \in \mathcal{P}$. Now take $B \in \mathcal{L}$, then the
preceding argument shows that $\mathcal{P} \subseteq \mathcal{L}_B$,
and again we may conclude that $\mathcal{L} \subseteq \mathcal{L}_B$. Thus
we have shown that $A \cap B \in \mathcal{L}$, provided $A, B \in \mathcal{L}$, so that
$\mathcal{L}$ is closed under finite intersections. Thus $\mathcal{L}$ is a Boolean algebra.

3.
$\mathcal{L}$ is a $\sigma$-algebra as well. It is enough to show that
$\mathcal{L}$ is closed under countable unions. But since
$$
\bigcup_{n \in \Nat} A_n = \bigcup_{n \in \Nat} \left(A_n \setminus \bigcup_{i=1}^{n-1}A_i\right),
$$
this follows immediately.
\EndProof

Consider an immediate and fairly typical application. It states that
two finite measures are equal on a $\sigma$-algebra, provided they are
equal on a generator which is closed under finite intersections. The
proof technique is worth noting: We collect all sets for which the
assertion holds into one family of sets and investigate its
properties, starting from an originally given set. If we find that the
family has the desired property, then we look at the corresponding
closure. To be specific, have a look at the  proof of the following statement.

\BeginLemma{are-equal}
Let $\mu, \nu$ be finite measures on a $\sigma$-algebra $\sigma({\cal B})$,
where ${\cal B}$ is a family of sets which
is closed under finite intersections. Then
$\mu(A) = \nu(A)$ for all $A\in\sigma({\cal B})$, provided $\mu(B) = \nu(B)$ for
all $B\in{\cal B}$. 
\EndLemma

\BeginProof
We have a look at all sets for which the assertion is true, and
investigate this set. Put
\begin{equation*}
  {\cal G} := \{A\in\sigma({\cal B}) \mid \mu(A) = \nu(A)\},
\end{equation*}
then ${\cal G}$ has these properties:
\begin{itemize}
\item ${\cal B}\subseteq{\cal G}$ by assumption.
\item Since ${\cal B}$ is closed under finite intersections,
  $S\in{\cal B}\subseteq{\cal G}.$
\item ${\cal G}$ is closed under complements.
\item ${\cal G}$ is closed under countable disjoint unions; in fact,
  let $\Folge{A}$ be a sequence of mutually disjoint sets in ${\cal
    G}$ and $A := \bigcup_{n\in\Nat} A_n$, then
  \begin{equation*}
    \mu(A) = \sum_{n\in\Nat} \mu(A_n) = \sum_{n\in\Nat} \nu(A_n) = \nu(A),
  \end{equation*}
hence $A\in{\cal G}$. 
\end{itemize}
But this means that ${\cal G}$ is a $\lambda$-class containing ${\cal
  B}$. But the smallest $\lambda$-class containing ${\cal G}$ is
$\sigma({\cal B})$ by Theorem~\ref{Pi-Lambda}, so that we have now
\begin{equation*}
  \sigma({\cal B}) \subseteq {\cal G} \subseteq \sigma({\cal B}),
\end{equation*}
the last inclusion coming from the definition of ${\cal G}$. Thus we
may conclude that ${\cal G} = \sigma({\cal B})$, hence all sets in
$\sigma({\cal B})$ have the desired property.
\EndProof
}{}

We obtain as a slight extension to Theorem~\ref{caratheodory}

\BeginTheorem{caratheodory-semiring}
Let ${\cal C}$ is a semiring over a set $S$ and $\mu: {\cal C}\to [0, \infty]$ 
monotone and countably additive.
\begin{enumerate}
\item There exists an extension of $\mu$ to a measure on the
  $\sigma$-algebra $\sigma({\cal C})$ generated by ${\cal C}$. 
\item If $\mu$ is \emph{$\sigma$-finite}, then the
  extension is uniquely determined.
\end{enumerate}
\QED
\EndTheorem

The assumption on $\mu$ being $\sigma$-finite is in fact necessary: 
\BeginExample{sigma-finite-is}
Let ${\cal S}$ be the semiring of all left open, right closed
intervals on $\Real$, and put
\begin{equation*}
  \mu(I) :=
  \begin{cases}
    0& \text{ if } I = \emptyset,\\
\infty, & \text{ otherwise.}
  \end{cases}
\end{equation*}
Then $\mu$ has more than one extension to $\sigma({\cal S})$. For
example, let $c>0$ and put $\nu_c(A) := c\cdot |A|$ with $|A|$ as the
number of elements of $A$. Plainly, $\nu_c$ extends $\mu$ for every
$c$. 
\EndExample
Consequently, the assumption that $\mu$ is $\sigma$-finite cannot be
omitted in order to make sure that the extension is uniquely
determined. 

\Subsubsection{$\mu$-Measurable Sets}
\label{sec:mu-measurable-sets}

But Carathéodory's approach gives even more than an extension to the
$\sigma$-algebra generated from a semiring. This is what we will
discuss next in order to find a connection with the discussion about
the Axiom of Choice. 

Fix for the time being an outer measure $\mu$ on $\PowerSet{S}$ which
we assume as finite. Call
$A\subseteq S$ a\emph{ \index{$\mu$-null}$\mu$-null} set iff we can find a $\mu$-measurable set
$A_1$  with $A \subseteq A_1$ and $\mu(A_1) = 0$. Thus a $\mu$-null
set is covered by a measurable set which has $\mu$-measure
$0$. Because
$
\mu(X\cap S\setminus A) \leq \mu(X)
$
for every $X\subseteq S$, and because an outer measure is monotone, we
conclude that each $\mu$-null set is itself $\mu$-measurable. In the
same way we conclude that each set $A$ which can be squeezed between
two $\mu$-measurable sets of the same measure (hence $A_1\subseteq A
\subseteq A_2$ with $\mu(A_1) = \mu(A_2)$ must be $\mu$-measurable,
because in this case $A\setminus A_1 \subseteq A\setminus A_2$ with
$\mu(A\setminus A_2) = 0$. Hence ${\cal C}_\mu$ is \emph{\index{$\sigma$-algebra!complete}complete} in the
sense that any $A$ which can be sandwiched in this way is a member of
${\cal C}_\mu$. 

This is a characterization of ${\cal C}_\mu$ using these ideas.

\BeginCorollary{char-c-mu}
Let ${\cal C}$ be an algebra over a set $S$ and $\mu: {\cal C}\to\pReal$ monotone and
countably additive with $\mu(\emptyset) = 0$. Then these statements
are equivalent for $A\subseteq S$:
\begin{enumerate}
\item \label{char-c-mu-item:1} $A\in{\cal C}_\mu$.
\item \label{char-c-mu-item:2} There exists $A_1, A_2\in\sigma({\cal
    C})$ with $A_1\subseteq A \subseteq A_2$ and $\mu(A_1) = \mu(A_2)$.
\end{enumerate}
\EndCorollary

\BeginProof
The implication \labelImpl{char-c-mu-item:2}{char-c-mu-item:1} follows from the
discussion above, so we will look at
 \labelImpl{char-c-mu-item:1}{char-c-mu-item:2}. But this is trivial.
\EndProof

Having a look at this development, we see that we can extend our
measure far beyond the $\sigma$-algebra which is generated from the
given semiring. One might suspect even that this extension process
gives us the whole power set of the set we started from as the domain
for the extended measure. That would of course be tremendously
practical because we then could assign a measure to each subset. But,
alas, if the Axiom of Choice is assumed, these hopes are
shattered. The following example demonstrates this. Before discussing
it, however, we define and characterize $\mu$-measurable sets on a
$\sigma$-algebra. 

If $\mu$ is a finite measure on $\sigma$-algebra ${\cal B}$, we can
define the outer measure $\mu^*(A)$ for any subset $A\subseteq S$ as
we did for functions on a semiring. But since the algebraic structure
of a $\sigma$-algebra is richer, it is not difficult to see that
\begin{equation*}
  \mu^*(A) = \inf\{\mu(B) \mid B\in {\cal B}, A\subseteq B\}.
\end{equation*}
This is so because a cover of the set $A$ through a countable union of
elements on ${\cal B}$ is the same as the cover of $A$ through an
element of ${\cal B}$, because the $\sigma$-algebra ${\cal B}$ is
closed under countable unions. In a similar way we can try to
approximate $A$ from the inside, defining the \emph{\index{inner
    measure}inner measure} through
\begin{equation*}
  \mu_*(A) :=  \sup\{\mu(B) \mid B\in {\cal B}, A\supseteq B\}.
\end{equation*}
So $\mu_*(A)$ is the best approximation from the inside that is
available to us. Of course, if $A\in{\cal B}$ we have
$
\mu^*(A) = \mu(A) = \mu_*(A),
$
because apparently $A$ is the best approximation to itself. 

We can perform the approximation through a sequence of sets, so we are
able to precisely fix the inner and the outer measure through  elements
of the $\sigma$-algebra.

\BeginLemma{approx-in-out}
Let $A\subseteq S$ and $\mu$ be a finite measure on the
$\sigma$-algebra ${\cal B}$. 
\begin{enumerate}
\item There exists $A^*\in{\cal B}$ such that $\mu^*(A) = \mu(A*)$.
\item There exists $A_*\in{\cal B}$ such that $\mu_*(A) = \mu(A_*)$.
\end{enumerate}
\EndLemma

\BeginProof
We demonstrate only the first part. For each $n\in\Nat$ there exists
$A_n\in{\cal B}$ such that $A\subseteq B_n$ and $\mu(B_n) < \mu(A) +
1/n$. Put $A_n := B_1\cap\dots\cap B_n\in{\cal B}$, then $A\subseteq A_n$,
$\mu(A_n) < \mu(A) +1/n$, and $\Folge{A}$ decreases. Let $A^* :=
\bigcap_{n\in\Nat} A_n\in{\cal B}$, then $\mu(A^*) =
\inf_{n\in\Nat}\mu(A_n) = \mu^*(A)$ by the second part of
Lemma~\ref{meas-monotone}, because $\mu(A_1) < \infty$.
\EndProof

The set $A^*$ could be called the measurable closure of $A$,
similarly, $A_*$ its measurable kernel. Using this terminology, we call a set $\mu$-measurable
iff its closure and its kernel give the same value. 

\BeginDefinition{mu-measurable}
Let $\mu$ be a finite measure on the
$\sigma$-algebra ${\cal B}$.  $A\subseteq S$ is called
\emph{\index{$\mu$-measurable}$\mu$-measurable} iff $\mu_*(A) = \mu^*(A)$. 
\EndDefinition

Every set in ${\cal B}$ is $\mu$-measurable, and ${\cal B}_\mu$ is the
$\sigma$-algebra of all $\mu$-measurable sets. 

The example which has been announced above shows us that under the
assumption of {\AC} not every subset of the unit interval is
$\lambda$-measurable, where $\lambda$ is Lebesgue measure. Hence we
will present a set the inner and the outer measure of which are
different.

\BeginExample{vitaly}
Define $\isEquiv{x}{y}{\alpha}$ iff $x - y$ is rational for $x, y\in[0, 1]$. Then $\alpha$
is an equivalence relation, because the sum of two rational numbers is a
rational number again. This is sometimes called  \index{Vitali's equivalence
relation}\emph{Vitali's equivalence
relation}. The relation $\alpha$ partitions the interval $[0, 1]$
into equivalence classes. Select from each equivalence class an
element (which we can do by {\AC}), denote the set of selected
elements by $V$. Hence $V\cap\Klasse{x}{\alpha}$ contains for each
$x\in[0, 1]$ exactly one element. We want to show that $V$ is not
$\lambda$-measurable, where $\lambda$ is  Lebesgue measure. 

The set $P := \Rational\cap[0, 1]$ is countable. Define
$
V_p := \{v + p \mid v \in V\},
$
for $p\in P$. If $p, q\in P$ are different, $V_p\cap V_q = \emptyset$,
since $v_1 + p = v_2 + q$ implies $v_1 - v_2 = q - p\in\Rational$,
thus $\isEquiv{v_1}{v_2}{\alpha}$, so $v_1$ and $v_2$ are in the same
class, hence $v_1 = v_2$, thus $p = q$, which is a contradiction. 

Put $A :=\bigcup_{p\in P} V_p$, then $[0, 1]\subseteq A \subseteq [0,
2]$: Take $x\in [0, 1]$, then there exists $v\in V$ with
$\isEquiv{x}{v}{\alpha}$, thus $r := x - v\in \Rational$, hence $x\in
V_r$. On the other hand, if $x\in V_r$, then $x = v + r$, hence $0
\leq x \leq 2$.

If $A$ is $\lambda$-measurable, then $\lambda(A) = 0$ is
impossible, because this would imply $\lambda([0, 1]) = 0$, since
$\lambda$ is monotone. Thus $\lambda(A) > 0$. But $\lambda(V_p) =
\lambda(V)$ for each $p$, so that $\lambda(A) = \infty$ by countable
additivity. But this contradicts $\lambda([0, 2]) = 2$. Hence $A$ is
not $\lambda$-measurable, which implies that $V$ is not
$\lambda$-measurable. 
\EndExample


\Subsection{Games}
\label{sec:games}

We have two players, \emph{\index{Angel}Angel} and
\emph{\index{Demon}Demon}, playing against each other. For simplicity,
we assume that \emph{\index{playing}playing} means offering a natural
number, and that the game ---~like True Love~--- never ends. Let $A$
be a set of infinite sequences of natural numbers, then the \index{game}game
$G_A$ is played as follows.  Angel starts with $a_0\in \Nat$, Demon
answers with $b_0\in \Nat$, taking Angel's move $a_0$ into
account. Angel replies with $a_1$, taking the game's history $\langle
a_0, b_0\rangle$ into account, then Demon answers with $b_1$,
contingent upon $\langle a_0, b_0, a_1\rangle$, and so on. Angel wins
this game, if the sequence $\langle a_0, b_0, a_1, b_1, \dots\rangle$
is a member of $A$, otherwise Demon wins.

Let us have a look at strategies. Define 
\begin{equation*}
  \Natt := \Nat^\Nat
\end{equation*}
as the set of all sequences of natural
numbers, and let 
\begin{equation*}
  \Seq := \bigcup\{\langle n_1, \dots, n_k\rangle \mid k\geq0, n_1,
  \dots, n_k\in\Nat\}
\end{equation*}
be the set of all finite sequences of natural numbers (with
$\epsilon$ as the empty sequence). For easier notation later on, we
define \index{appending an element}appending an element to a finite sequence by $\langle n_{1}, \dots, n_{k}\rangle\frown n := \langle n_{1},
\dots, n_{k}, n\rangle$.  $\Sequ$ and $\Seqg$ denote all
sequences of odd, resp., even, length, the empty sequence is denoted
by $\epsilon$, we assume $\epsilon\in\Seqg$. 

A \emph{\index{game!strategy}\index{strategy}strategy}
$\sigma$ for Angel is a map $\sigma: \Seqg\to\Nat$ which works in the
following way: $a_0 := \sigma(\epsilon)$ is the first move of Angel,
Demon replies with $b_0$, then Angel answers with $a_1 := \sigma(a_0,
b_0)$, Demon reacts with $b_1$, which Angel answers with $a_2 :=
\sigma(a_0, b_0, a_1, b_1)$, and so on. If Angel plays according to
strategy $\sigma$ and Demon's moves are given by $b := \langle b_0, b_1,
\dots\rangle\in \Natt$, then the game's events are collected in
$\sigma*b\in \Natt$; hence $\sigma*b = \langle a_0, b_0, a_1, b_1,
\dots\rangle$ with $a_{2k+1} = \sigma(a_0, b_0, \dots, a_{2k},
b_{2k}\rangle$ for $k\geq0$ and $a_0 = \sigma(\epsilon)$. Similarly, a
strategy $\tau$ for Demon is a map $\tau: \Sequ\to\Nat$, working in
this manner: If Angel plays $a_0$, Demon answers with $b_0 :=
\tau(a_0)$, then Angel plays $a_1$, to which Demon replies with
$b_1 := \tau(a_0, b_1, a_1)$, and so on. If Angel's moves are
collected in $a := \langle a_0, a_1, \dots\rangle$, and if Demon plays
strategy $\tau$, then the entire game is recorded in the sequence
$a*\tau$. Thus $a*\tau = \langle a_0, b_0, a_1, b_1, \dots\rangle$
with $b_k = \tau(a_0, b_0, \dots, a_k)$ for $k\geq0$. 

\BeginDefinition{win-strat}
$\sigma: \Seqg\to\Nat$ is a \emph{\index{strategy!winning}winning strategy for Angel} in game $G_A$ iff
\begin{equation*}
  \{\sigma*b \mid b\in \Natt\} \subseteq A,
\end{equation*}
$\tau: \Sequ \to\Nat$ is a \emph{winning strategy for Demon} in game $G_A$
iff
\begin{equation*}
  \{a*\tau \mid a\in \Natt\} \subseteq \Natt\setminus A.
\end{equation*}
\EndDefinition

It is clear that at most one of Angel and Demon can have a winning
strategy. Suppose that in the contrary both have one, say, $\sigma$
for Angel and $\tau$ for Demon. Then $\sigma*\tau\in A$, since
$\sigma$ is winning for Angel, and $\sigma*\tau\not\in A$, since
$\tau$ is winning for Demon. So this assumption does not make sense.

We have a look at \emph{\index{game!Banach-Mazur}Banach-Mazur games},
another formulation of games which is sometimes more convenient. Each
Banach-Mazur game can be transformed into a game which we have defined
above.
 
Before discussing it, it will be convenient to introduce some
notation. Let $a, b\in \Seq$, hence $a$ and $b$ are finite sequences
of natural numbers. We say that $a \preceq b$ iff $a$ is an
\emph{initial piece} of $b$ (including $a = b$), so there exists $c\in
\Seq$ with $b = ac$; $c$ is denoted by $b/a$. If we want to exclude
equality, we write $a \prec b$.

\BeginExample{banach-mazur}
The game is played over $\Seq$, a subset $B\subseteq\Natt$ indicates a
winning situation. Angel plays $a_0\in\Seq$, Demon plays $b_0$ with
$a_0\preceq b_0$, then Angel plays $a_1$ with $a_0b_0\preceq a_1$,
etc. Angel wins this game iff the finite sequence
$a_0b_0a_1b_1\dots$ converges to an infinite sequence $x\in B$.

We encode this game in the following way. $\Seq$ is countable by
Proposition~\ref{seq-to-nat-bij}, so write this set as $\Seq = \{r_n
\mid n\in\Nat\}$. Put 
\begin{equation*}
  A := \{\langle w_0, w_1, \dots \rangle \mid r_{w_0} \preceq
  r_{w_1} \preceq r_{w_2} \dots \text{ converges to a sequence in } B\}.
\end{equation*}
It is then immediate that Angel has a strategy for winning the
Banach-Mazur game iff it has one for winning that game $G_{A}$. 
\EndExample

\Subsubsection{Determined Games}
\label{sec:determined-games}

Games in which neither Angel nor Demon have a winning strategy are
somewhat, well, indeterminate and might be avoided. We see some
similarity between a strategy and the selection argument in {\AC},
because a strategy selects an answer among several possible choices,
while a choice function picks elements, each from a given set. This
intuitive similarity will be investigated now.

\BeginDefinition{det-game}
A game $G_A$ is called \emph{\index{game!determined}determined} iff
either Angel or Demon has a winning strategy.
\EndDefinition

Suppose that each game $G_A$ is determined, no matter what set
$A\subseteq\Natt$ we chose, then we can define a choice function for
countable families of non-empty subsets of $\Natt$. 

\BeginTheorem{det-impl-choice} 
Assume that each game is determined. Then there exists a choice function
for countable families of non-empty subsets of $\Natt$.  
\EndTheorem

\BeginProof
1.
Let ${\cal F} := \{X_n\mid n \in \Nat\}$ be a countable family with
$\emptyset\not= X_n\subseteq\Natt$ for $n\in \Nat$. We will define a
function $f: {\cal F}\to \Natt$ such that $f(X_n)\in X_n$ for all
$n\in \Nat.$ The idea is to play a game which Angel cannot win, hence
for which Demon demon has a winning strategy. To be specific, if Angel
plays $\langle a_0, a_1, \dots\rangle$ and Demon plays $b := \langle b_0,
b_1, \dots \rangle$, then Demon wins iff $b\in X_{a_0}$. Since by
assumption Demon has a winning strategy $\tau$, we then put 
\begin{equation*}
  f(X_n) := \langle n, 0, 0, \dots\rangle*\tau.
\end{equation*}

2.
Let us look at this idea. Put
\begin{equation*}
  A := \{\langle x_0, x_1, \dots\rangle\in \Natt \mid \langle x_1, \dots\rangle\not\in X_{x_0}\}.
\end{equation*}
Suppose that Angel starts upon playing $a_0$. Since
$X_{a_0}\not=\emptyset$, Demon can take an arbitrary $b\in X_{a_0}$
and plays $\langle b_0, b_1, \dots\rangle$. Hence Angel cannot win, so
$B$ has a winning strategy $\tau$. 

3.
Now look at $\langle n, 0, 0, \dots\rangle*\tau\not\in A$, because
$\tau$ is a winning strategy. From the definition of $A$ we see that
this is an element of $X_n$, so we have found a choice function indeed. 
\EndProof

The space $\index{$\Natt$}\Natt$ looks a bit artificial, just as a mathematical
object to play around with. But this is not the case. It can be
shown that there exists a bijection $\Natt\to\Real$ with
some desirable properties (we will not enter into this construction,
however). With this in mind, we state as a consequence

\BeginCorollary{cor-det-impl-choice}
Assume that each game is determined. Then there exists a choice function
for countable families of non-empty subsets of $\Real$.  \QED
\EndCorollary

Let us fix the existence of a winning strategy for either Angel or
Demon in an axiom, the \emph{\index{Axiom of Determinacy}Axiom of Determinacy}.

\AxiomBox{AD}{Each game is determined.}

Given Corollary~\ref{cor-det-impl-choice}, the relationship of the
Axiom of Determinacy to the Axiom of Choice is of interest. Does {\AD}
imply {\AC}? The hope of establishing this are shattered, however, by
this observation.

\BeginProposition{ac-shatters-dt}
If \emph{{\AC}} holds, there exists $A\subseteq\Natt$ such that $G_A$ is not
determined. 
\EndProposition

Before entering the proof, we observe that the
set of all strategies $S_A$ for Angel resp. $S_D$ for Demon has the
same cardinality as the power set $\PowerSet{\Nat}$ of $\Nat$. 

\BeginProof
0.
We have to find a set $A\subseteq\Natt$ such that neither Angel nor
Demon has a winning strategy for the game $G_A$. By {\AC}, the sets
$S_A$ resp. $S_D$ can be well-ordered, by the observation just made we
can write 
\begin{align*}
  S_A & = \{\sigma_\alpha \mid \alpha < \omega\},\\
S_D & = \{\tau_\alpha \mid \alpha < \omega\}.
\end{align*}

1.  We will construct now disjoint sets $X=\{x_\alpha \mid \alpha <
\omega\}\subseteq\Natt$ and $Y=\{y_\alpha \mid \alpha <
\omega\}\subseteq\Natt$ indexed by $\{\alpha \mid \alpha < \omega\}$,
which will help define the game. Suppose $x_\beta$ and $y_\beta$ are
defined for all $\beta < \alpha$. Then, because $\alpha$ is countable,
the sets $\{x_\beta \mid \beta < \alpha\}$ and $\{y_\beta \mid \beta <
\alpha\}$ are countable as well, and there are uncountably many
$b\in\Natt$ such that $\sigma_\alpha*b\not\in \{x_\beta \mid \beta <
\alpha\}$. Take one of them and put $y_\alpha := \sigma_\alpha*b$. For
the same reason, there are uncountably many $a\in \Natt$ such that
$a*\tau_\alpha\not\in\{y_\beta \mid \beta\leq\alpha\}$; take one of
them and put $x_\alpha := a*\tau_\alpha$.

2.  Clearly, $X$ and $Y$ are disjoint. Angel does not have a strategy
for winning game $G_X$. Suppose it has a winning strategy $\sigma$, so
that $\sigma = \sigma_\alpha$ for some $\alpha<\omega$. But $y_\alpha
= \sigma_\alpha*b\not\in X$ by construction, which is a
contradiction. One similarly shows that Demon cannot have a winning
strategy for game $G_X$. Hence this game is not determined.  \EndProof

\Subsubsection{Proofs Through Games}
\label{sec:proofs-through-games}

We will show now that games are a tool for proofs. The basic idea is
to attach a statement to a game, and if Angels has a strategy for
winning the game, then the statement is established, otherwise it is
not. Hence we have to encode the statement in such a way that this
mechanism can be used, but we have also to establish a scenario in
which to argue. The formulation chosen suggests that Angel has to have
a winning strategy for winning a game, which in turn suggests that we
assume a framework in which games are determined. But we have seen
above that this is not without conflicts when considering {\AC}. This
section is devoted to establish that every subset of the unit interval
is Lebesgue measurable, provided each game is determined. We have seen
in Example~\ref{vitaly} that {\AC} implies that there exists a set
which is not Lebesgue measurable. Hence ``it is natural to postulate
that Determinacy holds to the extent that it does not contradict the
Axiom of Choice'', as T. Jech writes in his massive treatise of Set
Theory~\cite[p. 628]{Jech}.

\paragraph{The Goal.}
\label{sec:goal}
We want to show that each subset of the unit interval is measurable,
provided each game is determined. This is based on the observation
that it is sufficient to establish that $\lambda_*(A) > 0$ or
$\lambda_*([0, 1]\setminus A) > 0$ for each and every subset $A\subseteq
[0, 1]$, where, as above, $\lambda$ is Lebesgue measure on the unit
interval. This is the reason:

\BeginLemma{inner-vs-outer}
Assume that there exists a subset of the unit interval which is not
$\lambda$-measurab\-le. Then there exists a subset $M\subseteq[0, 1]$
with $\lambda_{*}(M) = 0$ and $\lambda^{*}(M) = 1$. \QED
\EndLemma 

\paragraph{The Basic Approach.}
\label{sec:basic-approach}
Given an arbitrary subset $X\subseteq[0, 1]$, we will define a game
$G_{X}$ such that if there exists a winning strategy for Angel, then we can
find a measurable subset $A\subseteq X$ which has positive Lebesgue
measure (hence $\lambda_{*}(X) > 0$). If there exists, however, a
winning strategy for Demon, then we can find a measurable subset
$A\subseteq[0, 1]$ with positive Lebesgue measure such that $A\cap X =
\emptyset$ (hence $\lambda_{*}([0, 1]\setminus A) >0$). 

\paragraph{Little Helpers.}
\label{sec:little-helpers}

We need some preparations before we start. So let's get on with it
now as not to interrupt the flow of discussion later on.

\BeginLemma{converg-to-one-point}
Let $\Folge{F}$ be a sequence of non-empty subsets of the unit interval $[0, 1]$
such that
\begin{enumerate}
\item Each $F_n$ is a finite union of closed intervals.
\item The sequence is monotonically decreasing, hence $F_1 \supseteq
  F_1 \supseteq \dots$.
\item The sequence of \index{diameter}diameters $\diam{F_{n}} := \sup_{x, y\in F_n} |x - y|$ tends to zero.
\end{enumerate}
Then there exists a unique $p\in [0, 1]$ with
$
\{p\} = \bigcap_{n\in\Nat} F_n.
$
\EndLemma

\BeginProof
1.
It is clear from the last condition that there can be at most one
point in the intersection of this sequence. Suppose there are two
distinct points $p, q$ in this intersection, then $
\delta := |p - q| > 0$. But there exists some $n_0\in\Nat$ with
$\diam{F_m} < \delta$ for all $m\geq n_0$. This is a contradiction. 

2.
Assume that $\bigcap_{n\in\Nat}
F_n=\emptyset$. Put $G_n := [0, 1]\setminus F_n$, then $G_n$ is the
union of a finite number of open intervals, say 
$
G_n = H_{n, 1}\cup \dots \cup H_{n, k_n},
$
and 
$
[0, 1] \subseteq \bigcup_{n\in\Nat} G_n.
$
By the Heine-Borel Theorem~\ref{heine-borel} there exist a
finite set of intervals $H_{n_i, j_i}$ with $1 \leq i \leq r, 1 \leq
j_i \leq k_{n_i}$ such that 
$
[0, 1] \subseteq \bigcup_{i=1}^r H_{n_i, j_i}.
$
Because the sequence of the $F_n$ decreases, the sequence $\Folge{G}$
is increasing, so we find an index $N$ such that $H_{n_i,
  j_i}\subseteq G_N$ for $1 \leq i \leq r, 1 \leq
j_i \leq k_{n_i}$. But this means $[0, 1] \subseteq G_N$, thus $F_N=\emptyset$,
contradicting the assumption that all $F_n$ are non-empty. 
\EndProof

Another preparation concerns the convergence of an infinite product. 

\BeginLemma{inf-prod-conv}
Let $\Folge{a}$ be a sequence of real numbers with $0 < a_n < 1$ for
all $n\in\Nat$. Then the following statements are equivalent
\begin{enumerate}
\item $\prod_{i\in\Nat} (1-a_i) := \lim_{n\to\infty}\prod_{i=1}^n
  (1-a_i)$ exists.
\item $\sum_{n\in\Nat} a_n$ converges.
\end{enumerate}
\EndLemma

\BeginProof
One shows easily by induction on $n$ that
\begin{equation*}
  \prod_{i=1}^n (1-a_i) > 1 - \bigl(\sum_{1=1}^n a_n\bigr)
\end{equation*}
for $n \geq 2$. Since $0 < a_n < 1$ for all $n\in\Nat$, this implies the equivalence. 
\EndProof

This has an interesting consequence, viz., that we have a positive
infinite product, provided the corresponding series converges. To be
specific:

\BeginCorollary{inf-prod-conv-cor}
Let $\Folge{a}$ be a sequence of real numbers with $0 < a_n < 1$ for
all $n\in\Nat$. Then the following statements are equivalent
\begin{enumerate}
\item $\prod_{i\in\Nat} (1-a_i)$ is positive.
\item $\sum_{n\in\Nat} a_n$ converges.
\end{enumerate}
\EndCorollary

\BeginProof
1.
Put $Q_{k} :=\prod_{i=1}^{k} a_{i}, Q := \lim_{k\rightarrow\infty} Q_{k}$.  Assume that $\sum_{n\in\Nat} a_n$ converges, then there exists $m\in
\Nat$ such that $\sigma_{m} := \sum_{i=m}^{\infty} < 1$. Hence we have
\begin{equation*}
  \frac{Q_{n}}{Q_{m}} > 1 - (a_{m+1} + \dots + a_{n}) > 1 - \sigma_{m}
\end{equation*}
for $n > m$, so that  
\begin{equation*}
  Q = \lim_{k\rightarrow\infty} Q_{k} > Q_{m}\cdot(1-\sigma_{m}) > 0.
\end{equation*}
2.
On the other hand, if the series diverges, then we can find an index
$m$ for $N\in\Nat$ such that 
$
a_{1} + \dots + a_{n} > N
$ 
whenever $n > m$. Hence 
\begin{equation*}
  \prod_{n\in \Nat}\frac{1}{1-a_{n}} \leq \lim_{k\rightarrow\infty}
  \frac{1}{1 -(a_{1}+ \dots + a_{k})} = 0
\end{equation*}
\EndProof

This observation will be helpful when looking at our game.

\paragraph{The Game.}
\label{sec:game}
Before discussing the game proper, we set the stage. Fix a
sequence $\Folge{r}$ of reals such that 
$
\sum_{n\in\Nat} r_n < 1
$
and
$
1/2 > r_1 > r_2 > \dots.
$

Let $k\in\Nat$ be a natural number, and define ${\cal J}_k$ as the
collection of sets $S$ with these properties
\begin{itemize}
\item $S\subseteq[0, 1]$ is a finite union of closed intervals with
  rational end points.
\item The \emph{diameter} $\diam{S} = \sup_{x, y \in
    S} |x - y|$ of $S$ is smaller than $1/2 ^k$.
\item The Lebesgue measure $\lambda(S)$ of $S$ is $r_1\cdot
  {\dots}\cdot
  r_k$. 
\end{itemize}
Put ${\cal J}_{0} := \{[0, 1]\}$ as the mandatory first draw of Angel.
Note that ${\cal J}_{k}$ is countable for all $k\in\Nat$, so that
$\bigcup_{k\geq0} {\cal J}_{k}$ is countable as well by
Proposition~\ref{ac-implies-countable} (this was proved without using
{\AC}!).

The game starts. We fix $X\subseteq [0, 1]$ as the Great Prize; this is the set we
want to investigate. Angels starts with choosing the unit interval
$S_{0} := [0, 1]$, Demon chooses a set $S_{1}\in{\cal J}_{1}$, then
Angel chooses a set $S_{2}\in{\cal J}_{2}$ with $S_{2}\subset S_{1}
\subset S_{0}$, Demon chooses a set $S_{3}\subset S_{2}$ with
$S_{3}\in{\cal J}_{3}$, and so on. In this way, the game defines a decreasing
sequence $\Folge{S}$ of closed sets the diameter of which tends to
zero. By Lemma~\ref{converg-to-one-point} there exists exactly one
point $p$ with $p\in\bigcap_{n\in\Nat} S_{n}$. If $p\in [0,
1]\setminus X$, then Angel wins, if $p\in X$, then Demon wins. 

\paragraph{Analysis of the Game.}
\label{sec:analysis-game}
First note that we will not encode the game into a syntactic form
according to the definition of $G_{A}$. This would require much
encoding and decoding between the formal representation and the
informal one, so that the basic ideas might get lost. Since life is
difficult enough, we stick to the informal representation, trusting
that the formal one could easily be derived from it, and focus on the
ideas behind the game. After all, we want to prove something  through this game which is
not entirely trivial.

The game spawns a tree rooted at $S_{0} := [0, 1]$ with offsprings all those
elements $S_{1}$ of ${\cal J}_{1}$ with $S_{1} \subset S_{0}$. If we
are at node $S_{k}\in{\cal J}_{k}$, then this node has all elements
$S\in{\cal J}_{k+1}$ as offsprings for which $S\subset S_{k}$
holds. Consequently, the tree's depth will be infinite, because the
game continues forever. The offsprings of a node will be investigated
in a moment. 

We define for easier discussion the sets 
\begin{align*}
  {\cal W}_{k} & := \{\langle S_{0}, \dots, S_{k}\rangle
  \in\prod_{i=0}^{k} {\cal J}_{i}\mid S_{0}\supset S_{1}\supset
  \dots\supset S_{k}\},\\
{\cal W}^{*}& := \bigcup_{k\geq0} {\cal W}_{k}
\end{align*}
as the set of all paths which are possible in this game. Hence Angel
chooses initially $S_{0} = [0, 1]$, Demon chooses $S_{1}\in{\cal
  J}_{1}$ with $S_{1}\subset S_{0}$ (hence $\langle S_{0},
S_{1}\rangle\in {\cal W}_{1}$), so that $\langle S_{0}, S_{1},
S_{2}\rangle\in{\cal W}_{2}$, etc. ${\cal W}_{2n}$ is the set of all
possible paths after the $n$-th draw of Angel, and ${\cal W}_{2n+1}$
yields the state of affairs after the $n$-th move of Demon.

For an analysis of strategies, we will fix now $k\in\Nat$ and a map
$\Gamma: {\cal W}_{k}\to {\cal J}_{k+1}$ such that
$\Gamma(S_{0}, \dots, S_{k}) \subset S_{k}$, hence $\langle
S_{0}, \dots, S_{k}, \Gamma(S_{0}, \dots, S_{k})\rangle = \langle
S_{0}, \dots, S_{k}\rangle\frown \Gamma(S_{0}, \dots, S_{k})\in{\cal
  W}_{k+1}.$ Just to have a handy name for it, call such a map
\emph{admissible at $k$}.

\BeginLemma{zerl-next-state}
Assume $\Gamma$ is admissible at $k$. Given $\langle S_{0}, \dots, S_{k}\rangle\in{\cal W}_{k}$, there exists
$m\in\Nat$ and 
a finite sequence $T_{k+1, i}\in{\cal J}_{k+1}$ for $1 \leq i \leq m$ such that
\begin{enumerate}
\item $T_{k+1, i}\subset S_{k}$ for all $i$,
\item $\lambda\bigl(\bigcup_{i=1}^{m} \Gamma(S_{0},
\dots, S_{k}, T_{k+1, i})\bigr) \geq
  \lambda(S_{k})\cdot(1 - 2\cdot r_{k+1})$,
\item The sets $\Gamma(S_{0},
\dots, S_{k}, T_{k+1, 1}), \dots, \Gamma(S_{0},
\dots, S_{k}, T_{k+1, m})$ are mutually disjoint.
\end{enumerate}
\EndLemma

\BeginProof
The sets $T_{k+1, i}$ are defined by induction. Assume that $T_{k+1,
  1}, \dots, T_{k+1, j}$ is already defined for $j \geq 0$, put
\begin{equation*}
R_{j} := S_{k}\setminus\bigcup_{i=1}^{j}\Gamma(S_{0}, S_{1},
\dots, S_{k}, T_{k+1, i}).
\end{equation*}

Now we have two possible cases: either 
\begin{equation*}
(\ddag)\ \lambda(R_{j}) > 2\cdot \lambda(S_{k})\cdot r_{k+1}
\end{equation*}
or this
inequality is false. Note that $\lambda(S_{k}) =
r_{1}\cdot{\dots}\cdot r_{k}$, and $1/2 > r_{k} > r_{k+1}$, so that
initially $\lambda(R_{0}) = \lambda(S_{k}) > 2\cdot\lambda(S_{k})\cdot
r_{k+1}$. Now assume that $(\ddag)$ holds. Because $S_{k}$ is the
union of a finite number of closed intervals, and because $R_{j}$ does
not exhaust $S_{k}$, we conclude that $R_{j}$ contains a subset $P$
with diameter $\diam{P}\leq\diam{R_{j}}\leq2^{-(k+1)}$ such that
$\lambda(P) > \lambda(S_{k})$. We can select $P$ in such a way that it
is a finite union of intervals. Then there exists $T_{k+1, j+1}
\subseteq P$ which belongs to ${\cal J}_{k+1}$. Take it. Then the
first property is satisfied. 

This process continues until inequality $(\ddag)$ becomes
false, which gives the second property. Because
\begin{equation*}
\Gamma(S_{0}, \dots, S_{k}, T_{k+1, i})\subset
T_{k+1, i} \subset S_{k}\setminus \bigcup_{j=1}^{i-1}
\Gamma(S_{0}, \dots, S_{k}, T_{k+1, j}),
\end{equation*}
we conclude that the
sets $\Gamma(S_{0},
\dots, S_{k}, T_{k+1, 1}), \dots, \Gamma(S_{0},
\dots, S_{k}, T_{k+1, m})$ are mutually disjoint. 
\EndProof

Now let $\tau$ be a strategy for Demon, hence $\tau:
\bigcup_{k\geq0}{\cal W}_{2k+1}\to \bigcup_{k\leq0}{\cal J}_{2k}$ is a map such that
$\tau(S_{0}, \dots, S_{2k})\subset S_{2k}$. If the game's history at
time $k$ is given by the path $\langle S_{0}, \dots, S_{2k}\rangle$ with Angels having
played $S_{2k}$ as a last move, then the game continues with
$\tau(S_{0}, \dots, S_{2k})$ as Demon's next move,
so that the new path is just 
$
\langle S_{0}, \dots, S_{2k}\rangle\frown\tau(S_{0}, \dots, S_{2k}).
$ 

Let's see what happens if Angel selects the next move according to
Lemma~\ref{zerl-next-state}. Initially, Angels plays $S_{0}$, then
Demon plays $\tau(S_{0})$, so that the game's history is now
$\langle S_{0}\rangle\frown\tau(S_{0})$; let $T_{0, 1}, \dots, T_{0, m_{0}}$ be the
sets selected according to Lemma~\ref{zerl-next-state} for this
history, then the possible continuations in the game are $t_{i} :=
\langle S_{0}\rangle\frown\tau(S_{0})\frown T_{0, i}$ for $1\leq i \leq m_{0}$, so
that Demon's next move is $t_{i}\frown\tau(t_{i})$, thus
\begin{equation*}
K_{\tau}(\langle S_{0}\rangle\frown\tau(S_{0})) := \{\langle S_{0}\rangle\frown\tau(S_{0})\frown T_{0,
  i}\frown\tau\bigl(\langle S_{0}\rangle\frown\tau(S_{0})\frown T_{0, i})\bigr) \mid 1 \leq i
\leq m_{0}\}\in {\cal W}_{3}
\end{equation*}
describes all possible moves for Demon in this scenario. Given $\tau$,
this depends on $S_{0}$ as the history up to that moment, and on the
choice to Angel's moves according to Lemma~\ref{zerl-next-state}. To
see the pattern, consider Demon's next move. Take $t = \langle t_{0},
t_{1}, t_{2}, t_{3}\rangle\in
K_{\tau}(S_{0}\frown\tau(S_{0}))$, then $\tau(t)\in{\cal J}_{4}$ with $\tau(t)\subset
t_{3}$, and choose $T_{1, 1}, \dots, T_{1, m_{1}}$ according to
Lemma~\ref{zerl-next-state} as possible next moves for Angel, so that
the set of all possible moves for Demon given this history is an element of the
set 
\begin{equation*}
K_{\tau}(t) = K_{\tau}(\langle t_{1}, t_{2},
t_{3}\rangle\frown\tau(t_{1}, t_{2},
t_{3})) := \{t\frown T_{1, i}\frown \tau(t\frown T_{1, i}) \mid 1 \leq i \leq m_{1}\}\in{\cal W}_{5}.
\end{equation*} 

This provides a window into what is happening. Now let us look at
the broader picture. Denote by for $t\in{\cal W}_{n}$ by $J_{\tau}(t)$ the set 
$
\{t\frown T_{n, 1}, \dots, t\frown T_{n, m}\},
$
where $T_{n, 1}, \dots, T_{n, m}$ are determined for $t$ and $\tau$
according to Lemma~\ref{zerl-next-state}  as the set of all possible moves for Angel. Hence given history $t$,
$J_{\tau}(t)$ is the set of all possible paths for which Demon has to
provide the next move. Then put
\begin{equation*}
  J_{\tau}^{n} :=  \bigcup_{s_{2}\in J_{\tau}(\langle S_{0}\rangle\frown
    \tau(S_{0}))}
 \ \bigcup_{s_{4}\in J_{\tau}(s_{2}\frown
    \tau(s_{2}))}
\dots
\bigcup_{s_{2(n-1)}\in J_{\tau}(s_{2(n-2)}\frown
    \tau(s_{2(n-2)}))}
\ J_{\tau}\bigl(s_{2(n-1)}\frown\tau(s_{2(n-1)})\bigr)
\end{equation*}
with
\begin{equation*}
  J_{\tau}^{1} = J_{\tau}(\langle S_{0}\rangle\frown \tau(S_{0})).
\end{equation*}
Finally, define
\begin{equation*}
  A_{n} := \bigcup\{\tau(s_{2n}) \mid s_{2n}\in J_{\tau}^{n}\}.
\end{equation*}
Hence $J_{\tau}^{n}$ contains all possible moves of Angel at time
$2n$, so that $A_{n}$ tells us what Demon can do at time $2n+1$. These
are the important properties of $\Folge{A}$:

\BeginLemma{demon-can-do-vol}
We have for all $n\in\Nat$
\begin{enumerate}
\item $\lambda(A_{n}) \geq r_{1}\cdot\prod_{i=1}^{n} (1-2\cdot
  r_{2i})$
\item  $A_{n+1}\subset A_{n}$
\end{enumerate}
\EndLemma

\BeginProof
1.
The second property follows immediately from
Lemma~\ref{zerl-next-state}, so we will focus on the first
property. It will be proved by induction on $n$. We infer from Lemma~\ref{zerl-next-state} that the sets
$\tau(s_{2n})$ are mutually disjoint, when $s_{2n}$ runs through
$J_{\tau}^{n}$

2.
$n = 1$. 
We obtain immediately from Lemma~\ref{zerl-next-state} that 
\begin{align*}
  \lambda(A_{1}) & = \lambda\bigl(\bigcup\{\tau(s_{2}) \mid s_{2}\in
  J_{\tau}(\langle S_{0}\rangle\frown \tau(S_{0})\})\bigr)\\
& \geq r_{1}\cdot(1-2\cdot r_{2})
\end{align*}
(set $\Gamma := \tau$ and $k=1$).

2.
Induction step $n\rightarrow n+1$.  We infer from
Lemma~\ref{zerl-next-state} that
\begin{equation*}
  (\dag)\ \lambda\bigl(\bigcup\{\tau(s_{2(n+1)} \mid s_{2(n+1)}\in
  J_{\tau}(s_{2n})\}\bigr)
\geq 
\lambda(\tau(s_{2n+1}))\cdot(1-2\cdot r_{2(n+1)}).
\end{equation*}
Disjointness then implies
\begin{align*}
  \lambda(A_{n+1}) & = \sum_{s_{2n}\in J_{\tau}^{n}} \lambda\bigl(\bigcup\{\tau(s_{2(n+1)} \mid s_{2(n+1)}\in
  J_{\tau}(s_{2n})\}\bigr)\\
& \geq \sum_{s_{2n}\in
  J_{\tau}^{n}}\lambda(\tau(s_{2n}))\cdot(1-2\cdot r_{2(n+1)})
&&\text{(inequality }(\dag))\\
& = \lambda\bigl(\bigcup_{s_{2n}\in J_{\tau}^{n}}\tau(s_{2n+1})\bigr)\cdot(1 -
2\cdot r_{2(n+1)}) &&\text{(disjointness)}\\
& = \lambda(A_{n})\cdot (1-2\cdot r_{2(n+1)}) &&\text{(induction hypothesis)}\\
& \geq r_{1}\cdot\prod_{i=1}^{n+1}(1-2\cdot r_{2i})
\end{align*}
\EndProof

Now we are getting somewhere --- we show that we can find for every
element in $\bigcap_{n\in\Nat}A_{n}$ a strategy so that the moves of
Angel and of Demon \emph{converge} to this point. To be more specific:

\BeginLemma{exists-strategy}
Assume that Demon adopts strategy $\tau$. For every point
$p\in\bigcap_{n\in\Nat}A_{n}$ there exists for Angel a strategy
$\sigma_{p}$ with this property: If Angel plays $\sigma_{p}$ and Demon
plays $\tau$, then $\bigcap_{i=0}^{\infty}S_{i} = \{p\}$, where $S_{0},
S_{1}, \dots$ are the consecutive moves of the players.
\EndLemma

\BeginProof
The sets $s_{2n}\in J_{\tau}^{n}$ are mutually disjoint for fixed $n$,
so we find a unique sequence $s'_{2n}\in J_{\tau}^{n}$ for which $p\in
\tau(s'_{2n})$. Represent $s'_{2n} = \langle S_{0}, \dots,
S_{2n}\rangle$, and let $\sigma_{p}$ be a strategy for Angel such that  
$
\sigma_{p}(\langle S_{0}, \dots, S_{2n-1}\rangle\frown\tau( S_{0},
\dots, S_{2n-1})) = S_{2n}
$
holds. Thus $p\in \bigcap_{n\in\Nat} S_{n}$, if Angel plays
$\sigma_{p}$ and Demon plays $\tau$. 
\EndProof

Now let $\tau$ be a winning strategy for Demon, then
$
A := \bigcap_{n\in\Nat} A_{n}\subseteq [0, 1]\setminus X;
$
this is the outcome if Angel plays one of the strategy in
$\{\sigma_{p} \mid p \in A\}$. There may be other strategies for Angel than the one described above,
but no matter how Angel plays the game, we will end up in an element
not in $X$. This implies $\lambda(A) \leq \lambda_{*}([0, 1]\setminus
X)$. But we know from Lemma~\ref{demon-can-do-vol} that $\lambda(A)
\geq r_{1}\cdot \prod_{i=1}^{\infty}(1-2\cdot r_{2i}) > 0$ by
Lemma~\ref{inf-prod-conv} and it corollary, consequently, $\lambda_{*}([0, 1]\setminus
X) > 0$. If, however, Demon does not have a winning strategy, the
Angel has one, if we assume that the game is determined. The
argumentation is completely the same as above to show that
$\lambda_{*}(X) > 0$. 

Thus we have shown:
\BeginTheorem{det-vs-ac}
If each game is determined, then each subset of the unit interval is
$\lambda$-measurable.
\QED
\EndTheorem

We have seen that games are not only just for fun, but are a tool for
investigating properties of sets. In fact, one can define games for
investigating many topological properties, not all as laborious as
the one we have defined above. 

\Subsection{Wrapping it up}
\label{sec:wrapping-it-up}

This summarizes the discussion. Some hints for further information can
be found in the Bibliographic Notes. The Lecture
Note~\cite{Herrlich-Choice} by H. Herrlich and the list of its references
is a particularly rich rich bag of suggestions for further reading. The discussion in P. Taylor's book~\cite[p. 58]{Taylor} (``Although we, at the cusp of the century, now reject Choice ...'') is also worth looking at, albeit from a completely different angle. 

This is a small diagram indicating the dependencies discussed here.

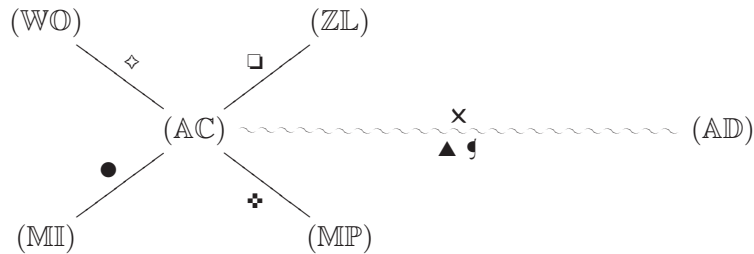
\begin{equation*}
  \xymatrix{
{\WO}\ar@{-}[dr]^{\text{\ding{71}}}
&&{\ZL}\ar@{-}[dl]_{\text{\ding{111}}}\\
&
{\AC}\ar@{-}[dl]_{\text{\ding{108}}}\ar@{-}[dr]_{\text{\ding{60}}}\ar@{-}@{~}[rrrrr]^{\text{\ding{53}}}_{\text{\ding{115}\ \ding{161}}}&&&&&{\AD}\\
{\MI} && {\MP}\\
}
\end{equation*}

The symbols provide a directory for the corresponding statements.

\begin{center}
  \begin{tabular}[h]{|c|l|}\hline\hline
    \ding{71} & Theorem~\ref{wo-equiv-ac}\\\hline
    \ding{111} & Proposition~\ref{zl-impl-ac}\\\hline
    \ding{108}& Theorem~\ref{pi-equiv-ac}\\\hline
    \ding{60}& Proposition~\ref{mp-equiv-zl}\\\hline\hline
    \ding{53}& Existence of a non-determined game under {\AC}, Proposition~\ref{ac-shatters-dt}\\\hline
\ding{115}& Choice function for countable families under {\AD},
Theorem~\ref{det-impl-choice} \\\hline
\ding{161}& Measurability of every subset of $[0, 1]$ under {\AD}, Theorem~\ref{det-vs-ac}\\\hline\hline
  \end{tabular}
\end{center}


\Subsection{Bibliographic Notes}
\label{sec:bibliographic-notes}

This chapter contains mostly classical topics. The proof of Cantor's
enumeration and its consequences for enumerating the set of all finite
sequences of natural numbers in taken
from~\cite{Kuratowski-Mostowski}, so is the discussion of
ordinals. Jech's representation~\cite{Jech} has been helpful as well,
so was~\cite{Goldrei}. The books by Davis~\cite{Davis-Engines} and by Aczel~\cite{Aczel-Infinity} contain some gripping
historical information on the subject of early set theory;
the monograph~\cite{Cantone-Omodeo-Policriti} discusses implications
for computing when the Axiom of Foundations (page~\pageref{lab-axiom-of-foundation}) is weakened.

Term rewriting is discussed in~\cite{Baader-Nipkow}; reduction systems
(Example~\ref{term-rewriting}) are central to it. Aumanns's
classic~\cite{Aumann}, unfortunately not as frequently used as this
valuable book should be, helped in discussing Boolean algebras, and
the proof for the general distributive law in Boolean algebras as well
as some exercises has been taken from~\cite{Birkhoff} and
from~\cite{Davey+Priestley}, see also~\cite{Stanley-1} for finite
lattices. The discussion on measure extension follows quite closely
the representation given in the first three chapters
of~\cite{Billingsley} with an occasional glimpse at~\cite{Elstrodt}
and the awesome~\cite{Bogachev}. Finally, games are introduced as
in~\cite[Chapter 33]{Jech}, see also~\cite{Jech-Choice}; the
game-theoretic proofs on measurability are taken
from~\cite{Mycielski-Determined}. Infinite products are discussed at
length in the delightful textbook~\cite{Bromwich}, see
also~\cite{Chrystal}. A general source for this chapter was the
exposition by H. Herrlich~\cite{Herrlich-Choice}, providing a
\emph{tour d'horizon}.

\newpage
\Subsection{Exercises}
\label{sec:exercises-set}
\Exercise{pairs-are-equal}{The Axiom of Pairs defines 
$
\langle a, b\rangle := \bigl\{\{a\}, \{a, b\}\bigr\},
$
see page~\pageref{sec:axioms-zfc}. Using the Axioms of ZFC, show that 
$
\langle a, b\rangle = \langle a', b'\rangle
$
iff 
$
a = a'
$
and 
$ 
b = b'.
$
}
\Exercise{ex-prop-invers}{
Show that $f: A \to B$ is injective iff $f^{-1}:
\PowerSet{B}\to\PowerSet{A}$ is surjective; $f$ is surjective iff
$f^{-1}$ is injective.
}
\Exercise{ex-divides}{Define $\leq_d$ on $\Nat$ as in
  Example~\ref{ord-div}. Show that $p$ is prime iff $p$ is a minimal
  element of $\Nat\setminus\{1\}$.}
\Exercise{ex-subset}{Order $S := \PowerSet{\Nat}\setminus\{\Nat\}$ by
  inclusion as in Example~\ref{ex-subset}. Show that the set $A :=
  \{\{2\cdot n, 2\cdot n+1\} \mid n\in \Nat\}$ is bounded in $S$; does
$A$ have a smallest lower bound?}
\Exercise{ex-knaster-tarski}{
Let $S$ be a set, $H: \PowerSet{S} \to\PowerSet{S}$ be an order
preserving map. Show that 
$
A := \bigcup\{X\in\PowerSet{S}\mid X\subseteq H(X)\}
$
is a fixed point of $H$, i.e., satisfies $H(A) = A$. Moreover, $A$ is
the greatest fixed point of $H$,i.e., if $H(Y) = Y$, then $Y\subseteq A.$
}

\Exercise{ex-banach-decomposition}{
Let $f: X\to Y$ and $g: Y \to X$ be maps. Using
Exercise~\ref{ex-knaster-tarski} show that there exists disjoint
subsets $X_{1}$ and $X_{2}$ of $X$ and disjoint subsets $Y_{1}$ and
$Y_{2}$ of $Y$ such that $X = X_{1}\cup X_{2}$, $Y = Y_{1}\cup Y_{2}$
and $\Bild{f}{X_{1}} = Y_{1}$, $\Bild{g}{Y_{2}} = X_{2}$. The map
$A \mapsto X\setminus\Bild{g}{Y\setminus\Bild{f}{A}}$ might be helpful.

This decomposition is attributed to \emph{\index{Banach decomposition}
  S. Banach}.}

\Exercise{ex-schroeder-bernstein-ancora}{
Use Exercise~\ref{ex-banach-decomposition} for a proof of the
Schröder-Bernstein-Theorem~\ref{Schr-Bernst}. 
}
\Exercise{ex-k-and-l}{Show that there exist for the bijection $J$ from
  Proposition~\ref{cantor-bijects} surjective maps $K: \Natn\to\Natn$
  and $L: \Natn\to\Natn$ such that $J(K(x), L(x)) = x, K(x) \leq x$
  and $L(x)\leq x$ for all $x\in \Natn$.}  
\Exercise{ex-pown-to-r}{
  Construct a bijection from the power set $\PowerSet{\Nat}$ to
  $\Real$ using the Schröder-Bernstein-Theorem~\ref{Schr-Bernst}. }
\Exercise{pairs-are countable}{ Show using the Schröder-Bernstein
  Theorem~\ref{Schr-Bernst} that the set of all subsets of $\Nat$ of
  size exactly $2$ is countable. Extend this result by showing that
  the set of all subsets of $\Nat$ of size exactly $k$ is
  countable. Can you show without {\AC} that the set of all finite
  subsets of $\Nat$ is countable?  }
\Exercise{ex-omega-is-ordinal}{Show that $\omega_{1} := \{\alpha \mid
  \alpha\text{ is a countable ordinal}\}$ is an ordinal
  (Proposition~\ref{omega-is-ordinal}). Show that $\omega_{1}$ is not
  countable.}  
\Exercise{ex-graph-color}{ An
  \index{graph!undirected}undirected graph ${\cal G} = (V, E)$ has
  nodes $V$ and (undirected) edges $E$. An edge connecting nodes $x$
  and $y$ be written as $\{x, y\}$; note $x\not=y$. A subgraph ${\cal
    G}' = (G', E')$ of ${\cal G}$ is a graph with $G'\subseteq G$ and
  $E'\subseteq E$. ${\cal G}$ is
  \emph{\index{graph!$k$-colorable}$k$-colorable} iff there exists a
  map $c: V\to \{1, \dots, k\}$ such that $c(x)\not=c(y)$, whenever
  $\{x, y\}\in E$ is an edge in ${\cal G}$. Show that ${\cal G}$ is
  $k$-colorable iff each of its finite subgraphs is $k$-colorable.  }
\Exercise{somen-ring}{Let $B$ be a Boolean algebra, and define $a
  \ominus b := (a\vee b)\wedge -(a\wedge b)$, as in
  Section~\ref{sec:prime-ideals}. Shown that $(B, \ominus, \cap)$ is a
  commutative ring.}  \Exercise{ex-set-1}{Complete the proof of
  Proposition~\ref{mp-equiv-zl} by proving that {\AC} $\Rightarrow$
  {\MP}.}  
\Exercise{ex-compl-prime-id}{Complete the proof of
  Lemma~\ref{compl-prime-id}.}  \Exercise{ex-compact-countable}{ Using
  the notation of Section~\ref{sec:comp-theor-prop}, show that
  $v^{*}\models \phi$ iff $\phi\in{\cal M}^{*}$ using induction on the
  structure of $\phi$.  } 
\Exercise{ex-graph-color-due}{ Do
  Exercise~\ref{ex-graph-color} again, using the Compactness
  Theorem~\ref{completeness-propositional}.  }
\Exercise{ex-ultrafilter-finite}{
Let ${\cal F}$ be an ultrafilter over an infinite set $S$. Show that if ${\cal
  F}$ contains a finite set, then there exists $s\in S$ such that
${\cal F}={\cal F}_{s}$, the ultrafilter defined by $s$, see Example~\ref{pointed-filter-0}.
}
\Exercise{ex-is-lattice}{ Consider this ordered set

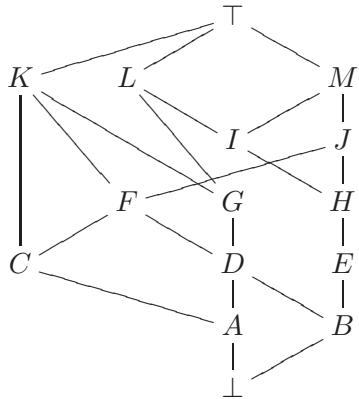
\smallBox{
\begin{equation*}
  \xymatrix@=10pt@C=25pt{
&&\top\ar@{-}[dll]\ar@{-}[dl]\ar@{-}[dr]\\
K\ar@{-}[ddd]\ar@{-}[ddrr]\ar@{-}[ddr]&L\ar@{-}[ddr]\ar@{-}[dr]&&M\ar@{-}[dl]\ar@{-}[d]\\
&&I\ar@{-}[dr]&J\ar@{-}[dll]\ar@{-}[d]\\
&F\ar@{-}[dl]\ar@{-}[dr]&G\ar@{-}[d]&H\ar@{-}[d]\\
C\ar@{-}[drr]&&D\ar@{-}[d]\ar@{-}[dr]&E\ar@{-}[d]\\
&&A\ar@{-}[d]&B\ar@{-}[dl]\\
&&\bot
}
\end{equation*}
}
\smallBox{
Discuss whether these values exist, and determine their values, if
they do:
  \begin{equation*}
  \begin{array}{lll}
  D\wedge E; & D \vee E; & (J \wedge L) \wedge K\\
  (L\wedge E) \wedge K; & L \wedge (E \wedge K); & C \wedge E\\
(C \vee D) \vee E; & \sup\{C, D, E\};& C\vee (D\vee E)\\
J\wedge(L\vee K); & (J\wedge L)\vee(J\wedge K); & C\vee G
  \end{array}
\end{equation*}
}
}

\Exercise{ex-covers}{Let $L$ be a lattice. An element $s\in L$ with
  $s\not=\bot$ is
  called \emph{\index{join irreducible}join irreducible} iff $s =
r\vee t$ implies $s = r$ or $s = t$. Element $t$
\emph{\index{covers}covers} element $s$ iff $s < t$, and if $s < v <
t$ for no element $v$. Show that if $L$ is a finite distributive
lattice, then $s$ is join irreducible iff $s$ covers exactly one
element.}
\Exercise{ex-finite-join-irred}{Let $P$ be a finite partially ordered
  set. Show that the down set $I\in {\cal D}(P)$ is join irreducible
  iff $I$ is a principal down set.}
\Exercise{ex-is-identif-join-irred}{
Identify the join-irreducible elements in $\PowerSet{S}$ for
$S\not=\emptyset$ and in the lattice of all open intervals $\bigl\{]a, b[
\mid a \leq b\bigr\}$, both ordered by inclusion. 
}

\Exercise{lattice-equiv-distr}{Show that in a lattice one distributive law implies
  the other one.
} 
\Exercise{ex-down-not-ideal}{Give an example for a down set in a lattice which is not an ideal.}
\Exercise{lattice-equiv-distr-1}{Show that in a distributive lattice
  $c\wedge x = c\wedge y$ and $c\vee x = x\vee y$ for some $c$ implies $x = y$.
}
\Exercise{ex-comm-subgroups}{
Let $G$ be a commutative group, written additively. Show that the
subgroups form a lattice under the subset relation.
} 
\Exercise{brouwerian-lattice}{ Assume that in lattice $L$ there exists
  for each $a, b\in L$ the relative \emph{\index{lattice!pseudo-complement}pseudo-complement} $b:a$ of $a$ in
  $b$; this is the largest element $x\in L$ such that $a\wedge x \leq
  b$. Show that a pseudo-complemented lattice is distributive.
  Furthermore, show that each Boolean algebra is
  pseudo-complemented. Lattices with pseudo-complements are called
  \emph{\index{lattice!Browerian}Browerian lattices.}} 
\Exercise{ex-complete-lattice}{
A lattice is called \emph{\index{lattice!complete}complete} iff it
contains suprema and infima for arbitrary subsets. Show that a bounded
partially ordered set $(L, \leq)$ is a complete lattice if the infimum
for arbitrary sets exists. Conclude that the set of all equivalence
relations on a set form a complete lattice under inclusion.}
\Exercise{ex-set-2}{Let
  $S\not=\emptyset$ be a set, $a\in S$. Compute
  for the Boolean algebra $B := \PowerSet{S}$ and the ideal 
  $I := \PowerSet{S\setminus\{a\}}$ the factor algebra
  $\Faktor{B}{I}$}
\Exercise{ex-spec-ord-char}{
Given a topological space $(X, \tau)$, the following conditions are equivalent for all $x, y\in X$.
\begin{enumerate}
\item\label{spec-ord-char:1} $\Closure{\{x\}} \subseteq \Closure{\{y\}}$.
\item\label{spec-ord-char:2} $x \in \Closure{\{y\}}$.
\item\label{spec-ord-char:3} $x\in U$ implies $y\in U$ for all open sets $U$.
\end{enumerate}
}
\Exercise{ex-set-3}{Characterize those ideals $I$ in a Boolean algebra
$B$ for which the factor algebra $\Faktor{B}{I}$ consists of exactly
two elements.}

 \Exercise{ex-set-alg}{%
Let $\emptyset\not={\cal A}\subseteq \PowerSet{S}$ be a finite family of sets with $S\in
{\cal A}$, say ${\cal A} = \{A_1, \dots, A_n\}$. Define 
$
A_T := \bigcap_{i\in T} A_i\cap\bigcap_{i\not\in T} S\setminus A_i
$
for $T\subseteq\{1,\dots, n\}$.
\begin{enumerate}
\item\label{ex-set-alg-item:1} ${\cal P} := \{A_T \mid \emptyset \not=
  T\subseteq\{1, \dots, n\}, A_T\not=\emptyset\}$ forms a partition of $S$.
\item \label{ex-set-alg-item:2} $\{\bigcup {\cal P}_0 \mid {\cal
    P}_0\subseteq{\cal P}\}$ is the smallest set algebra over $S$ which contains ${\cal A}$. 
\end{enumerate}
 }
\Exercise{ex-sequence-space}{
As in Example~\ref{coin-flip-1} on page~\pageref{coin-flip-1} let $X
:= \{0, 1\}^{\Nat}$ be the space of all infinite sequences. Put
\begin{equation*}
  {\cal C} := \{A \times \prod_{j>k}\{0, 1\} \mid k \in \Nat, A\in
  \PowerSet{\{0, 1\}^{k}}\}.
\end{equation*}
Show that ${\cal C}$ is an algebra. 
}

\Exercise{ex-sequence-space-cont}{
Let $X$ and ${\cal C}$ be as in Exercise~\ref{ex-sequence-space}. Show that
\begin{equation*}
  \mu\bigl(A \times \prod_{j>k}\{0, 1\}\bigr) := \frac{|A|}{2^{k}}
\end{equation*}
defines a  monotone and countably additive map $\mu: {\cal C}\to[0,
1]$ with $\mu(\emptyset) = 0$.
}

\Exercise{ex-subadditive-additive}{Show that a countably subadditive
  and monotone set function on a set algebra is additive.}

\newpage
\Subsection{Solutions}
\label{sec:solution-set}
\Solution{pairs-are-equal}{
$\langle a, b\rangle = \langle a', b'\rangle$ iff $\bigl\{\{a\}, \{a,
b\}\bigr\} = \bigl\{\{a'\}, \{a', b'\}\bigr\}$. If $\{a\} = \{a'\}$, we have $a = a'$, so that $\{a, b\} = \{a', b'\}$ implies $\{a, b\} = \{a, b'\}$. If $a = b$, this implies $a' = a = b'$, otherwise we obtain $b = b'$.
If, on the other hand,  $\{a\} = \{a', b'\}$, then
$a = a' = b'$, and then $\{a'\}= \{a, b\}$ implies $a = a' = b$.
In any case, $\langle a, b\rangle = \langle a', b'\rangle$ implies $a = a'$ and $b = b'$. The converse is trivial. 
}
\Solution{ex-prop-invers}{
If $f$ is injective, $X = \InvBild{f}{\Bild{f}{X})}$ for all
$X\subseteq A$, hence $f^{-1}$ is surjective. If $f$ is not injective,
we find $a, a'\in A$ with $a\not= a'$ and $b := f(a) = f(a')$. Hence
$\{a, a'\} = \InvBild{f}{\{b\}}$, so that neither $\{a\}$ nor $\{a'\}$
are inverse images under $f^{-1}$. Thus $f^{-1}$ is not surjective. 

If $f$ is surjective, $Y = \Bild{f}{\InvBild{f}{Y}}$ for all
$Y\subseteq B$, hence $\InvBild{f}{Y} = \InvBild{f}{Y'}$ implies 
$Y = \Bild{f}{\InvBild{f}{Y}} = \Bild{f}{\InvBild{f}{Y'}} =
Y'$. Conversely, if $f^{-1}$ is injective, we conclude from
$\InvBild{f}{\Bild{f}{A}} = \InvBild{f}{B}$ that $\Bild{f}{A} = B$.
}

\Solution{ex-divides}{If $p$ is prime, $p$ has no smaller divisor than
itself in $\Nat\setminus\{1\}$. If, on the other hand, $p$ is minimal
in $\Nat\setminus\{1\}$, and if $p$ can be written as $p = a \cdot b$
with $a, b\in \Nat\setminus\{1\}$, then both $a$ and $b$ are strictly
$\leq_d$-smaller than $p$.}
\Solution{ex-subset}{$B := \Nat\setminus\{1\}$ is an upper bound for
$A$, since $\{2\cdot n, 2\cdot n+1\}\subseteq B$ for all $n\in
\Nat$. $B$ is also the smallest upper bound,because each upper bound
must contain all $n\geq2$.}
\Solution{ex-knaster-tarski}{Let ${\cal Q} := \{X\in\PowerSet{S}\mid
  X\subseteq H(X)\}$. If $X\in {\cal Q}$, then $X\subseteq A$, hence
  $X\subseteq H(X)\subseteq H(A)$, so that $H(A)$ is an upper bound to
  ${\cal Q}$. This means $A\subseteq H(A)$, hence $A\in{\cal Q}$, so
  that $H(A)\subseteq A$. Hence $A$ is a fixed point of $H$. Let $Y$
  be another fixed point for ${\cal Q}$, then $Y\in{\cal Q}$, thus
  $Y\subseteq A$. 

This is a variant of the \emph{\index{Theorem!Knaster-Tarski}Knaster-Tarski Fixed point Theorem}.
}
\Solution{ex-banach-decomposition}{
Put $H(A) := X\setminus\Bild{g}{Y\setminus\Bild{f}{A}}$, then $H:
\PowerSet{X}\to\PowerSet{X}$ is order preserving. Let $X_{1} := A$ be the
maximal fixed point to $H$, and put $Y_{1} := \Bild{f}{A}$, then
$X\setminus A = \Bild{g}{Y\setminus Y_{1}}$, thus $X_{2} :=
X\setminus X_{1}$ and $Y_{2} := Y\setminus Y_{1}$ have the desired properties.
}
\Solution{ex-schroeder-bernstein-ancora}{
Let $f: A\to B$ and $g: B\to A$ be injective maps. Decompose $A =
A_{1}\cup A_{2}$ with disjoint $A_{1}, A_{2}$ and $B = B_{1}\cup
B_{2}$ with disjoint $B_{1}, B_{2}$ such that $\Bild{f}{A_{1}} =
B_{1}$ and $\Bild{g}{B_{2}} = A_{2}$ according to Exercise~\ref{ex-banach-decomposition}. Note that given $a\in A_{2}$
there exists  $b\in B_{2}$ with $g(b) = a$; because $g$ is injective,
$b$ is unique. Denote $b$ by
$g^{-1}(a)$. Then define $h: A\to B$ through
\begin{equation*}
  h(x) :=
  \begin{cases}
    f(x), & x \in A_{1},\\
g^{-1}(x), & x\in A_{2}.
  \end{cases}
\end{equation*}
It is plain that $h$ is injective. Let $b\in B_{1}$, then $b = f(a)$
for some $a\in A_{1}$; if $b\in B_{2}$, $g(b)\in A_{2}$, and $h(g(b)) =
b$ by construction. Thus $h$ is onto as well.
}
\Solution{ex-k-and-l}{Given $x\in \Natn$, there exists a unique $\langle a, b\rangle\in
\Natn\times\Natn$ so that $J(a, b) = x$. Define $K(x) := a$, then $K:
\Natn\to\Natn$ is well defined: if $J(a', b') = x$, then because $J$ is injective  $\langle a, b\rangle = \langle a',
b'\rangle$, in particular $a = a'$. It is also clear from the
construction that $K(x)\leq x$ holds. Since $K =
\pi_1\circ J^{-1}$ with $\pi_1$ as the projection to the first
component, and since both $\pi_1$ and $J^{-1}$ are surjective, $K$ is surjective. Similarly, $L(x) := b$ iff $J(a, b) =
x$ for some $a\in \Natn$. Then the desired properties for $L$ are
inferred in the same way. It is obvious that $J(K(x), L(x)) = x$
always holds.}
\Solution{pairs-are countable}{
Let 
$
N_{2} := \{\{a, b\}\mid a, b\in \Nat, a\not=b\}
$ 
all subsets of
$\Nat$ with exactly two elements, define 
$
P_{2} := \{\langle x, y\rangle \mid x, y\in \Nat, x < y\},
$
then $h: X \mapsto \langle\min
X, \max X\rangle$ is an injective map $N_{2}\to P_{2}$; the Cantor map
$J: P_{2}\to \Nat$ restricted to $P_{2}$ is injective as well by Proposition~\ref{cantor-bijects}, thus
$J \circ h: N_{2}\to \Nat$ is injective. On the other hand, $n \mapsto
\{n, n+1\}$ is injective $\Nat\to N_{2}$. 

Now let $k > 2$ and put
\begin{align*}
N_{k} & := \{A \subseteq \Nat \mid A\text{ has exactly $k$
  elements}\},\\
P_{k} & := \{\langle x_{1}, \dots, x_{k}\rangle\mid x_{1},
\dots, x_{k}\in \Nat, x_{1} < x_{2} < \dots < x_{k}\}.
\end{align*}
Define
$\gamma_{k}: N_{k}\to P_{k}$ inductively by $\gamma_{2} := h$, and
$\gamma_{k+1}(A) := \gamma_{k}(A\setminus\{\max A\})\frown \max
A$. Then $\gamma_{k}$ is an injection, so that
$\tau_{k}\circ\gamma_{k}: N_{k}\to \Nat$ is an injection, where
$\tau_{k}$ is defined in Proposition~\ref{nk-to-n}. On the other hand,
$n\mapsto \{n, n+1, \dots n+(k-1)\}$ is an injective map $\Nat\to
N_{k}$. 

By using Proposition~\ref{seq-to-nat-bij} it is shown that the set of
all finite subsets is countable without making use of {\AC}.
}
\Solution{ex-omega-is-ordinal}{Every element of $\omega_1$ is an
  ordinal, hence a set, thus~\ref{item:1} holds. Let
  $\alpha\in\omega_1$, then $\alpha = \{\zeta\mid \zeta <
  \alpha\}\subseteq\omega_1$, giving property~\ref{item:2}. Since each
  set of ordinals is well-ordered by $\in$, we see that
  property~\ref{item:3} holds. If finally $\emptyset\not= B \subseteq
  \omega_1$, then $B$ has a smallest element $\zeta$. If
  $\eta\in\zeta\cap B$, $\zeta$ would not be the smallest element of
  $B$, since $\eta\in\zeta$ would be strictly smaller. If $\omega_1$ is
  countable, we have $\omega_1\in\omega_1$, this contradicts
  Lemma~\ref{km-5}.}
\Solution{somen-ring}{The associate law both for $\ominus$ and $\cap$
  follows by direct computation. The neutral element of $\ominus$ is
  $\bot$, and $a\ominus a = \bot$, so each element is inverse to
  itself. $a \ominus b = b \ominus a$ is obvious. Thus $(B, \ominus)$
  is an Abelian group. Since $a\wedge(b\ominus c) = a\wedge b \ominus
  a\wedge c$ and $(b\ominus c)\wedge a = b\wedge a \ominus c\wedge a$,
we conclude that $(B, \ominus, \wedge)$ is a commutative ring.}
\Solution{ex-set-1}{Let ${\cal G}$ be a family of finite character,
  every chain ${\cal C}$ in ${\cal G}$ has an upper bound: We find for each finite set $\{x_1, \dots, x_n\}\subseteq
  \bigcup {\cal C}$ a set $C\in {\cal C}$ with $\{x_1, \dots,
  x_n\}\subseteq C$. Hence $\{x_1, \dots, x_n\}\subseteq \bigcup {\cal
    C}$, hence every finite subset of $\bigcup {\cal C}$ is in
  $\bigcup {\cal C}$. By Zorn's Lemma there exists a
  maximal element ${\cal M}$ for ${\cal G}$. Hence the assertion
  follows from Proposition~\ref{zl-impl-ac}.}
\Solution{ex-compl-prime-id}{Put $I := L\setminus F$.
  ``$\Rightarrow$''
  If $a\in I$ and
  $b\in I$, we have $a\in F$ or $b\in F$ is false, hence $a\vee b\in
  F$ is false, thus $a\vee b\in I$. $a\in I$ means
  $a\not\in F$, thus $b\leq a$ implies $b\not\in F$. Similarly,
  $a\wedge b\in I$ implies $a\in I$ and $b \in I$. Thus $I$ is an
  ideal. ``$\Leftarrow$'' is done in the same way.}

\Solution{ex-compact-countable}{Assume that $\gamma = x\in V$, then
  the assertion is trivial. Let the assertion be true for $\phi_{1}$
  and for $\phi_{2}$. 
 Then $v^{*}\models \phi_{1}\wedge\phi_{2}$ iff $v^{*}\models\phi_{1}$
 and $v^{*}\models\phi_{2}$. The induction hypothesis yields that this
 is equivalent to $\phi_{1}\in {\cal M}^{*}$ and $\phi_{2}\in{\cal
   M}^{*}$. Assume that $\phi_{1}\wedge\phi_{2}\not\in{\cal M}^{*}$
 then the construction of ${\cal M}^{*}$ implies that
 $\neg(\phi_{1}\wedge\phi_{2})\in{\cal M}^{*}$, since
 $\phi_{1}\wedge\phi_{2}$ is a formula, hence has to be considered at
 some time. But this implies $v^{*}(\phi_{1}\wedge\phi_{2}) = 0$,
 contradicting the assumption. The converse implication is
 trivial. Now assume that the assertion is true for formula $\phi$.
 If $v^{*}\models \neg\phi$, then $v^{*}(\phi) = 0$, hence
 $\phi\not\in{\cal M}^{*}$, thus, again by construction,
 $\neg\phi\in{\cal M}^{*}$.
}
\Solution{ex-ultrafilter-finite}{
Assume $A := \{s_{1}, \dots, s_{n}\}\in{\cal F}$ for some $n\in \Nat$. If
$\{s_{i}\}\not\in{\cal F}$ for all $i, 1 \leq i \leq n$, then
$S\setminus\{s_{i}\}\in{\cal F}$ for all $i$, hence $S\setminus
A\in{\cal F}$, a contradiction. 
}
\Solution{ex-is-lattice}{
  \begin{itemize}
    \item $D\wedge E = B; D \vee E = L; (J \wedge L) \wedge K = D
      \wedge K = D.$
\item $(L\wedge E) \wedge K = B\wedge B; L \wedge (E \wedge K) =
  L\wedge B = B;  C \wedge E = \bot.$
  \item $(C \vee D) \vee E = F\vee E = J;\ \sup\{C, D, E\} = \top; C\vee
    (D\vee E) = C\vee L = \top.$
    \item $J\wedge(L\vee K)= J\wedge \top = J; (J\wedge L)\vee(J\wedge
      K) = H\vee F = J; C\vee G = K.$
  \end{itemize}
}
\Solution{ex-covers}{
If $L$ is finite, $b = \sup\{a \in L \mid  a \leq b\}$. Thus if $s$ is
join irreducible, we can find exactly one element covering $s$. The
converse is trivial.
}
\Solution{ex-finite-join-irred}{
Assume $I := \{s \in P \mid s \leq t\} = I_{1}\cup I_{2}$ with $I_{1},
I_{2}\in {\cal D}(P)$. Assume $t\in I_{1}$, and let $s\in I_{2}$, then
$s\leq t$, hence $s\in I_{1}$, thus $I_{2}\subseteq
I_{1}$. Conversely, if $I = \bigcup_{t\in I} \{s \in P \mid s \leq
t\}$ is join irreducible, we find $t\in I$ such that $I = \{s \in P \mid s \leq
t\}$, since $P$ is finite. 
}

\Solution{ex-is-identif-join-irred}{
The singleton sets $\{x\}$ are join-irreducible in $\PowerSet{S}$. The
lattice of all open intervals in $\Real$ does not have any irreducible
elements.
}
\Solution{lattice-equiv-distr}{Assume that 
$
(x\vee y)\wedge z = (z\wedge y) \vee (z\wedge x)
$
holds, then
\begin{align*}
  (x\vee y)\wedge (x\vee z) & = \bigl((x\vee y)\wedge
  x\bigr)\vee\bigl((x\vee y)\wedge z\bigr)\\
& = x \vee \bigl((x\vee y)\wedge z\bigr)\\
& = x \vee \bigl((z\wedge y) \vee (z\wedge x)\bigr)\\
& = x \vee (y\wedge z)
\end{align*}
}
\Solution{ex-down-not-ideal}{Let $\langle x, y\rangle \leq \langle x', y'\rangle$ iff $x \leq x'$ and $y \leq y'$ for $\langle x, y\rangle, \langle x', y'\rangle\in \pReal\times\pReal$. Then $\{\langle 0, x\rangle \mid x \in \pReal\}\cup\{\langle x, 0\rangle\mid x \in \pReal\}$ is a down set. It is not an ideal, since it contains $\langle 0, 1\rangle$ and $\langle 1, 0\rangle$, but not $\langle1, 1\rangle = \langle 0, 1\rangle\vee\langle 1, 0\rangle$.}
\Solution{lattice-equiv-distr-1}{
  \begin{align*}
    x &= x\wedge(c\vee x)
&& = x\wedge (c\vee y)
&& = (x\wedge c) \vee (x\wedge y)\\
& = (y\wedge c) \vee (x\wedge y)
&& = y\wedge(c\vee x) 
&& = y\wedge(c\vee y) \\
& = y
  \end{align*}
}
\Solution{ex-comm-subgroups}{
The intersection $K\cap L$ of two subgroups $K$
and $L$ is a subgroup again. Put 
\begin{equation*}
  K+L := \{a + b \mid a\in K, b \in L\}.
\end{equation*}
This is a subgroup of $G$: if $g = a+b, g' = a'+b'\in K+L$, then $g -
g' = (a-a')+(b-b')\in K+L$. If is clear that $K\subseteq K+L,
L\subseteq K+L$, on the other hand, if $H\subseteq G$ is a subgroup
with $K\subseteq H, L\subseteq H$, then $K+L\subseteq H$, so that
$K+L$ is indeed the smallest subgroup containing $K$ and $L$. Hence
define
\begin{align*}
  K\wedge L & := K\cap L,\\
K\vee L & := K+L,
\end{align*}
then the set of all subgroups is a lattice. 

It could be noted that a very similar argument applies to the normal subgroups in
an arbitrary group (a subgroup $H$ is called \emph{\index{normal
    subgroup}normal} iff $\forall a\in G: aH = Ha$ holds).
}

\Solution{brouwerian-lattice}{Put 
$
d := (a\wedge b)\vee (a\wedge c)
$
and consider $d:a$. We have $a\wedge b \leq d$ and $a\wedge c\leq d$,
hence $b\leq d:a$ and $c \leq d:a$. Thus $b\wedge c \leq d:a$, which
implies $a\wedge(b\vee c)\leq a\wedge (d:a)\leq d = (a\wedge b)\vee
(a\wedge c)$. Because 
$(a\wedge b)\vee (a\wedge c) \leq a\wedge(b\vee c)$ always holds,
distributivity is established (see
Exercise~\ref{lattice-equiv-distr}). 

Define the pseudo-complement in a Boolean algebra through $b:a :=
-a\vee b$, then it is easily seen that the defining properties hold.
}
\Solution{ex-complete-lattice}{
The first assertion follows from the observation that 
\begin{equation*}
  \sup A = \inf\{b\in L\mid b\text{ is an upper bound to }A\}.
\end{equation*}
The set of all equivalence relations on a set $X$ has $\{\langle
x, x\rangle \mid x\in X\}$ as the smallest and $X\times X$ as the
largest element. 
The second assertion follows from the first one after having seen that
the intersection of an arbitrary set of equivalence relations on $X$
is an equivalence relation and its greatest lower bound. 
}
\Solution{ex-set-2}{Given $X, Y\in B$, we have
  $\isEquiv{X}{Y}{\sim_I}$ iff either both $X$ and $Y$ contain $a$ or
  both don't. Consequently, $\Faktor{B}{I} = \{\top, \bot\}$.}
\Solution{ex-spec-ord-char}{
\labelImpl{spec-ord-char:1}{spec-ord-char:2}: This follows from $x\in \Closure{\{x\}}$.

\labelImpl{spec-ord-char:2}{spec-ord-char:3}: If $x\in U$, but $y\not\in U$ for some open set $U$, we would have $\Closure{\{y\}}\subseteq X\setminus U$, contradicting the assumption.

\labelImpl{spec-ord-char:3}{spec-ord-char:1}: Let $y\in F$ for some closed set $F$, then $y\not\in X\setminus F$, hence $x\not\in X\setminus F$ by assumption, thus each closed set which contains $y$ contains $x$ as well. This implies  $\Closure{\{x\}} \subseteq \Closure{\{y\}}$

This indicates a first connection of topology and order. Define $x\leq
y$ iff $x\in \Closure{\{y\}}$, then this relation is certainly
reflexive and transitive; if $(X, \tau)$ has the additional property
that given $x\not= y$ there exists an open set containing exactly one
of them, then $\leq$ is also antisymmetric.  }
\Solution{ex-set-3}{$\Faktor{B}{I} = \{\top_I, \bot_I\}$ iff $I$ is a
  prime ideal. ``$\Rightarrow$'' A prime ideal $K$ extending $I$ is $K
  =\{x\in B \mid \Klasse{x}{\sim_I} \in \bot_I\} = I$ according to
  Theorem~\ref{extended-prime-ideal}. ``$\Leftarrow$'' A prime ideal
  is maximal by Lemma~\ref{prime-equiv-max}, hence the factor algebra
  must be trivial by Theorem~\ref{extended-prime-ideal}.}
\Solution{ex-set-alg}{The $A_T$ are mutually disjoint by construction,
  and their union is $S$, since $S\in {\cal A}$. Put $ {\cal C} :=
  \{\bigcup {\cal P}_0 \mid {\cal P}_0\subseteq{\cal P}\}, $ then
  $\emptyset\in{\cal C}$, and $S\setminus (\bigcup {\cal P}_0) =
  \bigcup \PowerSet{\{1, \dots, n\}}\setminus{\cal P}_0$ computes the
  complement of a set in ${\cal C}$. Since $\bigcup{\cal
    P}_0\cup\bigcup{\cal P}_1 = \bigcup ({\cal P}_0\cup{\cal P}_1)$
  and $\bigcup{\cal P}_0\cap\bigcup{\cal P}_1 = \bigcup ({\cal
    P}_0\cap{\cal P}_1)$, ${\cal C}$ is closed under the algebra
  operations. Since the algebra generated by ${\cal A}$ must contain
  ${\cal C}$, the assertion follows.  } \Solution{ex-sequence-space}{
  It is clear that $X\in {\cal C}$ and $\emptyset\in{\cal C}$
  hold. Let $D, E\in{\cal C}$, then $D = A \times \prod_{j>k}\{0, 1\},
  E = B \times \prod_{j>\ell}\{0, 1\}$ with $A\subseteq\{0, 1\}^{k},
  B\subseteq\{0, 1\}^{\ell}$. Assume $k\geq \ell$ (if $k\leq \ell$ the
  argument s the same), then $E = B' \times \prod_{j>k}\{0, 1\} $ with
  $B' := B\times\{0, 1\}^{k-\ell}\subseteq\{0, 1\}^{k}$, thus $ D\cap
  E = (A\cap B')\times\prod_{j>k}\{0, 1\}\in{\cal C}.  $ Hence ${\cal
    C}$ is closed under finite intersections. Since $ X\setminus
  (A\times\prod_{j>k}\{0, 1\} = \bigl(\{0, 1\}^{k}\setminus
  A\bigr)\times\prod_{j>k}\{0, 1\} $ for $A\subseteq \{0, 1\}^{k}$,
  ${\cal C}$ is closed under complementation as well.  }
\Solution{ex-sequence-space-cont}{ Assume that
\begin{equation*}
  A \times \prod_{j>k}\{0, 1\} =  A' \times \prod_{j>\ell}\{0, 1\},
\end{equation*}
with $k\leq\ell$, then 
$
A' = A\times\{0, 1\}^{\ell-k},
$
so that 
\begin{equation*}
\frac{|A'|}{2^{\ell}} =
\frac{|A|}{2^{k}}\cdot\frac{2^{\ell-k}}{2^{\ell-k}} =
\frac{|A|}{2^{k}}.
\end{equation*}
Thus $\mu$ is well-defined. $\mu(\emptyset) = 0$ is trivial; since
$A\subseteq B$ implies $|A| \leq |B|$ and $|A\cup B| = |A| + |B|$, if
$A$ and $B$ are disjoint, $\mu$ is monotone and additive. Countable
additivity is vacuously satisfied. 
}
\Solution{ex-subadditive-additive}{%
Let ${\cal C}$ be a set algebra, $\mu: {\cal C}\to[0, \infty]$ monotone and
additive. If $\bigcup_{n\in\Nat} A_n =: A\in {\cal C}$, then 
$
A = \bigcup_{n\in\Nat} B_n
$
with $B_1 := A_1$ and $B_{n+1} := A_{n+1}\setminus(\bigcup_{i=1}^n
A_i)$ (this is sometimes called the \emph{first entrance trick}). The
$B_n$ are mutually disjoint, and $B_n\in {\cal C}$, because ${\cal C}$
is an algebra. Thus
$
\mu(A) = \sum_{i=1}^\infty \mu(B_i) \leq \sum_{i=1}^\infty \mu(A_i).
$
}



\newpage
\addcontentsline{toc}{section}{References}

\newpage\addcontentsline{toc}{section}{Index}\printindex
\end{document}